\documentstyle[aaspp4,natbib,lscape]{article}
\bibpunct[]{(}{)}{;}{a}{}{,}

\received{25 April 25 1999}
\accepted{21 May 1999}

\lefthead{Heger, Langer \& Woosley}
\righthead{Presupernova Evolution of Rotating Massive Stars I}

\newcommand{\Lsun}{{\mathrm{L}_{\odot}}}
\newcommand{\Msun}{{\mathrm{M}_{\odot}}}
\newcommand{\MsunB}{{\mathbf{M_{\odot}}}}
\newcommand{\Rsun}{{\mathrm{R}_{\odot}}}
\newcommand{\clight}{{\mathrm{c}}} 
\newcommand{\cm}{{\mathrm{cm}}}
\newcommand{\km}{{\mathrm{km}}} 
\newcommand{\Sec}{{\mathrm{s}}}
\newcommand{\g}{{\mathrm{g}}} 
\newcommand{\rad}{{\mathrm{rad}}}
\newcommand{\yr}{{\mathrm{yr}}}

\newcommand{\K}{{\mathrm{K}}}
\newcommand{\erg}{{\mathrm{erg}}}

\newcommand{\MeV}{{\mathrm{MeV}}}

\newcommand{\rads}{{\rad\,\Sec^{-1}}}
\newcommand{\kms}{{\km\,\Sec^{-1}}}
\newcommand{\erggs}{{\erg\,\g^{-1}\,\Sec^{-1}}}
\newcommand{\ergs}{{\erg\,\Sec}} 

\newcommand{\junit}{{\cm^2\,\Sec^{-1}}}

\newcommand{\gccm}{{\g\,\cm^{-3}}} 
\newcommand{\ms}{{\mathrm{m}\Sec}}

\newcommand{\D}{{\mathrm d}}
\newcommand{\DD}{{\,\D\!\!\;}}
\newcommand{\Brak}[1]{{\left({#1}\right)}}
\newcommand{\SBrak}[1]{{\left[{#1}\right]}}
\newcommand{\CBrak}[1]{{\left\{{#1}\right\}}}
\newcommand{\DxDy}[2]{{\frac{\D{#1}}{\D{#2}}}}
\newcommand{\DxDyInd}[3]{{\Brak{\frac{\D{#1}}{\D{#2}}}_{\!\!{#3}}}}
\newcommand{\dxdy}[2]{{\frac{\partial{#1}}{\partial{#2}}}}
\newcommand{\dxdycz}[3]{{\Brak{\frac{\partial{#1}}{\partial{#2}}}_{\!\!{#3}}}}

\newcommand{\av}[1]{{\langle{#1}\rangle}}
\newcommand{\Frac}[2]{{\Brak{#1}/\Brak{#2}}}
\newcommand{\abs}[1]{{\left|{#1}\right|}}
\newcommand{\Min}[1]{{\min\CBrak{#1}}}
\newcommand{\Max}[1]{{\max\CBrak{#1}}}

\newcommand{\At}[2]{{\left.{#1}_{_{\phantom{|}}}\right|_{#2}}}

\newcommand{\isofont}[1]{{\mathsf{#1}}}
\newcommand{\isomass}[1]{{$\isofont{^{#1}}$}}
\newcommand{\isocharge}[1]{{$\isofont{_{#1}}$}}
\newcommand{\isotope}[3]{{\isocharge{#1}\isomass{#2}$\isofont{#3}$}}

\newcommand{\I}[2]{{\isotope{}{#1}{#2}}}
\newcommand{\El}[1]{{\I{}{#1}}}

\newcommand{\E}[1]{{\cdot10^{#1}}} 
\newcommand{\Ep}[1]{{10^{#1}}}
\newcommand{\IT}[2]{\isomass{#1}&\El{#2}} 

\renewcommand{\vec}[1]{{\mathbf #1}} 

\newcommand{\simgr}{\gtrsim}
\newcommand{\simle}{\lesssim}

\newcommand{\kB}{{k_{\mathrm B}}}

\newcommand{\CO}{CO}

\newcommand{\LP}{{L_{\mathrm{P}}}} 
\newcommand{\VP}{{V_{\!\mathrm{P}}}}
\newcommand{\SP}{{S_{\mathrm{P}}}} 
\newcommand{\rP}{{r_{\!\mathrm{P}}}}
\newcommand{\mP}{{m_{\mathrm{P}}}} 
\newcommand{\fP}{{f_{\mathrm{P}}}}
\newcommand{\fT}{{f_{\mathrm{T}}}}
\newcommand{\fmu}{{f_{\!\mathrm{\mu}}}}

\newcommand{\Rec}{{R_{\mathrm{e,c}}}}
\newcommand{\Ri}{{R_{\mathrm{i}}}}
\newcommand{\Ric}{{R_{\mathrm{i,c}}}}
\newcommand{\tauD}{{\tau_{\mathrm{D}}}}

\newcommand{\taudyn}{{\tau_{\mathrm{dyn}}}}

\newcommand{\ri}{{r_{\mathrm{i}}}} 
\newcommand{\ro}{{r_{\!\mathrm{o}}}}
\newcommand{\Dconv}{{D_{\mathrm{conv}}}}
\newcommand{\Dsem}{{D_{\mathrm{sem}}}}
\newcommand{\Drad}{{D_{\mathrm{rad}}}}
\newcommand{\vsem}{{v_{\mathrm{sem}}}}
\newcommand{\Dsemx}{{{D'}_{\mathrm{\!\!sem}}}}
\newcommand{\dinst}{{d_{\mathrm{inst}}}}
\newcommand{\HP}{{H_{\mathrm{P}}}} 
\newcommand{\HT}{{H_{\mathrm{T}}}}
\newcommand{\Nad}{{\nabla_{\!\mathrm{ad}}}}
\newcommand{\Nmu}{{\nabla_{\!\mathrm{\mu}}}}
\newcommand{\DDSI}{{D_{\mathrm{DSI}}}}
\newcommand{\RSH}{{R_{\mathrm{SH}}}}
\newcommand{\DSHI}{{D_{\mathrm{SHI}}}}
\newcommand{\asem}{{\alpha_{\mathrm{sem}}}}
\newcommand{\Pra}{{{\mathcal{P}_{\!\mathrm{r}}}}}
\newcommand{\Risa}{{R_{\mathrm{is,1}}}}
\newcommand{\Risb}{{R_{\mathrm{is,2}}}}
\newcommand{\cV}{{c_{\mathrm{V}}}} 
\newcommand{\cP}{{c_{\mathrm{P}}}}
\newcommand{\mup}{{\mu_{\mathrm{p}}}}
\newcommand{\mur}{{\mu_{\mathrm{r}}}}
\newcommand{\mi}{{m_{\mathrm{i}}}} 
\newcommand{\Zi}{{Z_{\mathrm{i}}}}
\newcommand{\vSSI}{{v_{\mathrm{SSI}}}}
\newcommand{\cs}{{c_{\mathrm{s}}}}
\newcommand{\HvSSI}{{H_{\mathrm{v,SSI}}}}
\newcommand{\DSSI}{{D_{\mathrm{SSI}}}}

\newcommand{\epsnu}{{\varepsilon_{\mathrm{\nu}}}}
\newcommand{\epsnuc}{{\varepsilon_{\mathrm{n}}}}

\newcommand{\vESO}{{v_{\mathrm{e}}}}
\newcommand{\vES}{{v_{\mathrm{ES}}}}
\newcommand{\vmu}{{v_{\mathrm{\mu}}}}
\newcommand{\tauKH}{{\tau_{\mathrm{KH}}}}

\newcommand{\tauKHx}{{\tau^*_{\mathrm{KH}}}}
\newcommand{\tauES}{{\tau_{\mathrm{ES}}}}
\newcommand{\tauB}{{\tau_{\mathrm{burn}}}}
\newcommand{\HvES}{{H_{\mathrm{v,ES}}}}
\newcommand{\DES}{{D_{\mathrm{ES}}}}
\newcommand{\vGSFO}{{v_{\mathrm{g}}}}
\newcommand{\vGSF}{{v_{\mathrm{GSF}}}}
\newcommand{\Hj}{{H_{\mathrm{j}}}}
\newcommand{\DGSF}{{D_{\mathrm{GSF}}}}
\newcommand{\HvGSF}{{H_{\mathrm{v,GSF}}}}
\newcommand{\nue}{{$\nu_{\mathrm{e}}$}}

\newcommand{\betap}{{$\beta^{+}$}}

\newcommand{\MZAMS}{{M_{\mathrm{ZAMS}}}}
 
\newcommand{\Jm}{{J\Brak{m}}}
\newcommand{\vZAMS}{{v_{\mathrm{ZAMS}}}}
\newcommand{\epot}{{\phi_{\mathrm{grav}}}}
\newcommand{\erot}{{e_{\mathrm{rot}}}}
\newcommand{\Erot}{{E_{\mathrm{rot}}}}
\newcommand{\Epot}{{E_{\mathrm{pot}}}}
\newcommand{\rhoc}{{\rho_{\mathrm{c}}}}
\newcommand{\Mi}{{M_{\mathrm{initial}}}}
\newcommand{\MHe}{{M_{\!\isofont{He}}}}
\newcommand{\MCO}{{M_{\!\isofont{\CO}}}}
\newcommand{\MFe}{{M_{\!\isofont{Fe}}}}

\newcommand{\Mf}{{M_{\mathrm{final}}}}
\newcommand{\Ji}{{J_{\mathrm{initial}}}}
\newcommand{\JHe}{{J_{\!\isofont{He}}}}
\newcommand{\JCO}{{J_{\!\isofont{\CO}}}}
\newcommand{\JFe}{{J_{\!\isofont{Fe}}}}

\newcommand{\jFe}{{j_{\!\isofont{Fe}}}}
\newcommand{\Jf}{{J_{\mathrm{final}}}}
\newcommand{\fc}{{f_{\!\mathrm{c}}}} 

\newcommand{\aMLT}{{\alpha_{\mathrm{MLT}}}}

\newcommand{\vconv}{{v_{\mathrm{conv}}}}
\newcommand{\vcrit}{{v_{\mathrm{crit}}}}

\newcommand{\Mdot}{{\dot{M}}}

\newcommand{\Jdot}{{\dot{J}}}

\newcommand{\jsurf}{{j_{\mathrm{surf}}}}

\newcommand{\wk}{{\omega_{\mathrm{Kep}}}} 
\newcommand{\N}{{\nabla}}

\newcommand{\Rbp}{{(\betap\nue)}}

\newcommand{\lSect}[1]{{\label{sec:#1}}}
\newcommand{\lFig}[1]{{\label{fig:#1}}}
\newcommand{\lEq}[1]{{\label{eq:#1}}}
\newcommand{\lTab}[1]{{\label{tab:#1}}}

\newcommand{\pFig}[1]{{\placefigure{fig:#1}}}
\newcommand{\pTab}[1]{{\placetable{tab:#1}}}

\newcommand{\Tabff}[1]{{\ref{tab:#1}}}
\newcommand{\Tab}[1]{{Table~\Tabff{#1}}}

\newcommand{\FigPan}[1]{{\footnotesize{#1}}}
\newcommand{\FIGFF}[2]{{\ref{fig:#2}\FigPan{#1}}}

\newcommand{\FIG}[2]{{Fig.~\FIGFF{#1}{#2}}}
\newcommand{\Fig}[1]{{\FIG{}{#1}}}

\newcommand{\FIGS}[2]{{Figs.~\FIGFF{#1}{#2}}}

\newcommand{\FigBB}[4]{{\FIGS{#1}{#2} and \FIGFF{#3}{#4}}}
\newcommand{\FigBb}[3]{{\FIGS{#1}{#2} and \FigPan{#3}}}
\newcommand{\FigB}[2]{{\FigBB{}{#1}{}{#2}}}
\newcommand{\FigCC}[6]{{\FIGS{#1}{#2}, \FIGFF{#3}{#4}, and \FIGFF{#5}{#6}}}

\newcommand{\FigCCc}[5]{{\FIGS{#1}{#2}, \FIGFF{#3}{#4} and \FigPan{#5}}}
\newcommand{\FigC}[3]{{\FigCC{}{#1}{}{#2}{}{#3}}}

\newcommand{\FigDdd}[6]{{\FIGS{#1}{#2} and \FigPan{#3}, \FIGFF{#4}{#5} and \FigPan{#6}}}

\newcommand{\Sectff}[1]{{\ref{sec:#1}}}

\newcommand{\App}[1]{{Appendix~\Sectff{#1}}}

\newcommand{\Sect}[1]{{Sect.~\Sectff{#1}}}

\newcommand{\Sects}[1]{{Sections~\Sectff{#1}}}
\newcommand{\Eqref}[1]{{\ref{eq:#1}}}

\newcommand{\Eqx}[1]{({Eq.~\Eqref{#1}})}
\newcommand{\Eqff}[1]{{(\Eqref{#1})}}
\newcommand{\Eq}[1]{{Eq.~\Eqff{#1}}}
\newcommand{\Eqs}[1]{{Eqs.~\Eqff{#1}}}

\newcommand{\Mod}[1]{{\texttt{#1}}}
\newcommand{\ModA}[1]{{Model   \Mod{#1}}}
\newcommand{\ModB}[2]{{Models  \Mod{#1} and \Mod{#2}}}
\newcommand{\ModC}[3]{{Models  \Mod{#1}, \Mod{#2}, and \Mod{#3}}}
\newcommand{\ModD}[4]{{Models  \Mod{#1}, \Mod{#2}, \Mod{#3}, and \Mod{#4}}}
\newcommand{\ModE}[5]{{Models  \Mod{#1}, \Mod{#2}, \Mod{#3}, \Mod{#4}, and \Mod{#5}}}
\newcommand{\ModF}[6]{{Models  \Mod{#1}, \Mod{#2}, \Mod{#3}, \Mod{#4}, \Mod{#5}, and \Mod{#6}}}

\newcommand{\ModCx}[3]{{(Models \Mod{#1}, \Mod{#2}, and \Mod{#3})}}

\newcommand{\pan}[1]{{\textbf{#1}}}
\newcommand{\Pan}[1]{{Panel~\pan{#1}}}
\newcommand{\Pans}[1]{{Panels~\pan{#1}}}
\newcommand{\PanRange}[2]{{Panels~\pan{#1}--\pan{#2}}}
\newcommand{\PanB}[2]{{\Pans{#1} and \pan{#2}}}
\newcommand{\PanC}[3]{{\Pans{#1}, \pan{#2}, and \pan{#3}}}

\newcommand{\Cite}[1]{{\cite[][]{#1}}}
\newcommand{\CiteA}[2]{{\cite[][#1]{#2}}}
\newcommand{\CITE}[2]{{\cite[#1][]{#2}}}

\newcommand{\CS}[2]{{\cite[see][#1]{#2}}}
\newcommand{\Csa}[1]{{\cite[see also][]{#1}}}
\newcommand{\Ce}[1]{{\cite[e.g.,][]{#1}}}
\newcommand{\Cse}[1]{{\cite[see, e.g.,][]{#1}}}

\begin{document}

\title{Presupernova Evolution of Rotating Massive Stars I: \\
       Numerical Method and \\
       Evolution of the Internal Stellar Structure}

\author{A.~Heger\altaffilmark{1,2},
        N.~Langer\altaffilmark{2,3}, \and
        S.~E.~Woosley\altaffilmark{1,2}}

\altaffiltext{1}{Astronomy and Astrophysics Department, 
    University of California, Santa Cruz, CA~95064}
\altaffiltext{2}{Max-Planck-Institut f\"ur Astrophysik, 
    Karl-Schawarzschild-Stra{\ss}e~1, 85740~Garching, Germany}
\altaffiltext{3}{Institut f\"ur Physik, 
    Am Neuen Palais~10, 14469~Potsdam, Germany}

\begin{abstract}
The evolution of rotating stars with zero-age main sequence (ZAMS)
masses in the range $8\,\Msun$ to $25\,\Msun$ is followed through all
stages of stable evolution.  The initial angular momentum is chosen
such that the star's equatorial rotational velocity on the ZAMS ranges
from zero to $\sim\!70\,\%$ of break-up.  The stars rotate rigidly on
the ZAMS as a consequence of angular momentum redistribution during
the pre-main sequence evolution.  Redistribution of angular momentum
and chemical species are then followed as a consequence of
Eddington-Sweet circulation, the Solberg-H{\o}iland instability, the
Goldreich-Schubert-Fricke instability, and secular and dynamic shear
instability.  The effects of the centrifugal force on the stellar
structure are included.  Convectively unstable zones are assumed to
tend towards rigid rotation and uncertain mixing efficiencies are
gauged by observations.  We find, as noted in previous work, that
rotation increases the helium core masses and enriches the stellar
envelopes with products of hydrogen burning.  We determine, for the
first time, the angular momentum distribution in typical presupernova
stars along with their detailed chemical structure.  Angular momentum
loss due to (non-magnetic) stellar winds and the redistribution of
angular momentum during core hydrogen burning are of crucial
importance for the specific angular momentum of the core.  Neglecting
magnetic fields, we find angular momentum transport from the core to
the envelope to be unimportant after core helium burning.  We obtain
specific angular momenta for the iron core and overlaying material of
$\Ep{16}\ldots\Ep{17}\,\ergs$.  These values are insensitive to the
initial angular momentum and to uncertainties in the efficiencies of
rotational mixing.  They are small enough to avoid triaxial
deformations of the iron core before it collapses, but could lead to
neutron stars which rotate close to break-up.  They are also in the
range required for the collapsar model of gamma-ray bursts.  The
apparent discrepancy with the measured rotation rates of young pulsars
is discussed.
\end{abstract}

\keywords{stars: rotation, supernovae, nucleosynthesis -- pulsars --
 hydrodynamics -- methods: numerical}


\section{Introduction}


The quantitative theory of stellar structure is more than 100 years
old {\Cse{Emd07}} and our understanding of the stellar interior has
improved dramatically during this time, especially since it became
possible to construct detailed stellar models with the help of
computers in the 1950's.  However, even today, our understanding of
many observable properties of massive stars ($\MZAMS \simgr 8 \Msun$,
$\log L/\Lsun \simgr 4$) remains rudimentary.  

Aside from comparatively minor uncertainties remaining in the
opacities and nuclear physics, the major frontiers in the study of
stars, and indeed stellar evolution in general, are proper treatments
of convection, mass loss, and rotation.  This paper is the first in a
series concerning the effects of rotation and angular momentum
transport on the evolution of stars massive enough that a single one
can become a supernova ($M \simgr 8\,\Msun$).

The first to recognize the importance of rotation for celestial bodies
was Sir Isaac Newton.  Early studies of rotating, self-gravitating,
incompressible fluids were carried out by McLaurin, Jakobi,
Poincar\'e, and Schwarzschild.  Additional important contributions to
the numerical treatment of rotating stars were provided by
{\cite{KT70}}.  {\cite{KMT70}} performed calculations taking these
effects into account using a simple model for angular momentum
transport.  Studies with artificial rotation laws were carried out by
{\cite{ES76}}.  In their pioneering work, {\cite{ES78}} considered
several rotationally induced instabilities, made order-of-magnitude
estimates for their efficiencies, and performed time-dependent stellar
evolution calculations of rotating massive stars up to the ignition of
carbon burning.  Later, {\cite{pin89}} introduced a parameterization
of the poorly known efficiencies of the rotationally induced transport
processes of {\cite{ES78}} and gauged them to solar models.  The
formalism we shall employ here is based largely upon these two works.
We differ, however, in using more recent data to calibrate the
uncertain efficiencies for angular momentum and composition transport
in this formalism and especially in following the stars past carbon
burning, all the way to the presupernova state.

Our formalism (discussed in detail, in $\S$2) is relatively simple
compared to others used in recent studies of rotation
during hydrogen and helium burning, e.g., {\cite{Lan92}},
{\cite{Den94}}, {\cite{ery94}}, {\cite{CZ92}}, {\cite{Zah92}},
{\cite{USS96}}, {\cite{tal97}}, {\cite{Meyn97}}, and {\cite{MZ98}},
but easier to understand and implement, and more easily extrapolable
to the late stages of stellar evolution.  Indeed our poor understanding,
especially during the late stages of stellar evolution, of both convection
and possible modifications to angular momentum transport by magnetic
fields (not considered in the present work nor in the papers cited
above) suggests that it is worth trying something simple first.

Most of the rotation physics described in {\Sect{rotphys}} can already
be found, with slight modifications, in previous papers.  However,
since this is the first in a series of papers, it will facilitate our
presentation to have all the relevant equations collected in one
place.  We also correct several errors.  In previous publications,
e.g. in the equation for the secular shear instability, and cast the
results in a consistent notation.

Following a summary of how we model various rotationally induced
instabilities (\Sect{rotphys}) and a discussion of the uncertain
parameters of the model (\Sect{calib}), we discuss the implementation
of this physics in the stellar codes in {\Sect{NumSolve}} and give an
overview of the initial models in {\Sect{InitialCond}}.  The evolution
during hydrogen burning and helium burning is discussed in
{\Sects{cHburn}} and {\Sectff{cHeBurn}}, respectively.  In
{\Sect{CmpOther}} we compare our results to the works of other
authors.  The late evolution is discussed in {\Sect{LateEv}} and the
final angular momentum distribution at the presupernova stage is given
in {\Sect{j:preSN}}.  Its implications are discussed in
{\Sect{youngPSR}} and  a summary and our conclusions are given in
{\Sect{SumConcl}}.

Discussion of the details of observable parameters (evolution in the
HR diagram, surface abundances, lifetimes) and presupernova
nucleosynthesis are deferred to future papers.


\section{Rotation and Mixing in Massive Stars}
\lSect{rotphys}

\subsection{Modification to the stellar structure equations}
\lSect{ModRot}

In rotating stars, centrifugal forces act on the matter and lead to
deviations from spherical symmetry.  For slow to moderate rotation
these deformations remain rotationally symmetric {\Cite{Tas78}}.  Only
if the rotational energy exceeds a notable fraction of the binding
energy of the star does genuine triaxial deformation result.

In this work we consider only the case of ``slow'' rotation, i.e.,
where no triaxial deformations are expected.  Some stars may reach
``critical'' rotation velocity ({\Sect{Omega}}) at their surfaces
during brief stages of their evolution {\Cite{HL98}}.  However, except
for possibly modifying the mass loss rate ({\Sect{Omega}}), this
affects only the very outermost layers and is not expected to have a
big influence on the results of this paper.

Even for slow rotation, the shapes of surfaces of constant pressure,
constant density, and constant temperature are affected by the
centrifugal potential and thus deviate from spherical symmetry.  The
momentum equation and the energy transport equation for spherically
symmetric stars must be modified to take this effect into account.

In this work, the centrifugal force is included following
{\cite{KT70}} in the approximation of {\cite{ES76}} and applied to the
hydrodynamic stellar structure equations {\Cite{Fli93}}.  In this
approach, mass shells correspond to isobars instead of spherical
shells.  Corrections are applied to the acceleration and the radiative
temperature gradient.  According to {\cite{Zah75}}, {\cite{CZ92}}, and
{\cite{Zah92}}, anisotropic turbulence acts much stronger on isobars
than in the perpendicular direction.  This enforces ``shellular''
rotation rather than cylindrical rotation {\Cite{MM97}}, and it sweeps
out compositional differences on isobars.  Therefore it can be assumed
that matter on isobars is approximately chemically homogeneous.
Together with the shellular rotation this allows us to retain a
one-dimensional approximation.  The specific angular momentum, $j$, of
a mass shell is treated as a local variable and the angular velocity,
$\omega$, is computed from the specific moment of inertia, $i$.  The
time-dependent angular momentum redistribution is discussed in
{\Sect{AngTrans}}, and its influence on transport processes in
{\Sect{RotMix}}.  We begin here by describing the modification to the
stellar structure equations of non-rotating stars {\Csa{ES76,MM97}}.

Let $\VP$ be the volume enclosed by a surface of constant pressure,
$P$, and $\SP:=\partial \VP$ its surface area.  Then its
``radius'', $\rP$, is defined as the radius of a sphere of the same
volume, $\VP=4\pi \rP^3/3$, and the equation of continuity becomes
\begin{equation}
  \dxdycz{\mP}{\rP}{t}=4\pi\rP^2\rho
  \;,
\lEq{mDefRot}
\end{equation}
where $\rho$ is the density and $\mP$ the mass enclosed by $\SP$.
For quantities varying on isobars, a mean value is
defined by
\begin{equation}
\av{\;\cdot\;} := \frac{1}{\SP} \oint_{\SP}\cdot\;\;\D\sigma\;,
\end{equation}
where $\DD\sigma$ is an element of isobaric surface area.  The
effective gravitational acceleration $\vec{g}$ is normal to $\SP$.
For the equation of momentum balance, one finds \Cite{ES76}
\begin{equation}
  \dxdycz{P}{\mP}{t}=
     -\frac{G \mP}{4\pi \rP^4} \fP
     -\frac{1}{4\pi \rP^2}\dxdycz{^2 \rP}{t^2}{\mP}
  \;,
  \lEq{MomBalRot}
\end{equation}
where $G$ is the gravitational constant, $P$, the pressure, $t$, the
time, and the inertia term (last term) is added here.  The influence
of rotation is described by the quantity $\fP$
\begin{equation}
  \fP := \frac{4\pi \rP^4}{G\mP\SP} 
               \av{g^{-1}}^{-1}
  \;,
\end{equation}
where $g:=\abs{\vec{g}}$.  The radiative temperature gradient then
takes the form
\begin{equation}
  \dxdycz{\ln T}{\ln P}{t} = 
	\frac{3\kappa}{16\pi a c G}
	\frac{P}{T^4}
	\frac{\LP}{\mP}
 	\frac{\fT}{\fP}
	\SBrak{1+\frac{\rP^2}{G \mP \fP}
              \dxdycz{^2 \rP}{t^2}{\mP}}^{\!-1}
\;,
\lEq{NabRadRot}
\end{equation}
where $\kappa$ is the opacity, $T$ the temperature, $a$ the radiation
constant, and $\LP$, the energy flux through $\SP$.  The last
factor on the right hand side is included to account for inertia as it
follows from the momentum equation {\Cite{Fli93}}, and
\begin{equation}
\fT := \Brak{\frac{4\pi \rP^2}{\SP}}^{\!2} 
             \Brak{\av{g}\av{g^{-1}}}^{-1}
  \;.
\end{equation}
For the derivation of these formulae and for a numerical evaluation of
$\fT$ and $\fP$, see {\cite{ES76}}.  The equations for $\fT$ and $\fP$
are solved iteratively with the stellar structure equations in order
to obtain consistent models {\Cite{ES76,Fli93}}.  In the rest of this
work, the subscript $P$ is omitted (except for $\fP$).

There is, in principle, an inconsistency between the assumption of
shellular rotation and the method described by {\cite{KT70}}, i.e.,
the assumption of shellular rotation does not generally lead to a
conservative potential as it does for a constant rotation rate on
cylinders, which is used by {\cite{KT70}}.  However, {\cite{MM97}}
show that replacing the average $\av{\;\cdot\;}$ by ``appropriate mean
values'', i.e., reinterpreting the quantities describing the stellar
structure as the mean values over the isobars, allows one to keep the
formalism of {\cite{KT70}} as a good approximation.

\subsection{Ordinary mixing in the absence of rotation}
\lSect{MixInst}

Compositional mixing is generally treated as a diffusive process and
implemented by solving the diffusion equation
\begin{equation}
  \dxdycz{X_n}{t}{m}=\dxdycz{}{m}{t} \SBrak{(4\pi r^2 \rho)^2  D
     \dxdycz{X_n}{m}{t}}+\DxDyInd{X_n}{t}{\mathrm{nuc}}
\lEq{Diff}
\;,
\end{equation}
where $D$ is the diffusion coefficient constructed from the sum of
individual mixing processes and $X_n$, the mass fraction of species
$n$.  The second term on the right hand side accounts for nuclear
reactions.  At the inner and outer boundary reflecting conditions
are used:
\begin{equation}
\At{\dxdycz{X_n}{m}{t}}{m=0}=0=\At{\dxdycz{X_n}{m}{t}}{m=M(t)}
\;.
\end{equation}
Mixing, burning, and mass loss are treated as separate, sequential
operations.  The different contributions to the diffusion coefficient,
$D$, are discussed in the following sections.  


\subsubsection{Convection and overshooting}
\lSect{Conv}

\pFig{shear}

Convection occurs when the temperature gradient exceeds the adiabatic
condition, as modified by any gradient in mean molecular weight, $\mu$
(\Fig{shear}).  That is, a stratification is stable against convection
if
\begin{equation}
  \Nad-\N+\frac{\varphi}{\delta}\Nmu \geq 0
  \lEq{Ledoux}
\end{equation}
{\Ce{KW91}}.  This is the so-called Ledoux criterion for convection.
Here the common definitions are used:
\begin{equation}
\Nad:=\dxdycz{\,\ln T}{\,\ln P}{\mathrm{ad}}
\;,\quad
\Nmu:=\DxDy{\,\ln \mu}{\,\ln P}
\;,\quad
\N:=\DxDy{\,\ln T}{\,\ln P}
\;,
\end{equation}
\begin{equation}
\delta:=-\dxdycz{\ln\rho}{\ln T}{\mu,P}
\;,\qquad
\varphi:=\dxdycz{\ln\rho}{\ln\mu}{P,T}
\;.
\end{equation}
The index ``ad'' stands here for ``at constant entropy {\emph{and}}
composition''.

The diffusion coefficient for composition mixing is treated according
to the mixing-length theory {\Cite{Vit53,Bom58}}:
\begin{equation}
\Dconv:=\aMLT\HP\vconv/3
\;,
\end{equation}
where $\vconv$ is the convective velocity.  The pressure scale-height
is defined for the hydrostatic case by
\begin{equation}
\HP:=-\DxDy{r}{\ln P}
  =
     \frac{P}{\rho g}
\;.
\end{equation}
The local gravitational acceleration is given by $g=Gm/r^2$.  In this
work a mixing-length parameter of $\aMLT=1.5$ {\Cite{Lan91B}} is used.

The mixing performed by convection is fast in comparison to most of
the other time-scales relevant for the stellar evolution.  It operates
on the local dynamical time-scale and usually manages to smooth out
any compositional inhomogeneities in the regions where it is active.
Only when the time-scale of thermonuclear burning becomes comparable
to that of convection, as, e.g., during central silicon and shell
oxygen burning, can notable gradients persist.

In the present work ``overshooting'' of the convection into the
convectively stable regime defined by {\Eq{Ledoux}} is neglected.  It
will be shown that rotation leads to mixing above the convective core
of massive stars during central hydrogen burning and thereby to the
formation of more massive helium cores later in the evolution.  In
order to obtain such mixing, large overshooting is often introduced in
literature {\Ce{CS91,sch92}}, but moderate rotation can lead to
similar effects.

\subsubsection{Semiconvection}
\lSect{Semiconv}

Semiconvection is a secular instability which can occur in
non-rotating stars.  According to a local, linear stability analysis by
{\cite{Kat66}}, it is an oscillatory instability which appears in
regions where an unstable temperature gradient is stabilized against
convection by a sufficiently large gradient in the mean molecular
weight ($\mu$-gradient), i.e., it lives in the regime
\begin{equation}
\frac{\varphi}{\delta}\Nmu \geq \N-\Nad \geq 0
\lEq{semiconv}
\end{equation}
{\CiteA{; and \Fig{shear}}{KW91}}.  Heat transfer between a
displaced mass element and its surrounding causes the growth of the
instability on the local thermal time-scale.

In the code STERN (see {\Sect{STERN}}), semiconvection is treated
following {\cite{LSF83}}.  The diffusion coefficient for this process
is computed from
\begin{equation}
\Dsem=\frac{\asem K}{6\cP\rho}
  \frac{\N-\Nad}{\Nad-\N+\frac{\varphi}{\delta}\Nmu}
\;,\quad
K=\frac{4acT^3}{3\kappa\rho}
\;,
\lEq{Dsem+K}
\end{equation}
where $K$ is the thermal conductivity and $\cP$ the specific heat
at constant pressure.  As proposed by {\cite{Lan91B}}, an efficiency
parameter of $\asem=0.04$ is adopted here.

In KEPLER (see {\Sect{KEPLER}}) semiconvection is computed from
{\Cite{WZW78,WW93}}
\begin{equation}
\Dsemx=\frac{1}{6}\aMLT^2\vsem\HP
\;,
\end{equation}
where the velocity $\vsem$ is determined through
\begin{equation}
\vsem=\sqrt{\Brak{\N-\Nad}\frac{P\delta}{g\rho^2}\DxDy{P}{r}}
\;.
\end{equation}
The diffusion coefficient is limited to a fraction $\asem$ of the
radiative diffusion coefficient
\begin{equation}
  \Drad=\frac{K}{\rho\cV}
\end{equation}
by means of
\begin{equation}
\Dsem=\frac{\asem\Drad\Dsemx}{\Dsemx+\asem\Drad}
\;.
\end{equation}
As usual, $\cV$ denotes the specific heat at constant volume.  In
this work a value of $\asem=10^{-4}$ is used in KEPLER, which results
in a comparable efficiency for semiconvection as the value used for
STERN {\Cite{Woo97:pc}}.

\subsection{Rotationally induced mixing}
\lSect{RotMix}

In this work, the mixing processes discussed in {\cite{ES78}} are
included in a parametric way, following the work of {\cite{pin89}}.
Five different processes are considered.  To account for the uncertain
mixing efficiency of each, they are weighed by efficiency factors
{\CiteA{; \Sect{calib}}{pin89}} and then added to the diffusion
coefficient, $D$, in the diffusion equation {\Eqff{Diff}}.

\subsubsection{Dynamical shear instability}
\lSect{DSI}

\pFig{DSI}

Dynamical shear instability occurs when the energy that can be gained
from the shear flow becomes comparable to the work that has to be done
against the gravitational potential for the adiabatic turn-over of a
mass element (``eddy'').  This means that it is stabilized by density
gradients.  Since there is no work required to mix on isobars, this
instability can work very efficiently on those {\CITE{horizontal
turbulence; }{Zah92}} and thus enforce rigid rotation horizontally
{\Cite{ES78,pin89}}.  Thus chemical inhomogeneities are smoothed on
isobars.  This, together with the so called baroclinic instability,
which also acts barotropic for shear on isobars on a dynamical time
scale {\Cite{Zah83}}, justifies the assumption of shellular
rotation and that the composition is only a function of the isobars
(\Sect{ModRot}).

The linear condition for stability is given by
\begin{equation}
\Ri:=\frac{\rho\delta}{P}\Brak{\Nad-\N+\frac{\varphi}{\delta}\Nmu}
	\Brak{g \DxDy{\ln r}{\omega}}^{\!2}>\Ric
    \approx\frac{1}{4}
\lEq{def:ri}
\end{equation}
for the case of a rotating fluid {\Cite{Zah74}}.  Here, $\omega$ is the
angular velocity, $\Ri$, Richardson number, and $\Ric$, its critical
value, about $1/4$.  Note that the term for $\Nmu$ in {\Eq{def:ri}}
was omitted in the original work by {\cite{ES78}} and {\cite{pin89}}.
The corresponding diffusion coefficient is computed from the spatial
extent of the unstable region $\dinst$, limited to a pressure
scale-height, and the local dynamical time-scale {\Csa{ES78}}:
\begin{equation}
\DDSI=\SBrak{\Min{\dinst,\HP} 
  \Brak{1-\Max{\frac{\Ri}{\Ric},0}}}^2
  /\taudyn
\;,
\end{equation}
where the dynamical time-scale is defined by
\begin{equation}
\taudyn:=\sqrt{r^3/(G\,m)}
\;.
\end{equation}
Furthermore, it is assumed that the instability is weaker when the
deviation from the Richardson criterion is smaller.  This is accounted
for by a factor $(1-\Ric/\Ri)^2$, which is limited to the range
$[0,1]$.  For $\Ri > \Ric$, the flow is assumed to be stable against
the dynamical shear instability and thus $\DDSI$ is set to $0$.

\subsubsection{Solberg-H{\o}iland instability}

\pFig{SHI}

The Solberg-H{\o}iland instability arises if an adiabatically
displaced mass element experiences a net force (the sum of gravity,
buoyancy and centrifugal force) that has components in the direction
of the displacement only.  {\cite{Was46}} gives a condition for the
stability against axisymmetric adiabatic perturbations of this kind.
It separates into two scalar conditions.   At the equator the condition
for stability in the vertical direction is 
\begin{equation}
\RSH:=\frac{g}{\rho}\SBrak{\Brak{\DxDy{\rho}{r}}_{\!\mathrm{ad}}
	-\DxDy{\rho}{r}}
	+\frac{1}{r^3}\DxDy{}{r}\Brak{r^2\omega}^2 \geq 0
\lEq{SHcA}
\end{equation}
{\Cite{Tas78,ES78}}.  If the specific angular momentum $j\sim r^2\omega$
is constant with $r$, the last term on the left-hand side vanishes and
the Ledoux criterion results --- not the Schwarzschild criterion as
stated by \cite{ES78}.  This can be seen by rewriting the condition
for stability 
\begin{equation}
\RSH:=\frac{g\delta}{\HP}\SBrak{\Nad-\N+\frac{\varphi}{\delta}\Nmu}
	+\frac{1}{r^3}\DxDy{}{r}\Brak{r^2\omega}^2 \geq 0
\lEq{SHcB}
\end{equation}
and comparing it with \Eq{Ledoux}.  If, on the other hand, the medium
is marginally stable to convection, the first term on the right
hand side vanishes and the Rayleigh criterion results {\Cite{Tas78,KW91}}.
Note that this instability only occurs in regions of
{\emph{decreasing}} specific angular momentum (\Fig{SHI}) and is
strongly suppressed in stable stratifications
($\N<\Nad+\frac{\varphi}{\delta}\Nmu$).  

The diffusion coefficient resulting from the Solberg-H{\o}iland
instability is estimated in a way similar to that for the dynamical
shear instability.  The extent of the unstable region, $\dinst$,
limited to the pressure scale-height, is used as the characteristic
length-scale, and the dynamical time-scale is used as characteristic
time-scale:
\begin{equation}
\DSHI=\SBrak{\Min{\dinst,\HP} \Brak{\frac{r\,\RSH}{g}}}^2/\taudyn
\;.
\end{equation}
Again, as for the dynamical shear instability, a factor of order unity
($r\,\RSH/g$) was introduced to smoothly turn on the instability as
the criterion for stability gets increasingly violated, and $\DSHI$ is
set to $0$ wherever the stability criterion is fulfilled.

\subsubsection{Secular shear instability}
\lSect{SSI}

The strict criterion for dynamical shear instability can be relaxed
considerably by allowing for thermal adjustment of radial
perturbations.  However, this process then operates only on a thermal
time-scale, and is therefore a secular process.  Gradients in the
mean molecular weight, which may inhibit the occurrence of the
instability, also have to be taken into account.

According to {\cite{ES78}}, the following two conditions have to be
violated simultaneously for this instability to set in (\Fig{shear}):
\begin{equation}
\Risa:=\frac{\Pra\,\Rec}{8}\frac{\rho\delta}{P}\Brak{\Nad-\nabla}
  \Brak{g\DxDy{\ln r}{\omega}}^{\!2}
  >\Ric
\end{equation}
{\Cite{Tow58,Zah75}} because of the relaxed condition for the
temperature gradient, and
\begin{equation}
\Risb:=\frac{\rho\varphi\Nmu}{P}
	\Brak{g\DxDy{\ln r}{\omega}}^{\!2}
	>\Ric
\lEq{SSIb}
\end{equation}
since the condition for the $\mu$-gradient is not relaxed.  The latter
formula follows from the physical arguments of {\cite{ES78}}, but
corrects an error in their Eq.~(10).  For the critical Reynolds
number, $\Rec$, a value of $2500$ is assumed in this work {\CITE{but
see also}{RZ99}}.  The Prandtl number, $\Pra$, is defined as the ratio
of the thermal diffusion time-scale to the angular momentum diffusion
time-scale, and is estimated according to {\cite{Tas78}}:
\begin{equation}
\Pra=
  \frac{\cV\Brak{\mup+\mur}}{\chi}
\;,
\end{equation}
where the coefficients of shear viscosity of the plasma and by 
radiation are computed according to 
\begin{equation}
\mup\approx0.406\,\frac{\sqrt{\mi (\kB T)^5}}
	{\Brak{\Zi e}^{4}\ln\Lambda}
\;,\qquad
\mur=\frac{4 a T^4}{15 c \kappa \rho}
\end{equation}
{\Cite{Spi62,Tas78}}, respectively.  The quantity $\Lambda$ is the
ratio of the cut-off length for ion collisions, which is taken as the
ratio of the Debye length, to the impact parameter for a $\pi/2$
deflection for Rutherford scattering of the ions, i.e.,
\begin{equation}
\Lambda=\frac{2}{3 e^3}\sqrt{
	\frac{\mi \Brak{\kB T}^{3}}{\pi \rho \Zi^5}}
\end{equation}
\CITE{for details, see}{Spi62}.  Here, $e$ is the charge of the
electron in e.s.u., $c$ the velocity of light, $\kB$ Boltzmann's
constant, $\Zi$ the charge number of the ion, and $\mi$ its mass.  It
should be noted that for burning phases beyond hydrogen burning, as
well as for helium, carbon, or oxygen stars, it is important to take
the $\Zi$-dependence of the plasma viscosity into account.  The
quantity $\Lambda$ enters only logarithmically and $\ln \Lambda$ is
$\sim25$.  At temperatures below the Fermi temperature, depending
somewhat on the chemical composition, the ion viscosity dominates over
the electron contribution.  For the evaluation of the formulae above,
complete ionization is assumed.

If magnetic fields and neutrinos are neglected, the thermal
conductivity is given by $\chi\approx K$ {\CITE{\Eq{Dsem+K};
}{Tas78}}.  The opacity, $\kappa$, used in this work takes into
account the energy transport by radiation as well as heat conduction
by degenerate electrons.  Following {\cite{ES78}}, the circulation
velocity associated with this process is computed from the time-scale
and the length-scale of the turbulent elements,
\begin{equation}
\vSSI=\sqrt{\frac{\nu}{\Rec}\DxDy{\omega}{\ln r}}
\;,
\end{equation}
limited to the adiabatic sound velocity, $\cs$.  The kinematic
viscosity, $\nu$, is given by {\Cite{Tas78}}
\begin{equation}
\nu=\frac{\mup+\mur}{\rho}
\;.
\end{equation}
For the typical length-scale the velocity scale height of the flow is
assumed,
\begin{equation}
\HvSSI:=\abs{\DxDy{r}{\ln \vSSI}}
\;,
\end{equation}
limited to the pressure scale height.  The resulting diffusion
coefficient is given by
\begin{equation}
\DSSI=\Min{\vSSI,\cs}
      \Min{\HvSSI,\HP}
      \Brak{1-\frac{\Max{\Risa,\Risb}}{\Ric}}^{\!2}
\;.
\end{equation}
Again, the instability is smoothly turned on with increasing
violation of the stability criteria (term in the last bracket).

In recent work, {\cite{MM96}},{\cite{Mae97}}, and {\cite{MZ98}}
reconsidered the interaction of thermal diffusivity, horizontal
turbulence (due to the baroclinic instability), and vertical shear.
An important conclusion that can be drawn from their work is that
$\mu$-gradients might not completely suppress the occurrence of the
shear instability, since the medium is already turbulent due to the
baroclinic instability.  Consequently, some mixing can occur
{\Cite{Mae97}}.  In the present work, we parameterize the efficiency
of the secular shear instability for chemical mixing and of the
$\mu$-gradients in suppressing its occurrence (\Sect{uncertMix}).

\subsubsection{Eddington-Sweet circulation}
\lSect{ES}

As first shown by {\cite{Zei24A,Zei24B}} for rigid rotation, and later
by {\cite{BK59}} for a general rotation law, a rotating star cannot be
in hydrostatic and radiative thermal equilibrium at the same time.
This is so because surfaces of constant temperature and constant
pressure do not coincide.  Consequently, large-scale circulations
develop.  Since inhomogeneities on isobars are quickly smoothed out by
the horizontal turbulence only the perpendicular ($\approx$ radial)
component of the circulation velocity is considered here, and the
process is approximated by diffusion along the radial coordinate.

\cite{Kip74} estimated the circulation velocity as
\begin{equation}
\vESO:=\frac{\Nad}
		{\delta\,\Brak{\Nad-\nabla}}
	\frac{\omega^2 r^3 l}{\Brak{G m}^2}
	\Brak{\frac{2 \Brak{\epsnuc+\epsnu} r^2 }{l}
		-\frac{2 r^2}{m}
		-\frac{3}{4 \pi \rho r}}
\;.
\lEq{vESO}
\end{equation}

In the presence of $\mu$-gradients, meridional circulation has to
work against the potential and thus might be inhibited or suppressed
{\Cite{Mes52,Mes53}}.  Formally, this can be written as a ``stabilizing''
circulation velocity, 
\begin{equation}
\vmu:=\frac{\HP}{\tauKHx}
      \frac{\varphi\Nmu}{\delta\,\Brak{\nabla-\Nad}}        
\lEq{vmu}
\end{equation}
{\Cite{Kip74,pin89}}, where 
\begin{equation}
\tauKHx:=\frac{G m^2}{r \Brak{l - m \epsnu}}
\lEq{tauKHx}
\end{equation}
is the local Kelvin-Helmholtz time-scale, used here as an estimate for
the local thermal adjustment time-scale of the currents
{\Cite{pin89}}.  The spatial extent of the currents is typically of
the order of the radius coordinate $r$.  Here, neutrino losses are
taken into account, because they reduce the thermal time-scale in the
late stages of the stellar evolution significantly.  Note that
$\epsnu$ is defined as the energy {\emph{generation}} rate due to
neutrino losses and therefore is negative.  This increases the
numerator in the definition of the local Kelvin-Helmholtz time-scale
and thus decreases $\tauKHx$.

For the evaluation of the diffusion coefficient, the sign of the
circulation velocity does not matter, but the stabilizing ``currents''
due to $\mu$-gradients always point in the direction opposite to the
meridional flow, thus resulting in a reduction of the effective
circulation velocity.  The velocity is then computed from
\begin{equation}
\vES:=\Max{\abs{\vESO}-\abs{\vmu},0}
\lEq{vES}
\end{equation}
{\CiteA{; and {\Fig{GSF}}}{ES78}}.  The diffusion coefficient is
calculated as the the product of the circulation velocity and a
typical length-scale for the circulation.  This is assumed to be the
minimum of the extent $\dinst$ of the instability and the velocity
scale-height
\begin{equation}
\HvES:=\abs{\DxDy{r}{\ln \vES}}
\end{equation}
{\Cite{ES78}}, i.e., 
\begin{equation}
\DES:=\Min{\dinst,\HvES}\vES
\;.
\end{equation} 

In recent work, {\cite{CZ92,Zah92,USS96,tal97,MZ98}} have discussed
several improvements to the theory of meridional circulation and its
interaction with the baroclinic instability.  In contrast to the
present work, their method requires the solution of a fourth order
differential equation in $\omega$, which is numerically very involved.
So far this method has only been used to investigate main sequence
stars.  An interesting result of these work for the Eddington-Sweet
circulation is that the stabilizing effect of $\mu$-gradients,
entering through $\vESO$ in $\vES$ ({\Eqs{vESO}} and {\Eqff{vES}}),
may be reduced (\Sect{uncertMix}).  The second important change to the
above estimate is that the interaction of the baroclinic instability
and the large-scale meridional reduces the mixing efficiency of the
Eddington-Sweet circulation in agreement with the numerical studies by
{\cite{pin89}} for the sun.  We consider these effects when we perform
an empirical calibration of the mixing efficiencies in {\Sect{calib}}.

\subsubsection{Goldreich-Schubert-Fricke instability}
\lSect{GSF}

\pFig{GSF}

{\cite{GS67}} and {\cite{Fri68}} performed an analysis of stability
against axisymmetric perturbations (GSF instability).  For the
inviscid limit {\mbox{($\Pra \ll 1$)}}, which can be well assumed in
the interior of stars, they derive two conditions for stability in
chemically homogeneous stars {\Cite{Kip69}}:
\begin{equation}
   \dxdy{j}{r} \geq 0
\qquad\mbox{and}\qquad
   \dxdy{\omega}{z} = 0
\;.
\lEq{GSFc}
\end{equation}
The first condition is the secular analogue to the Solberg-H{\o}iland
stability criterion {\Eq{SHcB}}, where the stabilization by the
temperature gradient is removed due to thermal conduction.  This is
similar to the relation between the secular and the dynamical shear
instability.  The second condition in {\Eq{GSFc}} is the analogue to
the Taylor-Proudman theorem for slowly rotating incompressible fluids
{\Cite{Kip74,Tas78}}.  If the rotational velocity depends on the
distance from the equatorial plane, i.e., the rotation profile is not
conservative, meridional flows will be driven.  Also in this case, the
buoyancy force, which acts to suppress the instability, can be removed
by heat conduction.  However, this occurs only on a thermal
time-scale.  Interestingly, the typical velocities for both the
above processes are quite similar {\Cite{Kip74}}.

Since the second condition of {\Eq{GSFc}} is in general in
contradiction with the shellular rotation law enforced by the
baroclinic instability, except for the case of solid body rotation,
the GSF instability will tend to enforce uniform rotation in
chemically homogeneous regions {\Cite{ES78}}.

The dependence of the GSF instability on differential rotation is
stronger than that of Eddington-Sweet circulation, and the large-scale
circulation velocity in the equatorial plane can be estimated by
\begin{equation}
\vGSFO=\frac{2 \HT r}{\Hj^2}
	\Brak{1+2\DxDy{\ln r}{\ln \omega}}^{-1}
	\vESO
      =\frac{2\HT}{\Hj}\,\DxDy{\ln \omega}{\ln r}
	\vESO
\end{equation}
{\Cite{ES78,JK70,JK71}}.  Here $\HT:=-\Frac{\D r}{\D \ln T}$ is the
temperature scale-height and $\Hj:=\Frac{\D r}{\D \ln j}$, the
scale-height of the angular momentum distribution.  The GSF
instability has the same $\mu$-dependence as Eddington-Sweet
circulation {\Cite{ES78}} and therefore the resulting circulation
velocity is computed in the same way, taking the stabilizing effect of
the $\mu$-gradient into account:
\begin{equation}
\vGSF:=\Max{\abs{\vGSFO}-\abs{\vmu},0}
\;.
\end{equation}
Again, the diffusion coefficient is determined from the circulation
velocity, $\vGSF$, and the minimum of the circulation velocity scale
height, $\HvGSF$, and the extent, $\dinst$, of the instability:
\begin{equation}
\DGSF:=\Min{\dinst,\HvGSF}\vGSF
\;,
\end{equation}
where we define in the same way as above
\begin{equation}
\HvGSF:=\abs{\DxDy{r}{\ln \vGSF}}
\;.
\end{equation}
{\Fig{GSF}} compares the parameter space in which the GSF and the
Eddington-Sweet instability operate.  For small angular velocity
gradients the Eddington-Sweet circulation dominates, while the GSF
instability becomes more important as the differential rotation
increases.  Note that for strong differential rotation the shear
instability also occurs (cf. {\Fig{shear}})

\subsection{Other instabilities}
\lSect{OtherInst}

The five instabilities discussed in the previous section are not
a complete list of all rotationally induced instabilities for
massive stellar evolution.  However, they appear to be the most
relevant ones, or at least the best understood.

For the ABCD-instability {\Cite{SKR84}} and the triply diffusive
instability {\Cite{KS83}}, no reliable estimates of efficiency
exist.  Furthermore, non-axisymmetric instabilities may also exist,
but are poorly investigated so far.

Another important issue is the interaction of the different
instabilities, and the interaction of rotation and rotationally
induced instabilities with the instabilities listed in
{\Sect{MixInst}}.  The interaction of the shear instabilities and the
Eddington-Sweet circulation has been investigated by, e.g.,
{\cite{CZ92,Zah92,USS96,MM97,Mae97,TZ97,tal97,MZ98}}, and
semiconvection has recently also been included by {\cite{Mae97}} and
{\cite{MZ98}}.  However, the effects of the interactions are not
large and therefore not taken into account in the present work.

Perhapes most importantly, we have neglected magnetic fields.
Magnetic fields might transport angular momentum by torques
{\CITE{$\sim r^3 B_{r}B_{\phi}$;}{Spr97}}, or cause instabilities by
magnetic buoyancy resulting from the winding up of magnetic field
lines by differential rotation.  This could be effective even if the
initial field strength is small {\Cite{Spr97,SP98}}.  Unfortunately,
little is known about either the strength of the initial field or the
efficiency of instabilities in amplifying the magnetic field.  The
Velikhov-Chandrasekhar instability depends only on the presence of
magnetic fields, not on their strength, but it is efficiently
suppressed by $\mu$-gradients {\Cite{Ach78,Spr97}}.  Detailed studies
of the action of magnetic fields inside stars must be left to future
investigations.

\subsection{Angular momentum transport}
\lSect{AngTrans}

Following {\cite{ES78}} and {\cite{pin89}}, we formulate the transport
of angular momentum as a diffusive process,
\begin{equation}
  \dxdycz{\omega}{t}{m}=\frac{1}{i}\dxdycz{}{m}{t}
     \SBrak{(4\pi r^2 \rho)^2 i\nu \dxdycz{\omega}{m}{t}} 
   - \frac{2\omega}{r}\dxdycz{r}{t}{m}
     \Brak{\frac{1}{2}\DxDy{\,\ln i}{\,\ln r}}
\lEq{AngDiff}
\end{equation}
{\Cite{ES78}}, where $\nu$ is the turbulent viscosity and $i$, the
specific angular momentum of a shell at mass coordinate $m$.  For a
spherical shell of constant density, inner radius $\ri$ and outer
radius $\ro$, the specific moment of inertia, $i$, is given by
$i=0.4\,\Frac{\ro^5-\ri^5}{\ro^3+\ri^3}$; for a thin shell
of radius $r$ this simplifies to $i=2/3\,r^2$.  The last term in
{\Eq{AngDiff}}, an advection term, accounts for contraction or
expansion of the layers at constant mass coordinate.  The factor in
the last bracket on the right hand side vanishes if the gyration
constant $k:=i/r^2$ does not depend on $r$.

{\Eq{AngDiff}} is essentially a diffusion equation for $\omega$
along the ``moment of inertia coordinate'',
\begin{equation}
  I(m):=\int_0^m i(m')\DD m'
\;,
\end{equation}
defined analogously to the mass coordinate {\Ce{KW91}}.  This equation
conserves angular momentum and leads to rigid rotation in a region of
extent $\ell$ whenever the diffusion time-scale, $\tauD :=
\ell^2/\nu$, is short in comparison to structural changes of the star.
Since the Eddington-Sweet circulation may redistribute angular
momentum by advection rather than by viscous stress {\Cite{Zah92}},
the equilibrium solution might deviate from rigid rotation assumed
here in regions where it is the dominant process.  However, for
consistency to {\cite{ES78,pin89}} and for simplification of the
numerical treatment we stick with the prescription outlined above.
Compared to {\cite{TZ97}} we get very similar results at the end of
central hydrogen burning (see {\Sect{CmpOther}}).

At the inner and outer boundary, reflecting conditions similar to those
given in {\Eq{Diff}} for the compositional mixing are used.
At the surface of the star, the angular momentum contained
in the layers which are lost due to stellar winds is removed from
the star (\Sect{AngLoss}).

The turbulent viscosity, $\nu$, is determined as the sum of the
convective and semiconvective diffusion coefficients, and those from
rotationally induced instabilities {\CiteA{; {\Sect{calib}}}{ES78}}.
In contrast to {\cite{ES78}} and {\cite{pin89}}, in the present work the
transport equation for angular momentum is solved for the entire star
as a whole.

Since the evolutionary time-scale of the star is in most cases much
longer than the convective time-scale, {\Eq{AngDiff}} results in rigid
rotation in those regions.  Unlike composition, which can show
significant gradients even inside convective regions due to burning
(e.g., during central silicon burning), angular momentum is locally
conserved, and therefore convective regions can more easily reach
rigid rotation than chemical homogeneity during hydrostatic burning
phases.  This, however, does not hold if the respective layers are
contracting or expanding rapidly.

The approximation that convection leads to rigid rotation rather than
constant specific angular momentum seems to be justified, at least if
the rotational period is long in comparison to the convective time
scale, and it may also hold for more rapid rotation if convective
blobs can be assumed to scatter elastically {\Cite{KNL95}}.  The
latitudinally averaged rotation rate of the solar convection zone
deviates from solid body rotation by less than $5\,\%$ {\Ce{ABC97}}.

\subsection{Enhanced mass loss due to rotation}
\lSect{MassLoss}
\lSect{Omega}

Mass loss from the stellar surface (``stellar winds'') significantly
affects the evolution of massive stars {\Cite{CM86}}.  In the present
work, the empirical mass loss rate of {\cite{NJ90}} is used.  For
Wolf-Rayet stars, the prescription of {\cite{Lan89B}} is applied.  The
uncertainties in these mass loss rates are considerable due to the
uncertainties in the observational data and their interpretation.

These mass loss rates are further modified to account for the effect
of stellar rotation according to {\cite{FA86}}
\begin{equation}
  \Mdot(\omega) := \Mdot(\omega=0) \times
        \left(\frac{1}{1-\Omega}\right)^{\xi}
\;,\qquad
\xi\approx0.43
\lEq{MassLossRot}
\;,
\end{equation}
where 
\begin{equation}
\Omega:=\frac{v}{\vcrit}
\;,
\lEq{Omega}
\end{equation}
is the ratio of the equatorial surface rotation rate to the critical
rotation rate defined by
\begin{equation}
\vcrit^2:=\frac{Gm}{r}\Brak{1-\Gamma}
\;.
\lEq{vcrit}
\end{equation}
The Eddington factor,
\begin{equation}
\Gamma:=\frac{\kappa L}{4 \pi c G m}
\;,
\lEq{Gamma}
\end{equation}
is evaluated only in the radiative part of the optical depth range
$\tau \in [2/3 , 100]$ {\Cite{Lam93,Lan97:LBV}}, where
$\tau(r)=\int_r^{\infty}\kappa\rho\DD r$ has the usual definition.

The quantitative result for the $\Omega$-dependence of the mass loss
rate obtained by {\cite{FA86}} was questioned by {\cite{OCG96}}, who
performed hydrodynamic simulations of the winds of rotating hot stars
including the effect of non-radial radiation forces and
gravity-darkening in the approximation of {\cite{Zei24A,Zei24B}}.
In any case, the latitude dependence of the surface properties
(temperature, radiation flux, etc.) of rapidly rotating luminous stars
is largely unknown as {\cite{Kip77}} showed in a generalization of the
von Zeipel theorem that they depend strongly on the details of the
internal rotation law {\Csa{Mae99}}.  However, the only crucial
ingredient for our model calculations, which is confirmed by
{\cite{OG97}}, is the fact that the latitudinally integrated mass loss
rate
increases strongly as the star approaches the $\Omega$-limit, so that
the star cannot exceed critical rotation, but rather loses more mass
and angular momentum {\Cite{Lan98}}.

\subsection{Angular momentum loss}
\lSect{AngLoss}

The loss of angular momentum from the surface due to stellar winds is
approximated by removing of the angular momentum along with the
surface layer, i.e., 
\begin{equation}
\Jdot=\Mdot\jsurf
\;,
\end{equation}
where $\jsurf$ is the latitudinally averaged 
specific angular momentum at the surface
of the star when the mass loss is assumed independent of latitude.

\section{Calibration of the mixing efficiencies}
\lSect{uncertMix}
\lSect{calib}

The diffusion coefficients used in this work are subject to
considerable uncertainties, as they result from order-of-magnitude
estimates of some of the relevant time- and length-scales.  Therefore,
efficiency factors of order unity are introduced, in order to
calibrate the diffusion coefficients with observational data.  This is
similar to the treatment of {\cite{pin89}}.

The first adjustable parameter is the ratio of 
the turbulent viscosity to the diffusion
coefficient, $\fc:=D/\nu$.  The contribution of the rotationally
induced instabilities to the diffusion coefficient is assumed to be
reduced by the factor $\fc$, while their full value enters the
turbulent viscosity,
\begin{equation}
D=\Dconv+\Dsem+\fc\Brak{\DDSI+\DSHI+\DSSI+\DES+\DGSF}
\;,
\end{equation}
\begin{equation}
\nu=\Dconv+\Dsem+\DDSI+\DSHI+\DSSI+\DES+\DGSF
\;.
\end{equation}

The second parameter, $\fmu\in[0,1]$, describes the sensitivity of the
rotationally induced mixing to $\mu$-gradients, i.e., $\Nmu$
is replaced by $\fmu\Nmu$.

In order to reproduce the surface {\I{7}{Li}} abundance in the sun,
{\cite{pin89}} introduced the factor $\fc\in[0,1]$.  They found a value
of $\fc=0.046$ for their best fit.  From theoretical work {\cite{CZ92}}
found a similar value, $\fc=1/30$, for the combined action of shear and
meridional circulation.  This is the value chosen for most of the
models presented in this work (cf. {\Tab{InitModels}}).

\pFig{M-CNOHe-gauge}
\pFig{fc-gauge}

The best observational probe of rotationally induced mixing in
stars is the evolution of the
surface composition during central hydrogen burning.  While lithium
and boron are depleted early during this phase {\Cite{VLL96,FLV96}},
since they are destroyed at relatively low temperatures, {\I{14}{N}},
is only produced at higher temperature, i.e., much deeper
inside the star.  Therefore, an increase of nitrogen at the surface
should be accompanied by an decrease of carbon ({\I{12}{C}}) or, in
the case of even deeper mixing, oxygen ({\I{16}{O}}), which is
destroyed at even higher temperatures.

An enrichment of nitrogen of order $2\ldots3$ is observed for evolved
stars of about $10$ to $20\,\Msun$ {\Cite{GL92,Her94,vra98}}.  Since
observations can only give the projected rotation rate and are also
restricted to low projected rotational velocities (\Cite{GL92,vra98}),
only a qualitative comparison with our models is possible.

The processing of carbon to nitrogen which occurs at core hydrogen
ignition does not introduce large $\mu$-gradients.  Therefore, the
occurrence of a surface nitrogen enrichment and carbon depletion is
rather insensitive to $\fmu$.  In contrast, any enrichment of helium
in O stars {\Cite{her92,HVM98}} strongly restricts $\fmu$.
Unfortunately, helium abundances are hard to measure and
correspondingly uncertain {\Cite{Her94}}.

For purposes of calibration, we computed evolutionary sequences for
solar metallicity stars in the mass range $4\,\Msun$ to $60\,\Msun$
through core hydrogen burning, adopting a typical zero-age main
sequence rotational velocity of $\sim 200\,\kms$
{\Cite{Sle70,Fuk82,Lang91,Hal96,Pen96}}.  {\Fig{M-CNOHe-gauge}} shows
the surface values of helium, carbon, nitrogen, and oxygen at core
hydrogen exhaustion as function of the initial stellar mass for
various combinations of $\fmu$ and $\fc$.

A value of $\fmu=0.05$ reproduces an
enhancement of nitrogen by a factor of $2$ to $3$ in the mass range
$10\,\Msun$ to $20\,\Msun$, and results in a surface helium mass
fraction of $\sim40\,\%$ for the $60\,\Msun$ star, while the
enrichment remains quite small for stars below $20\,\Msun$.  For
$\fmu=0.01$, nitrogen and helium are clearly enriched too much for
stars below $30\,\Msun$.  On the other hand, the nitrogen enrichment
might be too low for values of $\fmu\ge0.1$.  Certainly, for
$\fmu=0.25$ and $\fmu=1.0$ the nitrogen abundance for the most massive
stars ($30\,\Msun\ldots60\,\Msun$) is inconsistent with the
observations.  The same is true for the helium abundances.

\pTab{InitModels}

In summary, $\fmu=0.05$ seems to be the best value (provided
$\fc=1/30$; see above).  This set of parameters is used in the present
work for the models whose name ends with ``{\Mod{B}}''
(\Tab{InitModels}).  The consequences of a variation of $\fc$ (for
fixed $\fmu=0.05$) is shown in {\Fig{fc-gauge}} for a $12\,\Msun$
star.  For small values of $\fc$ the nitrogen abundance is too low,
while for large values, helium becomes quite high.

As discussed above, too much surface enrichment occurs with $\fc=1/30$
for small values of $\fmu$ ($\simle0.01$).  Nevertheless, it is
interesting to investigate the case where $\mu$-gradients are
completely neglected, since the calibration of $\fc$ and $\fmu$ is not
unambiguous, and different combinations might result in similar
surface enrichments.  The surface abundance, however, are the only
clear observational constraint, while the degree of internal mixing is
not directly observable.  Therefore, a second parameter set of
$\fc=0.01$ and $\fmu=0$ is also used.  The resulting surface
abundances (displayed as thick grey line in {\Fig{M-CNOHe-gauge}})
show quite similar enrichments.  Models with this choice of $\fc$ and
$\fmu$ do {\emph{not}} carry a ``{\Mod{B}}'' at the end of their name
(\Tab{InitModels}).  A value of $\fc=0.01$ for $\fmu=0$ is also
supported by calibrations of the lithium, beryllium and boron surface
abundance for the sun by {\cite{Fli93}}.

\section{Numerical solution}
\lSect{NumSolve}

Two different numerical codes were used here to follow the stellar
evolution.  We now briefly describe each.

\subsection{STERN}
\lSect{STERN}

The STERN code is a pseudo-Lagrangian, implicit hydrodynamic code
{\Cite{lan88}}, based on the ``G\"ottinger stellar evolution code''.
For numerical solution, relative mass coordinate $q:=m/M$ is used
instead of the the mass coordinate $m$, which allows to reserve the
distribution of computational grid in the presence of mass loss.

The equation of state includes radiation, ionization, relativistic
electron degeneracy, and electron-positron pairs.  Ions are
treated as a Boltzmann gas {\Cite{EL86}}.

The chemical evolution due to thermonuclear burning is traced by $35$
isotopes: {\I{}n}, {\I{1,2}H}, {\I{3,4}{He}}, {\I{6,7}{Li}},
{\I{7,9}{Be}}, {\I{8,10,11}B},{\I{11,12,13}C},{\I{12,14,15}N},
{\I{16,17,18}O}, {\I{19}F}, {\I{20,21,22}{Ne}}, {\I{23}{Na}},
{\I{24,25,26}{Mg}}, {\I{26,27}{Al}},{\I{28,29,30}{Si}}, and
{\I{56}{Fe}}.  Except for {\I{19}F}, {\I{26}{Al}}, and {\I{56}{Fe}},
reactions between them are solved in a $32$ isotope network.  These
reaction rates are also used to determine the nuclear energy
generation rate.  The {\El{Ne}}/{\El{Na}} and {\El{Mg}}/{\El{Al}}
hydrogen-burning cycles are solved separately using a $13$ isotope
network including {\I1H}, {\I{18}O}, {\I{19}F}, {\I{20,21,22}{Ne}},
{\I{23}{Na}}, {\I{24,25,26}{Mg}}, {\I{26,27}{Al}}, {\I{28}{Si}}, and
{\I{16}{O }} {\Cite{Braun97}}.  The neutrino losses are determined
according to {\cite{MKI85}}.

The reaction networks are solved separately for each zone between the
individual stellar structure integration time-steps.  This allows for
subcycling of the reaction network with fine time-steps wherever
needed. 

\subsection{KEPLER}
\lSect{KEPLER}

In the KEPLER code {\Cite{WZW78,WWF84,WW88}} the equation of state
includes a crude treatment of Coulomb corrections, beyond what is used
in STERN {\Cite{WZW78}}.  A $19$-isotope network is employed through
oxygen burning, including the elements {\I1H}, {\I3{He}}, {\I4{He}},
{\I{12}C}, {\I{14}N}, {\I{16}O}, {\I{20}{Ne}}, {\I{24}{Mg}},
{\I{28}{Si}}, {\I{32}S}, {\I{36}{Ar}}, {\I{40}{Ca}}, {\I{44}{Ti}},
{\I{48}{Cr}}, {\I{52}{Fe}}, {\I{54}{Fe}}, {\I{56}{Ni}} and neutrons
and protons from photodisintegration.  Silicon burning is followed
using a quasi-equilibrium network of $137$ isotopes, in which
subgroups of elements are treated in nuclear statistical equilibrium
while reactions between these subgroups are considered explicitly.
Beyond silicon burning full nuclear statistical equilibrium is
assumed.  A more detailed description of the reaction networks in
KEPLER can be found in {\cite{WZW78}}.  However, an improvement of the
treatment of hydrogen burning has been implemented (\App{KepImpr}).

For the present work, angular momentum has been added to KEPLER as a
new local variable, and rotationally induced mixing processes
incorporated according to {\Sect{RotMix}}.  However, because changes
to the structural model calculations on KEPLER would be difficult, the
modifications to the momentum balance and the energy transport
({\Sect{ModRot}}) applied in STERN are not included into KEPLER.  The
same opacities {\Cite{IR96}} used in STERN are also included in KEPLER
(an update to previous versions of the code), which allows for more
consistency between the two calculations.  For temperatures above
$10^9\;\K$ the opacities used in KEPLER are still chiefly due to
electron scattering with corrections due to relativity and degeneracy
{\Cite{WZW78}}.

As outer boundary conditions a finite (or zero) boundary pressure is
often utilized in KEPLER.  The radius of the photosphere is determined
as the location where an optical depth of $2/3$ is reached.  This
treatment of the outer boundary condition, but also the mass loss, is
less accurate than that implemented in STERN {\Cite{Heg98}}.  For this
reason, the stellar evolution from the pre-main sequence until a
central temperature of $10^9\,\K$, i.e., before central neon ignition,
is followed by STERN, and the rest of the evolution until core
collapse by KEPLER.  At this stage of evolution, the total mass lost
in its remaining lifetime ($\simle100\,\yr$) prior to core collapse is
negligible.  The stellar envelope, and therefore the outer appearance
of the star, hardly changes.  However, stellar models followed form
the pre-main sequence using KEPLER give results similar to those
obtained by STERN.


\section{Initial models}
\lSect{InitialCond}

The initial model for the calculations presented in this work is that
of a fully convective, rigidly rotating (following our assumption that
convection does lead to rigid rotation) pre-main sequence star.  In the
Hertzsprung-Russell (HR) diagram such stars are located on their
Hayashi line.  These models are constructed from the Lane-Emden
equation {\Ce{KW91}} with a polytropic index of $n=3/2$.  Typically,
initial stellar radii around $1\,000\,\Rsun$ are used.  This kind of
initial condition is for computational convenience only and is not
intended to reproduce the true pre-main sequence evolution
{\Csa{BM94,BM96}}.

The influence of rotation on the stellar structure is negligible in
the initial models, but it becomes more important when the stars
contract towards central hydrogen ignition.  On the zero-age main
sequence (ZAMS) close-to rigid rotation establishes throughout the
star, mainly through the action of Eddington-Sweet circulation
(\Sect{ES}) and the Gold\-reich\--Schu\-bert\--Fricke instability
(\Sect{GSF}).  These processes are sufficiently efficient in the early
stellar evolution that rigid rotation is established virtually
independent of the initial angular momentum distribution assumed.
Almost no angular momentum is lost before the star reaches the main
sequence.

\pTab{InitAbuSTERN}

\pTab{InitAbuKEPLER}

All models in this work use an approximately solar initial chemical
composition with a mass fraction of all elements heavier than helium
(``metals'') of $Z=0.02$.  The mass fractions of hydrogen and helium
are set to $X=0.7$ and $Y=1-X-Z=0.28$, respectively.  In STERN
(\Sect{STERN}), the abundance ratios of the isotopes within each of
these groups are chosen to have the solar system meteoritic abundance
ratios according to {\cite{GN93}} (see {\Tab{InitAbuSTERN}}).
Calculations performed with the KEPLER code (\Sect{KEPLER}) start
on the pre-main sequence with a relative distribution of the metals
according to {\cite{AG89}} as given in {\Tab{InitAbuKEPLER}}.

For the main set of models in this work the initial angular momentum
is determined such that the stars reach a rotational velocity of
$\sim200\,\kms$ on the ZAMS.  This is a typical observed value for
stars in the mass range $8\,\Msun\ldots25\,\Msun$
{\Cite{Sle70,Fuk82,Hal96,Pen96,how97}}.  It corresponds to $\sim
35\,\%$ of their ``critical'' rotation speed (\Sect{Omega}).  Also
models with different initial rotation rates are computed, in order to
investigate the influence of this parameter on the evolution of
massive stars (see {\Tab{InitModels}}).

\section{Central hydrogen burning}
\lSect{cHburn}

\subsection[Chemical mixing: the example of $20\,\Msun$ stars]
{Chemical mixing: the example of $\mathbf{20\,\MsunB}$ stars}
\lSect{IntHydroMix}

\pFig{m-X-20AB}

In {\Fig{m-X-20AB}} the internal profiles of the most abundant
isotopes in a non-rotating star and two rotating $20\,\Msun$ models
are compared at core hydrogen exhaustion.  Convection causes flat
profiles in the innermost few solar masses.  Small convective and/or
semiconvective regions (similar to {\ModA{D15}} in {\Fig{D15cnv}})
cause steps in the profile above the convective core.

In the non-rotating case no mixing occurs in the envelope.  In
contrast, the rotating models mix thermonuclear processed matter into
the envelope.  If no inhibition of rotationally induced
instabilities by $\mu$-gradients is assumed an extended gradient in
helium (along with other species) reaches from the upper edge of the
convective core up to the surface ({\ModA{E20}} in
{\FIG{A}{m-X-20AB}}).  Due to the increase of the mean molecular
weight in the whole envelope, as a consequence of helium enrichment,
the mass of the hydrogen-depleted core of {\ModA{E20}} is about
$1.5\,\Msun$ larger than in the non-rotating case.

The dominant rotationally induced mixing process during central
hydrogen burning is Eddington-Sweet circulation.  It is fast
enough to keep the whole star close to rigid rotation
(\Sect{MSjtrans}), and thus renders shear instabilities unimportant.
The GSF instability remains one to two orders of magnitude less
efficient than the Eddington-Sweet circulation.  The $\mu$-gradients
above the convective core
in {\ModA{E20}} (see also {\Fig{E15cnv}}) are strong enough to
suppress the occurrence of extended semiconvective structures.  The
secular shear instability occurs only in a small layer close to the
surface, and never contributes significantly to the mixing.

\lSect{MixHmu}

If $\mu$-gradients {\emph{are}} taken into account for the
rotationally induced instabilities ({\ModA{E20B}};
{\FigBB{B}{m-X-20AB}{}{E15Bcnv}}), the $\mu$-gradient which forms at
the upper edge of the convective core is not smoothed out fast enough,
but instead almost completely chokes off any mixing between core and
envelope quite early during core hydrogen burning.  Therefore, below
$m=10\,\Msun$ the composition of {\ModA{E20B}} remains quite similar
to that of {\ModA{D20}}.  The higher concentration of carbon in
{\ModA{E20B}}, however, shows the occurrence of {\emph{some}} mixing
early on.

Above the ``barrier'' due to the $\mu$-gradient ($\mu$-barrier) mixing
is efficient (see the small slope of the composition profiles in the
envelope of {\ModA{E20B}}; {\FIG{B}{m-X-20AB}}), and stronger than for
{\ModA{E20}}, since the efficiency for compositional mixing is assumed
to be $\fc=1/30$ in {\ModA{E20B}} instead of $\fc=1/100$ for
{\ModA{E20}}.

The relative contributions of the different rotationally induced
mixing processes above the $\mu$-barrier are similar in
{\ModB{E20}{E20B}}, except that close to the $\mu$-barrier the GSF
instability becomes important in {\ModA{E20B}}.  Within the
$\mu$-barrier, almost all rotationally induced mixing is suppressed
and the mixing is dominated by semiconvection.  The secular shear
instability is
inhibited by the $\mu$-gradient.  

Strong angular velocity gradients at the boundaries of convective layers 
cause, in principle, layers where the shear flow can overcome  
the stabilizing effect of the $\mu$-gradients.  However, they are
too thin to be resolved in the present calculations.

\subsection{Transport of angular momentum}
\lSect{MSjtrans}

\pFig{m-wjJ53-E15AB}

Similar to chemical mixing, the transport of angular momentum depends
strongly on the inhibition of rotationally induced mixing by
$\mu$-gradients.  {\Fig{m-wjJ53-E15AB}} compares the internal angular
velocity profile of two $15\,\Msun$ stars ({\ModB{E15}{E15B}}) which
were computed with different values of $\fmu$ (\Sect{calib}).

In {\ModA{E15}}, the difference between surface and core angular
velocity remains less than $30\,\%$ during core H-burning.  The
over-all decrease of the rotation rate by a factor of $\sim3$ is
caused by two effects: mass loss from the surface, which carries away
$\sim40\,\%$ of the initial angular momentum, and the expansion of the
stellar envelope, which increases the total moment of inertia by a
factor of $\sim2$.  At the same time, the stellar core contracts.  The
persistence of almost rigid rotation during core hydrogen burning
implies transport of angular momentum from the core to the envelope.
This is confirmed by {\FIG{C}{m-wjJ53-E15AB}} which shows a decrease
of the core specific angular momentum with time (see also
{\FIG{E}{m-wjJ53-E15AB}}).  Because of its small radial extent, the
core contains only a small fraction of the total angular momentum of
the star (\FigBb{E}{m-wjJ53-E15AB}{F}).  For higher mass loss rates,
i.e., for more massive stars, the spin-down (decrease of $\omega$) is
dominated by the mass loss, while at lower mass it is dominated by the
expansion of the envelope.

{\FIG{B}{m-wjJ53-E15AB}} shows that the inhibition of rotational
mixing leads to differential rotation during core hydrogen burning.
The ratio of the core to envelope angular velocity in {\ModA{E15B}}
becomes $\sim4$ at core hydrogen exhaustion.  The envelope rotates
slightly faster than in {\ModA{E15}} since the star loses only
$20\,\%$ of the initial total angular momentum, i.e., about half as
much as {\ModA{E15}}.  This is due to the lower luminosity of
{\ModA{E15B}} during core hydrogen burning --- due to less efficient
chemical mixing (\Sect{IntHydroMix}) --- and consequently about
$60\,\%$ less mass loss than in {\ModA{E15}}.
{\FigBB{D}{m-wjJ53-E15AB}{F}{m-wjJ53-E15AB}} show that in
{\ModA{E15B}} the core angular momentum is constant throughout core
hydrogen burning.

\lSect{J53def}

{\FigBb{E}{m-wjJ53-E15AB}{F}} compare the angular momentum
distribution of {\ModA{E15}} and {\ModA{E15B}} at various evolutionary
stages using the variable $\Jm/m^{5/3}$, with $J\Brak{m}:=\int_0^m
j\Brak{m'}\DD m'$.  Since for a rigidly rotating body of constant
density, $\rho_0$, the angular momentum $\Jm$ enclosed by the mass
coordinate $m$ is
\begin{equation}
J\Brak{m} =\frac{3\omega k}{5}\Brak{\frac{3}{4\pi\rho_0}}^{\!2/3} 
	    m^{5/3} 
       \propto m^{5/3}
\;
\lEq{J53def}
\end{equation}
the curves in {\FigBb{E}{m-wjJ53-E15AB}{F}} are more or less flat.
The evolution of $\Jm$ illustrates the transport of angular momentum
throughout stellar evolution.  $\Jm$ drops when angular momentum is
transported through the mass shell $m$.  If no transport angular
momentum through the mass shell $m$ occurs, $\Jm$, and also
$\Jm/m^{5/3}$, remain constant.  Furthermore, following a line of
constant $J$ from one evolutionary stage to a subsequent one shows to
what mass coordinate angular momentum has been transported in the star
during the time between the two evolutionary stages.  We will refer
more to {\FigBb{E}{m-wjJ53-E15AB}{F}} in the discussion of the angular
momentum transport during the later evolutionary phases.

\subsection{The influence of the initial rotation rate}
\lSect{inflji}

\pFig{m-wjJ53-GF15B}

{\Fig{m-wjJ53-GF15B}} shows the evolution of angular velocity and
specific angular momentum in {\ModB{G15B}{F15B}}, which both contain
the inhibition of rotational mixing due to $\mu$-gradients.  The
latter model initially has three times more angular momentum than the
first.

While this difference of a factor of three in the rotation rate is
conserved in the envelope throughout core hydrogen burning, it becomes
much smaller in the cores.  The faster rotation of {\ModA{F15B}}
sustains the transport of angular momentum out of the core for a
longer time than in {\ModA{G15B}}, where the core angular momentum is
almost completely conserved (\Fig{m-wjJ53-GF15B}).  That is, the
angular momentum is less efficiently trapped in the fast rotating
{\ModA{F15B}} than in the {\ModB{G15B}{E15B}}
(\FigB{m-wjJ53-E15AB}{m-wjJ53-GF15B}).  This feedback process leads to
a convergence of the core rotation rates.  We note already here that
this convergence persists during the later burning stages and leads to
very similar iron core angular momenta for a wide range of initial
rotation rates (cf.~{\Sect{j:preSN}} below).

The stronger core angular momentum depletion in faster rotating models
occurs simultaneously with rotationally induced mixing across the
$\mu$-barrier: The masses of the convective cores at the end of
central hydrogen burning are $2.4\,\Msun$, $2.5\,\Msun$, $2.6\,\Msun$,
and $2.8\,\Msun$ for {\ModD{D15}{G15B}{E15B}{F15B}}, respectively.
However, {\ModA{E15}}, where the $\mu$-barrier was assumed to be
inefficient, has a core of about $3.5\,\Msun$.  Thus, even for very
rapid rotation the assumption of $\mu$-barriers inhibiting rotational
mixing strongly restricts the core growth due to rotation.  (see also
{\FigC{D15cnv}{E15cnv}{E15Bcnv}}).

\section{Central helium burning}
\lSect{cHeBurn}

After core hydrogen exhaustion, the models become red supergiants
(except for {\ModA{H12B}} which first burns helium as a as a blue
supergiant for some time) and their extended hydrogen-rich envelopes
become convective.  The pulsational properties of these envelopes have
been discussed by {\cite{heg97}} and the evolution of the surface
rotation rates, especially during blue loops, by {\cite{HL98}}.  In
the following, we investigate the evolution of the cores, using
the $15\,\Msun$ models as example.

\pTab{tauES}

The importance of rotation in the post main sequence evolution can
be estimated from {\Tab{tauES}}, which compares the 
Eddington-Sweet time-scale
\begin{equation}
  \tauES\sim\tauKH\Brak{\frac{\wk}{\omega}}^{\!2}
  \;,\qquad
  \wk:=\sqrt{Gm/r^3}
  \;.
\end{equation}
{\Cite{Zah92}} in the cores of our {\ModB{E15}{E15B}} during the
various burning stages with the respective nuclear time scales.  For
the amount of differential rotation in our models, the characteristic
time-scale for mixing due to the GSF instability (\Sect{GSF}) is
comparable to the Eddington-Sweet time-scale.

The core hydrogen burning phase is the only one where mixing and
nuclear time scale are comparable.  During core helium burning, the
mixing time scale is one or two orders of magnitude larger than the
nuclear time-scale, which may still allow for some effects of
rotational mixing.  The later phases are too short to allow for any
rotationally induced mixing in the cores; note however, that at the
core boundaries some effects of rotational mixing may still be
possible in case of strong gradients in the angular velocity
(cf.~\Sect{hydMix} below).

An energetic limit to the amount of mixing due soley to shear
instabilities can be obtained by comparing the rotational energy of
the core with the potential energy required to lift processed matter
from the upper edge of the convective core to the hydrogen-burning
shell source {\Cite{Heg98}}.  For a typical value of $\omega/\wk=0.05$
and a difference in the mean molecular weight of fully ionized carbon
relative to helium of $\sim0.3$ (oxygen would be even heavier), an
enrichment of carbon by at most $\simle0.5\,\%$ is possible.  This
assumes the carbon to be homogeneously distributed throughout the
radiative layer and that all the rotational energy of the core is used
to supply the buoyancy energy.  Note that this limit does not apply to
instabilities which tap the energy flux in the star like the
Eddington-Sweet circulation.

\subsection{Chemical mixing}
\lSect{HeMix}

\pFig{m-X-DE15AB}

In a non-rotating star using the Ledoux criterion for convection
(\Sect{Semiconv}) prevents the growth of the convective helium core
that would occur if the Schwarzschild criterion were assumed.
Instead, several convective regions, separated by semiconvective
layers, form above the convective core
({\FigBB{C}{m-X-DE15AB}{}{D15cnv}}).  In the rotating models with
$\fmu=0.05$ (e.g., {\ModA{E15B}} in
{\FigBB{B}{m-X-DE15AB}{}{E15Bcnv}}) the shear across the
semiconvective layers is not strong enough to overcome the stabilizing
$\mu$-gradient, even for the fast rotating {\ModA{F15B}}.

If rotationally induced mixing is assumed to be insensitive to
$\mu$-gradients (i.e., $\fmu=0$; {\ModA{E15}} in
{\FigBB{A}{m-X-DE15AB}{}{E15cnv}}) the dynamical shear instability
operates in the semiconvective regions and dissolves them, similar to
the case of Schwarzschild convection.  In this case, the rotational
mixing leads to considerably more massive helium cores.  The
resulting higher burning temperatures in the cores lead to lower
central carbon-to-oxygen ratios at core helium exhaustion.

An interesting issue is the mixing (and angular momentum transport) in
the radiative helium layer between the convective core and the
hydrogen-burning shell.  If the products of helium burning could be
mixed upward into the hydrogen-burning shell, {\emph{primary}}
production of {\I{14}{N}} could occur.  If hydrogen were transported
down into the helium-burning center, a much stronger than normal
s-process could result and build up more heavy or neutron-rich
elements.  On the other hand, strong instabilities in this region
could also lead to a significant slowing-down of the core.

The dominant mixing process present in this layer is Eddington-Sweet
circulation, with some contribution from the GSF instability.  During
the early stages of core helium burning of models with $\fmu=0$ (e.g.,
{\ModA{E15}}), the secular shear instability dominates slightly over
the Eddington-Sweet circulation at the upper edge of the helium core.
Towards central helium exhaustion, the mixing is dominated by the GSF
instability.  In the case of $\fmu=0.05$, the secular shear
instability is suppressed by $\mu$-gradients. 

{\FIG{B}{m-X-DE15AB}} illustrates that some mixing occurs during core
helium burning: A gradient in {\I{12}C} and {\I{16}O} extends from the
convective core up to the edge of the helium core.  In this model, the
increase in {\I{12}C} or {\I{16}O} is not sufficient to result in any
significant primary nitrogen production in the hydrogen burning
shell.  Even though this effect is not notably more pronounced in the
initially faster rotating {\ModA{F15B}} --- due to the convergence of
the core rotation rates; cf.~\Sect{inflji} --- or for the different
initial masses investigated here, such a primary nitrogen production
appears possible in more favorable conditions, e.g., for higher
mixing efficiencies or at lower metallicity.

In {\ModA{E15}} (\FIG{A}{m-X-DE15AB}) the rotation of the helium core
is slower, and the {\I{12}C} and {\I{16}O} gradients are much steeper,
leveling off to the CNO equilibrium values a few tenths of a solar
mass above the convective core.  In the non-rotating {\ModA{D15}}
(\FIG{C}{m-X-DE15AB}), no enrichment of {\I{12}C} and {\I{16}O}
appears at all above the outermost semiconvective layer of the
convective core.

Even though the strong entropy gradient at the location of the
hydrogen-burning shell suppresses rotational mixing between the helium
core and the hydrogen burning shell, some mixing occurs due to the
large angular velocity gradient.  This can be seen in {\Fig{E15Bcnv}}:
The tail of the energy generation rate at the lower bound of the
hydrogen-burning shell source in {\ModA{E15B}} penetrates into the
helium core, i.e., some hydrogen is mixed downward.  Since the protons
burn quite fast as they are mixed deeper inside the helium core, they
cannot reach the central convective region.  However, some protons may
survive and get mixed into the convective helium shell later on
(\Sect{hydMix}).  In {\ModA{E15}} (\Fig{E15cnv}), where the core is
rotating slower, and also in the non-rotating {\ModA{D15}}
(\Fig{D15cnv}), this feature is not found.

\pFig{E15Bcnv}

\subsection{Transport of angular momentum}
\lSect{HeAngTrans}

After core hydrogen exhaustion, the stars undergo a phase of major
restructuring as the core contracts and the envelope expands.  This
leads to a spin-up of the core
({\FigDdd{A}{m-wjJ53-E15AB}{B}{A}{m-wjJ53-GF15B}{B}}) and a spin-down
of the envelope.  At the same time, the convective envelope grows in
mass and its bottom approaches the helium core.  A steep rise in the
specific angular momentum occurs at the bottom of the rigidly rotating
envelope that persists throughout core helium burning and beyond
({\FigDdd{C}{m-wjJ53-E15AB}{D}{C}{m-wjJ53-GF15B}{D}}).  The entire
helium core stays close to rigid rotation during central helium
burning ({\FigDdd{A}{m-wjJ53-E15AB}{B}{A}{m-wjJ53-GF15B}{B}}).

Up to core helium exhaustion, the specific angular momentum of the
helium core drops appreciably with time
(\FigB{m-wjJ53-E15AB}{m-wjJ53-GF15B}).  Three processes contribute to
this effect.  First, angular momentum is removed from the core during
the star's restructuring phase between core hydrogen depletion and
helium ignition.  Second, the core grows in mass due to hydrogen shell
burning and engulfs regions of lower specific angular momentum
(\FigBb{C}{m-wjJ53-E15AB}{D}).  The reasons for the low specific
angular momentum above the core are secular shear instabilities, the
{\emph{first dredge-up}}, and short-lived convective regions which
temporarily extend down to mass coordinates smaller than the final
helium core mass.  The regions of outwards decreasing specific angular
momentum are not Solberg-H{\o}iland unstable due to strong stabilizing
entropy and composition gradients.  Third, some angular momentum is
transported from the helium core into the envelope through the
hydrogen-burning shell.  The models of the ``\Mod{B}'' series lose
less angular momentum during the restructuring phase because of the
inhibiting effect of the $\mu$-gradients, but more during central
helium burning, due to their considerably faster rotation.

The relative loss of angular momentum in {\ModC{G15B}{E15B}{F15B}}
during helium burning increases with the initial amount of angular
momentum left at the end of central hydrogen burning.  Consequently
all three models end up with very similar core angular momenta and
rotation rates (\Fig{m-wjJ53-GF15B}), about three times that of
{\ModA{E15}} (\Fig{m-wjJ53-E15AB}).

\section{Comparison with previous work}
\lSect{CmpOther}

In contrast to {\cite{KMT70}}, who investigated rapidly rotating
$9\,\Msun$ stars ($v\simgr400\,\kms$), our models do not become
secularly unstable at the end of central helium burning, since,
according to our assumptions, $\mu$-barriers are less efficient in
suppressing angular momentum transport compared to {\cite{KMT70}}.

{\cite{ES78}} followed the evolution of $7\,\Msun$ and $10\,\Msun$
stars with a ZAMS rotational velocity of $\sim200\,\kms$ using
essentially the same method as in the present work, except for some
improvements in the input physics of the individual processes applied
here {\CITE{{\Sect{MixInst}} and}{ES78}}.  They used the Schwarzschild
criterion for convection, however, and did not include mass loss.  In
their work, the $\mu$-barrier above the convective hydrogen-burning
core suppressed mixing and transport of angular momentum almost
completely.  Therefore, their stellar cores lose very little angular
momentum during central hydrogen burning.  Although we use $\fmu<1$
and the inhibiting effect of the $\mu$-gradients is smaller, a similar
$\mu$-barrier forms during central hydrogen burning.  However, we
obtain some mixing between the core and the envelope early during core
hydrogen burning, some angular momentum loss from the core to the
envelope, and in most cases some enrichment of the surface with
H-burning products.  In an earlier work, {\cite{ES76}} disregarded
rotationally induced angular momentum transport, but imposed various
rotation laws.  In this case an even more extreme result was obtained:
all models reached critical rotation before carbon ignition.

{\cite{ery94}} considered turbulent diffusion according to
{\cite{Zah83}} in their computation of a rotating $20\,\Msun$ stars
with a metallicity of $Z=0.008$.  They found a surprisingly large
surface {\I{14}N} enrichment at the end of core hydrogen burning of
more than $2\,\%$ by mass.  Since the CNO cycle conserves the total
mass of the CNO isotopes, this result appears implausible and cannot
be reproduced in the present work.

{\cite{MM97}} used a prescription for the Eddington-Sweet circulation
according to {\cite{Zah92}}, and a modified Richardson criterion to
account for thermal effects {\Cite{Mae95,MM96}}.  They computed the
hydrogen-burning evolution of stars from $9\,\Msun$ to $60\,\Msun$ and
found a strong inhibiting effect of the $\mu$-gradients on the
rotational mixing, which resulted in stronger differential rotation at
core hydrogen exhaustion than found in our models of the ``\Mod{B}''
series.  Their models did not show any surface enrichment of helium.

These models were superseded by those of {\cite{Meyn97}}, who computed
the main sequence evolution of $20\,\Msun$ and $40\,\Msun$ stars,
using improved physics of rotationally induced mixing as discussed by
{\cite{MM97}}, and {\cite{Mae97A}}.  The inhibiting effect of
$\mu$-gradients on shear mixing and Eddington-Sweet circulation was
strongly reduced in the new formulation.  {\citeauthor{Meyn97}}'s
rotating $20\,\Msun$ model showed a larger envelope helium enrichment
than a comparable models of {\ModA{E20}}, and a similar mass of the
hydrogen-depleted core.  The physics used in {\cite{Meyn97}} has been
revised again by {\cite{MZ98}} for a more consistent treatment of
$\mu$-gradients.  Models with this prescription are not yet available.

{\cite{tal97}} followed the main sequence evolution of $9\,\Msun$
stars with ZAMS rotation rates of $100\,\kms$ and $300\,\kms$ until
end of central hydrogen burning, using the prescription for the
Eddington-Sweet circulation by {\cite{Zah92}}.  The helium enrichment
in the envelope showed a smooth profile, similar to our
{\ModB{E08}{E10}}.  At core hydrogen exhaustion, their models showed
steep composition gradients close to the stellar surface.  This may have
resulted from the low mass loss assumed in their calculation in
combination with inefficient mixing close to the surface.  With a
slightly larger mass loss rate, their rapidly rotating model would
have a much stronger surface enrichment.  Due to the downward
advection of angular momentum by meridional circulation in the theory
of {\cite{Zah92}}, {\cite{tal97}} found a somewhat stronger envelope
differential rotation compared to our {\ModB{E08}{E10}}, but a
comparable one to, e.g., {\ModA{E12B}}.  We conclude that this
downward advection is not a strong effect, which may justify its
neglect in the present work.

Summarizing, the prescription for rotationally induced mixing used in
the model of {\cite{KMT70}} corresponds roughly to $\fmu=\fc=\infty$
in terms of the present formulation.  {\cite{ES78}} used about
$\fmu=\fc=1$.  Neither work obtained any surface enrichment during
core hydrogen burning due to the strong inhibiting effect of
$\mu$-gradients {\Cite{MM97}}.  The recent picture of interacting
Eddington-Sweet circulation, anisotropic turbulence and shear
instabilities {\Cite{CZ92,MZ98}} has been continuously improved in the
last years {\Ce{USS96,Mae97A,TZ97,tal97}}.  The most recent work in
this series, {\cite{MZ98}}, includes an improved treatment of
compositional gradients, but this type of description for rotationally
induced mixing is complex, computationally expensive, and has not yet
been successfully tested for post-hydrogen burning stars.  However,
the results obtained in earlier work {\Cite{tal97,Meyn97}} for massive
main sequence stars are not significantly different from those of the
present work.

\section{Late evolution until core collapse}
\lSect{LateEv}

After core helium exhaustion, the carbon-oxygen core ({\CO} core)
contracts and subsequently phases of carbon, neon, oxygen, and silicon
central convective and shell burning follow inside this core.

{\Tab{FinalModels}} gives some key parameters of the final models: the
final mass of the star, the masses of the helium, {\CO}, and iron
cores, and the angular momenta contained in theses cores.  For the
iron core additionally the average specific angular momentum is given.
{\ModF{D10}{D12}{E08}{G12B}{E12B}{F12B}} develop degenerate
neon-oxygen cores and central neon burning starts off-center.  Due to
the computational difficulties (and expense) these models were not
followed until core collapse, but stopped during neon or oxygen shell
burning or even before neon ignition.  In {\ModA{D10}} even carbon
burning ignites off-center.  {\ModB{D10}{E08}} experience a dredge-up
of almost the entire helium shell by the convective envelope.
Therefore the final helium core is small and has little angular
momentum.  The remaining helium shell above the {\CO} core at the
point where the calculation is ended is only a few hundreths of a
solar mass.  The masses of the helium cores before the dredge-up are
$2.25$ and $2.1\,\Msun$, respectively.  {\ModA{E25}} loses its
hydrogen-rich envelope during central helium burning and becomes a
Wolf-Rayet star.  Strong Wolf-Rayet mass loss sets in and further
decreases the mass of the star.  It ends up with only $5.45\,\Msun$ at
the time of core collapse and very little angular momentum.

Note that in {\Tab{FinalModels}} the masses of some of the helium
cores in the non-rotating models are apparently larger than those of
the slowly rotating models of the ``\Mod{B}'' series.  This is an
artifact due to the criterion used to measure the mass of the helium
core.  We define the helium core by the mass coordinate at which the
hydrogen mass fraction drops below $\Ep{-3}$.  In the rotating models
of the ``\Mod{B}'' series, the hydrogen gradient at the top of the
helium core is significantly shallower.  If instead a hydrogen mass
fraction of $\Ep{-2}$ is chosen, the helium core masses are similar or
even larger for the rotating models.  Note that the size of the {\CO}
core is defined in a similar way: by the mass coordinate at which the
helium mass fraction drops below $\Ep{-3}$.

\subsection{Chemical mixing in the helium shell}
\lSect{hydMix}

After core helium exhaustion, the {\CO} core contracts and the burning
of helium continues in a shell.  At the same time, the outer layers of
the helium-rich shell cool down and the hydrogen shell source goes
out.  Since this implies a reduced entropy barrier, rotationally
induced mixing through the hydrogen-helium interface can now operate
more efficiently.  The protons which are mixed downward into the
helium shell do not burn immediately.  When later the helium-burning
driven convective shell extends upwards, it dredges these protons down
into the hot, helium-burning layers (see {\Fig{E15Bcnv}}).  This
mechanism can open new channels of nucleosynthesis.  This will be
investigated in more detail in forthcoming papers {\CS{ for a first
report}{lan99:CosV}}.

\subsection{Chemical mixing inside the {\CO} core}

During the final remaining stellar burning phases, rotational mixing
inside the {\CO} core is unimportant.  The strongest instabilities are
again the Eddington-Sweet circulation and the GSF instabilities, at
about same order of magnitude.  From {\Tab{tauES}} it can be seen that
their time scale is too long in order to be significant.

Also, the mixing of {\emph{traces}} of material into regions
of neighboring burning phase is not expected to introduce
qualitatively new nucleosynthesis channels, since all
abundant nuclear species in one burning phase are anyway present
in the neighboring one (e.g., mixing traces of oxygen into neon
burning is not exciting).

In the fast rotating cores of the models of the ``{\Mod{B}}'' series,
secular shear instabilities arise above several of the central and
shell convection zones for a limited time, but they do not become
efficient enough to cause any noticeable mixing.

\subsection{Transport of angular momentum}
\lSect{lateJtrans}

As for the chemical mixing, rotational mixing cannot effectively
remove angular momentum from the core during the late burning stages.
In particular, transport is too inefficient to keep the {\CO} core in
rigid rotation.  Strong differential rotation occurs
({\FigDdd{A}{m-wjJ53-E15AB}{B}{A}{m-wjJ53-GF15B}{B}}).  At this time,
the only instability capable of enforcing rigid rotation is
convection.  Since the radii of the lower boundaries of the major
convection zones of carbon, oxygen, and silicon burning are much
smaller than that of their upper edges, large differences in the
specific angular momentum exist between the bottom and the top of the
convection zone.  Thus, angular momentum is mainly carried outwards.
The typical signature of such a convection zone is a steep drop of the
specific angular momentum at its bottom, accompanied by a large
increase at its top (\FigC{m-wjJ53-E15AB}{m-wjJ53-GF15B}{jj15B}).

Convection zones that subsequently overlap can transport angular
momentum efficiently over scales larger than their individual extent.
This is most efficient when the lower boundary of a convective shell
overlaps with the upper boundary of a preceding convection zone.  For
the models investigated in this work, such an overlap occurs rather
infrequently ({\App{convDiag}}).  Subsequent shells, which are
driven by nuclear burning, tend to form their lower boundary at the
upper edge of a preceding shell, where the fuel for their burning is
not yet depleted.  The most prominent example of this is the
sequence of carbon-burning shells (e.g., {\Fig{E15cnv}}).
Exceptions occur only for some of the late carbon burning shells, and 
for the oxygen burning shells in {\ModB{G15B}{E15B}}. 

Convective angular momentum transport does not operate across the
boundary of the {\CO} core.  These cores retain their angular momentum
after core helium exhaustion.  Some redistribution, mainly due to
convection, occurs inside the cores.  For example,
{\FIG{E}{m-wjJ53-E15AB}} shows that after core helium exhaustion in
{\ModA{E15}}) no angular momentum is transported through the shells at
$m\approx 3.4\,\Msun$ and $m\approx5.1\,\Msun$, i.e., the boundaries
of the helium and the {\CO} core, respectively.

In models with more rapidly rotating cores,
({\ModC{G15B}{E15B}{F15B}};
{\FigCCc{F}{m-wjJ53-E15AB}{E}{m-wjJ53-GF15B}{F}}), the helium core
does lose some angular momentum, even though its upper boundary (at
$m\approx3.7\,\Msun$) remains a significant barrier for angular
momentum transport as indicated by the spike in
{\FIG{F}{m-wjJ53-E15AB}}.  The loss of angular momentum from the
helium core is correlated with the mixing of hydrogen into the helium
shell described in {\Sect{hydMix}}.  No significant angular momentum
was transported across the boundary of the {\CO} core in any of the
models ({\FigDdd{E}{m-wjJ53-E15AB}{F}{E}{m-wjJ53-GF15B}{F}}).

\subsection{Stability to triaxial deformations}

\pFig{m-phi}

As described in {\Sect{ModRot}}, the approximations employed in this
work are limited to slow rotation in the sense that no triaxial
deformation appear.  In the KEPLER code, the influence of the
centrifugal forces on the structure is completely neglected.  However,
when models from calculations with STERN, where centrifugal forces
{\emph{are}} included, are continued by KEPLER at a central
temperature of $\Ep9\,\K$, the evolution usually proceeds smoothly,
i.e., these forces {\emph{are not}} important at this late stage of
evolution.  On the ZAMS, the rotational energy $\Erot$ of the star is
negligible in comparison to its gravitational binding energy $\Epot$
for all models, even for those which are close to critical rotation at
their surface ($\Erot/\abs{\Epot}\ll1\,\%$).

However, in the course of their evolution the stellar models contract
and --- as outlined in {\Sect{lateJtrans}} above --- the transport of
angular momentum out of the core is inhibited or slow, with the
consequence of rapidly rotating cores
({\FigDdd{A}{m-wjJ53-E15AB}{B}{A}{m-wjJ53-GF15B}{B}}, and
{\Tab{jev}}).  For local angular momentum conservation in a shell with
given specific angular momentum $j$, the ratio of angular velocity to
Keplerian angular velocity scales as
\begin{equation}
\frac{\omega}{\wk}=\frac{j}{k\sqrt{Gmr}}\sim r^{-1/2}
\;,\qquad
k\approx2/3
\lEq{awawk}
\;.
\end{equation}
The ratio of the specific rotational energy to the gravitational
potential is then given by
\begin{equation}
\frac{\erot}{\epot}=\frac{1}{2}\Brak{\frac{\omega}{\wk}}^{\!2}
\;.
\end{equation}
This ratio is displayed for several $15\,\Msun$ pre-collapse models 
in {\Fig{m-phi}}.  

A uniformly rotating, self-gravitating, incompressible, and inviscid
fluid (McLaurin spheroid) becomes secularly unstable to triaxial
deformations when the ratio of rotational to gravitational potential
energy
\begin{equation}
\frac{\Erot}{\Epot}=\frac{1}{2}\int_0^m\omega^2(m')\DD m' \left/ 
   \int_0^m\wk^2(m')\DD m'\right.
\end{equation}
exceeds $\sim0.1375$ {\Ce{OB73,Tas78}}.  If this ratio exceeds
$\sim0.26$, the object becomes dynamically unstable to non-axisymmetric
instabilities and fission may occur {\Cite{OT69,OB73}}.  
The stars simulated in the present work are well below these limits
even at the pre-collapse stages ($\Erot/\Epot\simle30\,\%$ of the
critical value) and therefore no triaxial instabilities arise.

\section{Angular momentum prior to core collapse}
\lSect{j:preSN}

\pFig{jj15B}

\pFig{m-J53-presn-GEF}

Our model sequences are terminated at the onset of core collapse, defined by
the infall velocity inside the iron core exceeding $9\E{2}\,\kms$.  At
this stage of evolution the investigated stars typically have central
densities of $\simle1\E{10}\,\gccm$.
From the previous discussions it is clear that the distribution of
angular momentum in the star at onset of core collapse strongly
reflects its recent convective structure.  
{\Fig{jj15B}} shows the distribution of the specific angular momentum
at the pre-collapse stage of $15\,\Msun$ stars with different initial
rotation rates {\ModCx{G15B}{E15B}{F15B}}.  

These three models show a very similar final angular momentum distribution
(cf. also {\FIG{A}{m-J53-presn-GEF}}), due to a similarity entire
chemical structure.  The reason for this is the convergence
of the core rotation rates, i.e. their independence from the initial
rotation rates, already during hydrogen and helium burning, as outlined
in {\Sect{inflji}}. 

In contrast, {\ModA{E15}} has much less
angular momentum left in the core (see also {\Tab{FinalModels}}).  It
grows a larger helium and {\CO} core due to the lack of sensitivity to
$\mu$-gradients. 

\pTab{FinalModels}

The total angular momentum in the final models is dominated by that of
the envelope (\Tab{FinalModels}).  {\ModC{G15B}{E15B}{F15B}} show that
for initially faster rotation, a slightly larger helium core
results (\Tab{FinalModels}) and therefore the stars become more
luminous.  This in turn causes more mass and angular momentum
loss, which can, for the rapid rotators or for more massive stars, 
decreases the total angular momentum by a larger factor
({\FigBB{B}{m-J53-presn-GEF}{B}{m-J53-presn}}). 

\pFig{m-J53-presn}

\pFig{m-J53-presn4}

At hydrogen ignition, the total and even the mean specific angular
momentum of models with a given surface rotational velocity are larger
for larger initial masses (\Tab{FinalModels}).  On the contrary, the
final total angular momentum decreases for larger initial mass
({\Tab{FinalModels}} and {\Fig{m-J53-presn}}).  This trend is only
interrupted between $12$ and $15\,\Msun$ since our models with
initial masses of $12\,\Msun$ or less undergo a blue loop during core
helium burning which leads to an additional strong angular momentum
loss {\Cite{HL98}}.  As in the limit of vanishing mass loss, the angular
momentum of our models is conserved, the decrease of the total angular
momentum for higher initial masses is solely due to the increase of
the mass loss rate for larger initial masses.

The total angular momentum of the helium and {\CO} cores increases with
the the initial stellar mass of our models ({\Tab{FinalModels}}).
However, this trend is much weaker for the specific angular momentum
of the helium cores, the specific angular momenta of the {\CO} cores
even decreases a little with increasing initial mass.  This illustrates that
angular momentum transport from the core into the envelope is stronger
for larger cores.

Finally, we find that the specific angular momenta of the iron cores
are rather insensitive to the initial mass and rotation rate
(\Tab{FinalModels} and {\Fig{m-J53-presn4}}), due to the convergence
of the core rotation rates discussed in {\Sect{inflji}}.  In the
models with $\fmu = 0$, angular momentum transport was efficient and
final values of $\jFe\simeq 6\E{15}\,\junit$ are found.  The value for
{\ModA{E25}} is significantly lower since its {\CO} core was spun down
in a Wolf-Rayet phase.  The models with $\fmu = 0.05$ all end up with
$\jFe\simeq 1.2\E{15}\,\junit$.

Note that, unless the iron core that forms after central silicon
burning is already large enough to collapse, one or more subsequent
phases of silicon shell burning occur until the critical iron core
mass is exceeded.  The sizes of these shells depend on the details of
the preceding evolution.  As a result, the iron core mass does not
necessarily increase monotonically with initial mass or rotation.  For
example, the iron core of {\ModA{D20}} is larger than those of the
$20\,\Msun$ models of the ``\Mod{B}'' series.

\section{Implications for young pulsars and supernovae}
\lSect{youngPSR}

\pTab{jev}

{\Tab{jev}} shows, for times during the evolution, the specific
angular momentum contained in the innermost $1.7\,\Msun$ (the mass of
the iron core at core collapse) for {\ModA{E20}}.  Due to the
continuous contraction of the central region of the star, it spins up
and gets closer to critical rotation {\Eqx{awawk}}.  If the
pre-collapse value of the specific angular momentum is applied to a
neutron star with an assumed radius of $12\,\km$, it would rotate with
$90\,\%$ of Keplerian rotation (\Tab{jev}).  {\ModA{E20}} has the
largest iron core mass of all our models (\Tab{FinalModels}), and a
lower core specific angular momentum than the models computed with
$\fmu = 0.05$.  Those models have even much more angular momentum in
the collapsing iron core than a neutron star can possibly carry
($\omega/\wk\propto jm^{-1/2}$).  This much angular momentum would
certainly be important in the dynamics of core collapse, and it is
expected that significant deviations from spherical symmetry will
arise {\Cite{Ims95,aks97,ZM97,RMR98}}.

\subsection{Comparison with observed young pulsars}
\lSect{pulsCmp}

\pTab{jNS}

At $90\,\%$ of Keplerian angular velocity, the neutron star 
which might form in the collapse of
the iron core of {\ModA{E20}} would have a rotation period of
$1\,\ms$ (\Tab{jev}). 
In {\Tab{jNS}} the periods of the four known young neutron stars
associated with supernova remnants {\Cite{mas98}} are given along with
their specific angular momentum (with the same assumption regarding
moment of inertia as above).  Comparing this to the specific angular
momentum in the iron cores found in the pre-collapse models in
{\Tab{FinalModels}}, we see that the iron cores of our models have
roughly $\sim20$ to $100$ times more specific angular momentum than
found in these neutron stars.  Triaxial deformations and gravitational
radiation would result --- even during the explosion.  Still it might
be expected that the resulting neutron stars would spin much faster
than observed.

However, the observed ``young'' neutron stars have ages of several
hundred years.  They might have spun much faster immediately
after their formation.  In fact, it has been proposed recently 
that rapidly
rotating hot neutron stars are spun down on a time-scale of 
one year by r-mode oscillations and 
accompanying emission of gravitational radiation.
These oscillations are supposed to cease at spin periods
compatible with those observed in the young neutron stars 
{\Cite{LOM98,owe98}}.

Alternatively, an important angular momentum transport mechanism might
be missing in our models (see also {\Sect{OtherInst}}).  {\cite{SP98}}
have assumed, without computing detailed models, that the winding up
of weak magnetic fields by differential rotation can cause enough
Maxwell stress to keep the entire star in uniform rotation until the
end of central carbon burning.  This scenario approaches the problem of
the young neutron star periods from the other side: It implies initial
spin periods of $\sim100\,\Sec$.  Since this is much larger than
observed, they employed off-center ``kicks'' during the supernova
explosion to spin them up to the observed rotation rates.  This
scenario is speculative at present, since neither the evolution of
magnetic instabilities in the stellar interior nor the neutron star
kicks have been adequately investigated.

\subsection{Formation of Kerr black holes?}

\pFig{m-jLSO-E15AB}

If the large angular momenta obtained for the iron cores in this work
pose a problem for pulsars, they are very favorable for the collapsar
model for $\gamma$-ray bursts {\Cite{Woo93}}.  If the cores of the
stars would collapse to a black hole, the angular momentum calculated
here would be enough to support matter in a stable disk outside
{\Cite{ST83, Nov97}}.  This is indicated in {\Fig{m-jLSO-E15AB}},
where the distribution of the specific angular momentum at the
pre-collapse stage of the two $20\,\Msun$ {\ModB{E20}{E20B}} is shown.
Thin dashes and dash-dotted lines indicate the specific angular
momenta of the last stable orbit around a non-rotating and a maximum
rotating black hole with a mass equal to the mass coordinate.  If the
matter in the star has more angular momentum than necessary to get
into the last stable orbit, an accretion disc must form, and
efficiently transform gravitational binding energy into heat, up to
$42.3\,\%$ of the rest mass for a maximum rotating black hole
{\Cite{ST83, Nov97}}.  Note that the specific angular momentum
displayed in {\Fig{m-jLSO-E15AB}} is the latitudinal average over a
shell.  Its actual value at the equator is higher than that by
$50\,\%$, while it is zero at the pole.  Therefore matter might fall
in almost freely along the rotation axis, while it hits the
centrifugal barrier at the equator.  In case a prompt supernova
explosion fails and a black hole forms instead of a neutron star, this
might be a mechanism for an efficient energy source for supernovae or
even a $\gamma$-ray burst {\Cite{Woo93,PWF98,MW99}}.

\section{Summary and conclusions}
\lSect{SumConcl}

We have presented the first complete numerical simulation of the
evolution of rotating stars from the ignition of nuclear burning until
the supernova stage.  Emphasis has been placed on the modification of
the evolution induced by rotation.  This includes an examination of
the transport processes responsible for redistributing each angular
momentum and composition and the resultant changes that occur in the
stellar structure and nucleosynthesis.  The distribution of angular
momentum in the presupernova stage is of particular interest.


Two different one-dimensional hydrodynamic stellar evolution codes
were modified to include angular momentum as a new local variable.
The effects of centrifugal forces on the stellar structure were
treated in latitudinally averaged way.  Rotationally induced
instabilities were included (\Sect{MixInst}): secular and dynamic
shear instabilities, the Solberg-H{\o}iland instability, the
Eddington-Sweet circulation, and the Goldreich-Schubert-Fricke
instability.  The uncertain parameters of rotationally induced mixing
were calibrated using observational constraints on the surface
abundances (\Sect{calib}).  Observed surface enrichments with
CNO-processed matter were reproduced for stars in the mass range from
$5\,\Msun$ to $60\,\Msun$, for typical initial stellar rotation rates.
Stellar mass loss and its dependence on the surface rotation rate were
also taken into account (\Sect{MassLoss}).

The evolution of stars of approximately solar composition in the mass
range from $10\,\Msun$ to $25\,\Msun$ was modeled up to iron core
collapse, the immediate presupernova stage.
Models that used different assumptions regarding the stabilizing
effect of gradients in the mean molecular weight on rotationally
induced instabilities were computed and compared.  Observations 
indicate that gradients in the mean molecular weight inhibit
rotationally induced mixing much less than in the pioneering models of
{\cite{ES78}}.  This conclusion is also supported by recent
investigations of the physics of meridional circulations, shear
instabilities, and semiconvective mixing {\Cite{MZ98}}.

\subsection{Internal stellar structure}

During central hydrogen burning, the products of the burning are
mixed into the stellar envelope and new fuel is supplied to the
convectively burning stellar core by rotationally induced mixing.
Since this mixing proceeds on a time-scale comparable to the
thermonuclear time-scale of hydrogen burning, a gradient of
processed matter builds up inside the radiative envelope.  The
processed matter has a higher mean molecular weight, $\mu$, than the
pristine matter of the star, and therefore a gradient of the mean
molecular weight results.

If rotationally induced mixing occurs by processes that depend
sensitively upon these gradients, they act as a barrier
($\mu$-barrier), and mixing between the core and the envelope is
inhibited.  Exactly when this inhibition becomes important depends
on the initial angular momentum of the star.  The amount of mixing
that occurs between the core and the envelope is affected accordingly.
Mixing inside the envelope also increases for larger initial angular
momentum, since the dominant mixing process, Eddington-Sweet
circulation, has an efficiency that increases as the square of the
stellar rotation rate.  

As the evolution of the star proceeds to later stages, the time-scale
for rotationally induced mixing becomes too long in comparison to the
evolutionary time-scales to constitute an important source of
large-scale mixing.  Also, the mixing is not able to dissolve the
molecular weight barrier which forms in the core during central helium
burning.  In general, rotationally induced mixing does not strongly
affect the stellar structure after central helium ignition.  The
evolution of the star from this point until core collapse is similar
to that of a non-rotating star of same structure at this time, except
for the differences in the nucleosynthesis discussed below.

For models where rotationally induced mixing is assumed to be
insensitive to gradients in the mean molecular weight, no
$\mu$-barrier inhibits the mixing.  This affects the mixing between
the core and the envelope during central hydrogen burning.  The
$\mu$-barrier in the superadiabatic part of the core during central
helium burning is eroded by shear instabilities.  Consequently, the
convective core can grow unhindered.  As a result, the helium cores
are more massive, corresponding to non-rotating stars with about
$25\,\%$ higher initial mass.  Inside this helium core, the {\CO} core
is also larger than that of a non-rotating star with same helium core
mass.  Towards the end of central helium burning, fresh helium is
mixed into the convective core both by the continuing growth of this
core and by rotationally induced mixing.  The fresh helium
preferentially converts carbon into oxygen instead of producing new
carbon by the triple-alpha process.  This reduces the carbon abundance
in the core.  Except for this, the effect of rotationally induced
mixing is small after helium ignition for the reason outlined above.
In particular, also in this case, the hydrogen burning shell
constitutes an efficient barrier for mixing processes --- indeed even
more efficient, because the core rotates slower as in the case where
$\mu$-gradients were considered (see below).  A consequence of the
enlarged cores is that the limit on the initial stellar mass for core
collapse supernovae is somewhat smaller for higher initial rotation
rates.

\subsection{Angular momentum}

At central hydrogen ignition, the stars establish almost uniform
rotation.  If a molecular weight barrier forms as hydrogen burning
progresses, angular momentum is trapped inside the core and
differential rotation results, with up to a factor of $\sim3$
variation in the rotation rate between the core and the envelope.  If
$\mu$-barriers are unimportant for the rotationally induced mixing,
the stars stay close to rigid rotation until the end of central
hydrogen burning.  Since this barrier forms later in the faster
rotators, stars having different initial rotation rates may end up
with similar specific angular momenta in the core at the end of
central hydrogen burning
(\FigCCc{D}{m-wjJ53-E15AB}{C}{m-wjJ53-GF15B}{D}).  Due to angular
momentum transport during core helium burning, they may become even
more similar in the pre-collapse stage ({\Fig{jj15B}}).  Some angular
momentum gets lost from the core during the restructuring that occurs
after core hydrogen exhaustion, but during central helium burning the
hydrogen-burning shell constitutes an efficient barrier that inhibits
the transport of angular momentum out of the core.  Even so, the
average angular momentum of the core may decrease somewhat, since it
grows into regions with lower specific angular momentum on top of it.
The helium core itself stays close to uniform rotation
(\FigBb{A}{m-wjJ53-E15AB}{B}).  During central helium burning
rotationally induced mixing processes already become slow compared to
the evolution, and after core helium exhaustion they do not cause any
relevant transport of angular momentum.

Only convective processes are rapid enough to notably redistribute
angular momentum during the late stages of stellar evolution.  Within
the assumptions made, rigid rotation results in convective regions,
transporting angular momentum from their bottom to their top.
Subsequent phases of convective central and shell burning stages give
some outward transport of angular momentum inside the carbon-oxygen
core.  Since none of the convective shells penetrates through the
outer boundary of the {\CO} core, the angular momentum remains trapped
inside (\FigCCc{F}{m-wjJ53-E15AB}{E}{m-wjJ53-GF15B}{F}).  The outer
boundary of the helium core constitutes a similar barrier.

This has interesting consequence for the final angular momentum in the
core.  First, the different convective burning shells leave their
fingerprint not only in the chemical composition, but also in the
angular momentum distribution: a spiky profile results at the onset of
core collapse (\Fig{jj15B}).  The high peaks correspond to the upper
edges of the most recently active convection zones and the deep
valleys to their bottoms.  Shells of similar composition tend to
rotate almost rigidly.  Second, even the slowest rotating core of the
Type~II supernova progenitor stars considered here would result in a
neutron star rotating close to break-up if angular momentum were
conserved during the collapse.  This is not necessarily in
contradiction with observations of young neutron stars in supernova
remnants, even though the fastest of these rotates much slower.  These
pulsars are already hundreds of years old, and recent theoretical
investigations of hot, newly born neutron stars indicate they may spin
down to the observed rotation rates within about a year by emitting
gravitational waves {\Cite{LOM98}}.  The electro-magnetic radiation
emanating from pulsars is trapped inside the supernova ejecta during
that time, but the gravitational radiation of these very young neutron
stars might become detectable in the future {\Cite{owe98}}.

\acknowledgments

We are grateful to T.A.~Weaver for helpful discussions and aid with
KEPLER.  This work was supported by the Deutsche Forschungsgemeinschaft
through grants La~587/15 and La~587/16, by the the National
Science Foundation (AST 97-31569), and by the Alexander von Humboldt
Foundation.  AH was, in part, supported by a ``Doktorandenstipendium
aus Mitteln des 2.~Hochschulprogramms''.

\newpage

\begin{appendix}

\section{Improvement of hydrogen burning in the KEPLER code}
\lSect{KepImpr}

The treatment of nuclear burning in KEPLER was improved by taking into
account neutrino losses and corrections to the energy released during
hydrogen burning.


In general, until oxygen burning, KEPLER calculates the nuclear energy
generation rate by simple subtraction of the total nuclear binding
energy of the composition before and after a time-step.  However,
during hydrogen burning weak interactions are present, with
implications for the energy generation which do not occur in other
pre-oxygen burning stages.  In particular, the mass difference of
{\El{n}} and {\El{p+e^{-}}} of $782.3\,\MeV\,\clight^{-2}$ has to be
taken into account.  This reduces the energy release per {\I{4}{He}}
formed by twice that value, i.e., from $28.296\,\MeV$ for the binding
energy of two protons and two neutrons, to $26.731\,\MeV$ for the
total difference in rest mass.  That is a relative reduction of the
energy generation rate of hydrogen burning by $5.85\,\%$.


Additionally, within the CNO main cycle the weak reaction
{\I{13}{N}}\Rbp{\I{13}{C}} releases in average $0.71\,\MeV$ in the
form of neutrinos, {\I{15}{O}}\Rbp{\I{15}{N}} releases $1.0\,\MeV$,
and the reaction {\I{17}{F}}\Rbp{\I{17}{O}} of the secondary CNO
cycle, which has a probability of only $\sim10^{-4}$ relative to the
main cycle, $0.94\,\MeV$ {\Cite{Cla68}}.  On average, $1.71\,\MeV$ are
lost due to neutrinos for each four {\I{1}{H}} burned, reducing the
effective energy release by another $6.4\,\%$ from $26.731\,\MeV$ to
about $24.97\,\MeV$ per {\I{4}{He}} formed.  KEPLER was also corrected
for the appropriate neutrino losses of the three {\El{pp}} chains, but
they do not contribute in the high-mass stars of the present work.


These two effects sum up to a total reduction of the energy release of
hydrogen burning by $13.32\,\%$.  However, this alters 
the structure of zero-age main sequence stars only slightly, since the
energy generation rate for CNO hydrogen burning depends on a high
power of the temperature ($16$ to $18$) and thus a slight increase of
the central temperature compensates the lower energy generation rate.
The nucleosynthesis, as far as the ratios of the CNO isotopes are
concerned, are not altered notably by the higher temperature.

The lower total energy release of hydrogen burning reduces the
lifetime of the star during central hydrogen burning by about the same
factor.  The convective mixing in the core is not affected by the
shorter duration of central hydrogen burning, since the convective
mixing time-scale is still much shorter.  For mixing processes which
act on a time-scale comparable to that of central hydrogen
burning, e.g., semiconvection, the reduction of the main sequence
lifetime can become noticeable.  It is particularly important for
rotationally induced mixing and mass loss from the stellar surface.
Mass loss rates mostly depend only on stellar surface properties
(\Sect{MassLoss}) and thus, to first approximation, the total mass
loss scales with the evolutionary time-scale.  The stellar angular
momentum loss from the surface is also altered by this and the amount
of products of hydrogen burning exposed to the surface of the star
depends on both the time available for the rotationally induced mixing
processes to transport them to the surface and the amount of matter
removed from the surface.  Furthermore, a higher angular momentum loss
from the surface can introduce more shear and correspondingly more
mixing.

After termination of core hydrogen burning, hydrogen continues
burning in a shell above the helium core.  Since the stellar structure
determines the rate at which energy has to be released by hydrogen
burning, more hydrogen is burned in order to release the same amount
of energy using the corrected energy generation rate.  This increases
the rate at which the helium core grows and thus its final size.

In all models computed with KEPLER in the present work, the corrected
treatment of hydrogen burning was employed.

\section{Diagrams for the evolution of the internal stellar structure}
\lSect{convDiag}
\lSect{ExplConv}

The ``Kippenhahn diagrams'' of this Appendix show the convective
structure and the nuclear energy generation rate as a function of time
for selected model sequences.  In these plots the time-axis gives the
logarithm of the time left until core collapse (in $\yr$).  
From the infall velocities occurring in the last computed models and
from the radius at which the maximum infall velocity occurs
($\sim10^3\,\kms$ and $10^3\,\km$, respectively) we estimated
$10^{-7.5}\,\yr$ ($\sim1\,\Sec$) for the time until ``core bounce''.

The $y$-axis gives the (interior) mass coordinate, $m$, in units of
the solar mass.  The total mass of the star is indicated by a thick
solid line and shows the mass loss from the stellar surface.  Note
that because of the way the time-axis is chosen, the slope of this
curve does not directly correspond to the mass loss rate.

Diagonal hatching indicates convective regions.  In order to better
visualize the boundaries of convection zones they are framed by thin
lines.  Semiconvection is indicated by narrow cross-hatching.

The net nuclear contribution to the energy generation rate including
neutrino losses are shown as grey shading.  Increasingly darker grey
scale levels correspond to increasing orders of magnitude of the
energy generation rate.  The lightest grey shows regions with an
energy generation rate of $\Ep{-1}\,\erggs$ or more, the next darker
shade of grey underlies regions with an energy generation rate of
$\Ep{0}\,\erggs$ or above, and so forth.

The sequence of central convective burning phases is (from left to
right) hydrogen, helium, carbon, neon, oxygen, and silicon burning.

\end{appendix}

\pFig{D15cnv}

\pFig{E15cnv}

\pFig{E15Bcnv}


\clearpage
\onecolumn

\begin{table}
\centering
\begin{tabular}[t]{r@{}ll|r@{}ll|r@{}ll}
\hline\hline
\multicolumn{2}{r}{isotope} &
\multicolumn{1}{c|}{mass fraction} &
\multicolumn{2}{r}{isotope} &
\multicolumn{1}{c|}{mass fraction} &
\multicolumn{2}{r}{isotope} &
\multicolumn{1}{c}{mass fraction} \\ 
\hline
\noalign{\smallskip}
\IT{ 1}{H } & $7.00\E{- 1}$ & \IT{14}{N } & $1.07\E{- 3}$ & \IT{24}{Mg} & $5.60\E{- 4}$ \\
\IT{ 3}{He} & $3.78\E{- 5}$ & \IT{15}{N } & $3.92\E{- 6}$ & \IT{25}{Mg} & $7.09\E{- 5}$ \\
\IT{ 4}{He} & $2.80\E{- 1}$ & \IT{16}{O } & $1.04\E{- 3}$ & \IT{26}{Mg} & $7.81\E{- 5}$ \\
\IT{ 6}{Li} & $6.69\E{-10}$ & \IT{17}{O } & $3.95\E{- 6}$ & \IT{27}{Al} & $6.24\E{- 5}$ \\
\IT{ 7}{Li} & $9.63\E{- 9}$ & \IT{18}{O } & $2.08\E{- 5}$ & \IT{28}{Si} & $7.08\E{- 4}$ \\
\IT{ 9}{Be} & $1.72\E{-10}$ & \IT{19}{F } & $3.89\E{- 7}$ & \IT{29}{Si} & $3.58\E{- 5}$ \\
\IT{10}{B } & $1.04\E{- 9}$ & \IT{20}{Ne} & $1.77\E{- 3}$ & \IT{30}{Si} & $2.37\E{- 5}$ \\
\IT{11}{B } & $4.92\E{- 9}$ & \IT{21}{Ne} & $4.31\E{- 6}$ & \IT{56}{Fe} & $1.37\E{- 3}$ \\
\IT{12}{C } & $3.62\E{- 3}$ & \IT{22}{Ne} & $1.29\E{- 4}$ & 	                        \\
\IT{13}{C } & $4.03\E{- 5}$ & \IT{23}{Na} & $3.60\E{- 5}$ & 	                        \\
\hline
\end{tabular}
\caption{
Initial isotopic mass fractions for the models computed with the STERN
code (\Sect{STERN}).  They are taken from {\cite{GN93}}.  The
initial abundances of the radioactive isotopes {\I7{Be}}, {\I8B},
{\I{11}C}, {\I{12}N}, and {\I{26}{Al}}, and that of {\I2H} are set to
zero.
\lTab{InitAbuSTERN}}
\end{table}

\clearpage

\begin{table}
\centering
\begin{tabular}[t]{r@{}ll|r@{}ll|r@{}ll}
\hline\hline
\multicolumn{2}{r}{isotope} &
\multicolumn{1}{c|}{mass fraction} &
\multicolumn{2}{r}{isotope} &
\multicolumn{1}{c|}{mass fraction} &
\multicolumn{2}{r}{isotope} &
\multicolumn{1}{c}{mass fraction} \\ 
\hline
\noalign{\smallskip}
\IT{ 1}{H } & $7.0000\E{ -1}$ & \IT{16}{O } & $1.0175\E{ -2}$ & \IT{36}{Ar} & $1.0176\E{ -4}$ \\ 
\IT{ 3}{He} & $2.9798\E{ -5}$ & \IT{20}{Ne} & $1.8545\E{ -3}$ & \IT{40}{Ca} & $6.9515\E{ -5}$ \\
\IT{ 4}{He} & $2.7997\E{ -1}$ & \IT{24}{Mg} & $7.3366\E{ -4}$ & \IT{48}{Cr} & $3.1177\E{ -6}$ \\
\IT{12}{C } & $3.2467\E{ -3}$ & \IT{28}{Si} & $8.1332\E{ -4}$ & \IT{52}{Fe} & $1.9180\E{ -5}$ \\
\IT{14}{N } & $1.1732\E{ -3}$ & \IT{32}{S } & $4.5056\E{ -4}$ & \IT{54}{Fe} & $1.3594\E{ -3}$ \\
\hline
\end{tabular}
\caption{
Initial isotopic mass fractions for the models computed with the
KEPLER code (\Sect{KEPLER}).  The initial abundances of the
radioactive isotopes {\I{44}{Ti}} and {\I{56}{Ni}}, which are also in
the network, are set to zero.
}
\lTab{InitAbuKEPLER}
\end{table}

\clearpage

\begin{table}
\centering
\newcommand{\phA}{\phantom{.00}}
\begin{tabular}[t]{lrrrrrl}
\hline\hline
& $\Mi$
& $\Ji$
& $\vZAMS$
\\
\raisebox{1.5ex}[0pt]{model} 
& $\Msun$
& $10^{52}\,\ergs$
& $\kms$
& \raisebox{1.5ex}[0pt]{$\fc$} 
& \raisebox{1.5ex}[0pt]{$\fmu$} 
& \raisebox{1.5ex}[0pt]{evolution followed until}
\\
\hline
\Mod{D10 } & 10 &    0 &   0 &    - &    - & carbon shell burning \\
\Mod{D12 } & 12 &    0 &   0 &    - &    - & off-center neon burning\\
\Mod{D15 } & 15 &    0 &   0 &    - &    - & core collapse\\
\Mod{D20 } & 20 &    0 &   0 &    - &    - & core collapse\\
\Mod{D25 } & 25 &    0 &   0 &    - &    - & core collapse\\
\hline 
\Mod{E08 } &  8 & 0.53 & 205 & 0.01 &    0 & carbon shell burning\\
\Mod{E10 } & 10 & 0.80 & 207 & 0.01 &    0 & core collapse\\
\Mod{E12 } & 12 & 1.10 & 206 & 0.01 &    0 & core collapse\\
\Mod{E15 } & 15 & 1.60 & 206 & 0.01 &    0 & core collapse\\
\Mod{E20 } & 20 & 2.50 & 201 & 0.01 &    0 & core collapse\\
\Mod{E25 } & 25 & 3.50 & 205 & 0.01 &    0 & core collapse\\
\hline 
\Mod{G12 } & 12 & 0.55 & 100 & 0.01 &    0 & core hydrogen exhaustion\\
\hline 
\Mod{F12 } & 12 & 1.65 & 327 & 0.01 &    0 & core hydrogen exhaustion\\
\hline 
\Mod{E12B} & 12 & 1.10 & 206 & 1/30 & 0.05 & off-center neon burning\\
\Mod{E15B} & 15 & 1.60 & 206 & 1/30 & 0.05 & core collapse\\
\Mod{E20B} & 20 & 2.50 & 201 & 1/30 & 0.05 & core collapse\\
\hline 
\Mod{F12B} & 12 & 1.65 & 328 & 1/30 & 0.05 & carbon shell burning\\
\Mod{F15B} & 15 & 2.40 & 323 & 1/30 & 0.05 & core collapse\\
\Mod{F20B} & 20 & 3.75 & 307 & 1/30 & 0.05 & core collapse\\
\hline 
\Mod{G12B} & 12 & 0.55 &  99 & 1/30 & 0.05 & carbon shell burning\\
\Mod{G15B} & 15 & 0.80 & 102 & 1/30 & 0.05 & core collapse\\
\Mod{G20B} & 20 & 1.25 & 103 & 1/30 & 0.05 & core collapse\\
\hline 
\Mod{H12B} & 12 & 2.20 & 474 & 1/30 & 0.05 & carbon shell burning\\
\Mod{H15B} & 15 & 3.20 & 457 & 1/30 & 0.05 & core hydrogen exhaustion\\
\Mod{H20B} & 20 & 5.00 & 425 & 1/30 & 0.05 & core helium burning\\
\hline
\end{tabular}
\caption{
Parameters of the model sequences.  Shown are the model name (left
column), the initial mass $\Mi$, the initial angular momentum $\Ji$,
the equatorial rotational velocity at the surface at central hydrogen
ignition $\vZAMS$, and two parameters of rotationally induced mixing,
$\fc$ and $\fmu$.  The last column gives the final evolutionary stage
to which the models are evolved.
\lTab{InitModels}}
\end{table}

\clearpage

\begin{table}
\centering
\begin{tabular}{l||llll|llll}
\hline\hline
central 
& \multicolumn{4}{c|}{\Mod{E15}}
& \multicolumn{4}{c}{\Mod{E15B}}
\\ \cline{2-9}
burning 
& $\tauKHx$
& $\omega/\wk$
& $\tauES$
& $\tauB$
& $\tauKHx$
& $\omega/\wk$
& $\tauES$
& $\tauB$
\\
phase 
& $\yr$
& 
& $\yr$
& $\yr$ 
& $\yr$
& 
& $\yr$
& $\yr$ 
\\
\hline
hydrogen & $5\E{4} $ & $5\E{-2}$ & $2\E{7}$ & $1\E{7} $ & $5\E{4} $ & $5\E{-2}$ & $2\E{7}$ & $1\E{7} $ \\
helium   & $5\E{4} $ & $2\E{-2}$ & $1\E{8}$ & $1\E{6} $ & $7\E{4} $ & $6\E{-2}$ & $2\E{7}$ & $1\E{6} $ \\
carbon   & $4\E{4} $ & $3\E{-2}$ & $4\E{7}$ & $4\E{2} $ & $6\E{4} $ & $8\E{-2}$ & $9\E{6}$ & $4\E{3} $ \\
oxygen   & $1\E{1} $ & $5\E{-2}$ & $3\E{3}$ & $1\E{0} $ & $3\E{1} $ & $1\E{-1}$ & $3\E{3}$ & $5\E{0} $ \\
silicon  & $3\E{-1}$ & $8\E{-2}$ & $5\E{1}$ & $1\E{-2}$ & $3\E{-1}$ & $2\E{-1}$ & $1\E{1}$ & $3\E{-2}$ \\
\hline
\end{tabular}
\caption{
Values of the Kelvin-Helmholtz time-scale $\tauKHx$
{\Eqx{tauKHx}}, the ratio of angular velocity to the Keplerian
angular velocity $\omega/\wk$, the Eddington-Sweet circulation
time-scale $\tauES$ {\Cite{Zah92}}, and the burning time-scale $\tauB$ for
the core region during the major nuclear burning phases, for models
{\Mod{E15}} (left) and {\Mod{E15B}} (right).
\lTab{tauES}}
\end{table}

\clearpage

\begin{landscape}
\begin{table}
\notetoeditor{make table sideways}
\newcommand{\fA}{{^a}}
\newcommand{\fB}{{^b}}
\newcommand{\fC}{{^c}}
\begin{tabular}{lrrlrrrrrrrrr}
\hline\hline
& \multicolumn{1}{r}{$\Mi $}
& \multicolumn{1}{r}{$\Mf $}
& \multicolumn{1}{r}{$\MHe$}
& \multicolumn{1}{r}{$\MCO$}
& \multicolumn{1}{r}{$\MFe$}
& \multicolumn{1}{r}{$\vZAMS$}
& \multicolumn{1}{r}{$\Ji $}
& \multicolumn{1}{r}{$\Jf $}
& \multicolumn{1}{r}{$\JHe$}
& \multicolumn{1}{r}{$\JCO$}
& \multicolumn{1}{r}{$\JFe$}
& \multicolumn{1}{r}{$\jFe$}
\\
\raisebox{1.5ex}[0pt]{model}
& \multicolumn{1}{r}{$\Msun$}
& \multicolumn{1}{r}{$\Msun$}
& \multicolumn{1}{r}{$\Msun$}
& \multicolumn{1}{r}{$\Msun$}
& \multicolumn{1}{r}{$\Msun$}
& \multicolumn{1}{r}{$\kms$}
& \multicolumn{1}{r}{$\ergs$}
& \multicolumn{1}{r}{$\ergs$}
& \multicolumn{1}{r}{$\ergs$}
& \multicolumn{1}{r}{$\ergs$}
& \multicolumn{1}{r}{$\ergs$}
& \multicolumn{1}{r}{$\junit$}
\\
\hline
\Mod{D10 } & $10$ &  $9.61$ & $1.24\fA$
                                     & $1.22$ & $-\fB$ &   $0$ &          $0$ &          $0$ &          $0$ &          $0$ &          $0$ &          $0$ \\
\Mod{D12 } & $12$ & $11.42$ & $2.85$ & $1.42$ & $-\fB$ &   $0$ &          $0$ &          $0$ &          $0$ &          $0$ &          $0$ &          $0$ \\
\Mod{D15 } & $15$ & $13.55$ & $3.82$ & $1.77$ & $1.33$ &   $0$ &          $0$ &          $0$ &          $0$ &          $0$ &          $0$ &          $0$ \\
\Mod{D20 } & $20$ & $16.31$ & $5.68$ & $2.31$ & $1.64$ &   $0$ &          $0$ &          $0$ &          $0$ &          $0$ &          $0$ &          $0$ \\
\Mod{D25 } & $25$ & $18.72$ & $7.86$ & $3.11$ & $1.36$ &   $0$ &          $0$ &          $0$ &          $0$ &          $0$ &          $0$ &          $0$ \\
\hline 
\Mod{E08 } & $ 8$ &  $7.65$ & $1.38\fA$
                                     & $1.35$ & $-\fB$ & $205$ & $5.30\E{51}$ & $2.49\E{51}$ & $2.40\E{49}$ & $2.36\E{49}$ & $-\fB      $ & $-\fB      $ \\
\Mod{E10 } & $10$ & $ 9.23$ & $2.84$ & $1.78$ & $1.36$ & $207$ & $8.00\E{51}$ & $2.09\E{51}$ & $8.04\E{49}$ & $3.89\E{49}$ & $2.13\E{49}$ & $7.87\E{15}$ \\
\Mod{E12 } & $12$ & $10.35$ & $3.63$ & $2.37$ & $1.34$ & $206$ & $1.10\E{52}$ & $1.29\E{51}$ & $1.20\E{50}$ & $5.32\E{49}$ & $1.50\E{49}$ & $5.63\E{15}$ \\
\Mod{E15 } & $15$ & $10.86$ & $5.10$ & $3.40$ & $1.46$ & $206$ & $1.60\E{52}$ & $1.38\E{51}$ & $2.30\E{50}$ & $1.15\E{50}$ & $1.86\E{49}$ & $6.40\E{15}$ \\
\Mod{E20 } & $20$ & $11.00$ & $7.71$ & $5.01$ & $1.73$ & $201$ & $2.50\E{52}$ & $7.15\E{50}$ & $3.95\E{50}$ & $1.82\E{50}$ & $1.92\E{49}$ & $5.58\E{15}$ \\
\Mod{E25 }$\fC$	     	       	             	      	      	  	     	 	        	       		      		     	        
           & $25$ & $ 5.45$ & $5.45$ & $4.07$ & $1.69$ & $205$ & $3.50\E{52}$ & $1.40\E{50}$ & $1.40\E{50}$ & $7.52\E{49}$ & $1.05\E{49}$ & $3.12\E{15}$ \\
\hline 
\Mod{G12B} & $12$ & $11.32$ & $2.68$ & $1.41$ & $-\fB$ & $ 99$ & $5.50\E{51}$ & $3.07\E{51}$ & $1.15\E{50}$ & $4.18\E{49}$ & $-\fB      $ & $-\fB      $ \\
\Mod{G15B} & $15$ & $13.46$ & $3.63$ & $1.79$ & $1.34$ & $102$ & $8.00\E{51}$ & $3.99\E{51}$ & $2.36\E{50}$ & $6.80\E{49}$ & $3.28\E{49}$ & $1.23\E{16}$ \\
\Mod{G20B} & $20$ & $16.03$ & $5.55$ & $2.61$ & $1.38$ & $103$ & $1.25\E{52}$ & $3.49\E{51}$ & $5.33\E{50}$ & $1.24\E{50}$ & $3.13\E{49}$ & $1.14\E{16}$ \\
\hline 
\Mod{E12B} & $12$ & $11.25$ & $2.72$ & $1.46$ & $-\fB$ & $206$ & $1.10\E{52}$ & $4.92\E{51}$ & $1.29\E{50}$ & $4.94\E{49}$ & $-\fB      $ & $-\fB      $ \\
\Mod{E15B} & $15$ & $13.26$ & $3.69$ & $1.89$ & $1.40$ & $206$ & $1.60\E{52}$ & $6.96\E{51}$ & $2.73\E{50}$ & $8.77\E{49}$ & $4.07\E{49}$ & $1.46\E{16}$ \\
\Mod{E20B} & $20$ & $15.19$ & $5.71$ & $2.69$ & $1.38$ & $201$ & $2.50\E{52}$ & $5.10\E{51}$ & $6.36\E{50}$ & $1.55\E{50}$ & $3.47\E{49}$ & $1.26\E{16}$ \\
\hline 
\Mod{F12B} & $12$ & $10.93$ & $3.04$ & $1.61$ & $-\fB$ & $328$ & $1.65\E{52}$ & $3.43\E{51}$ & $1.72\E{50}$ & $5.98\E{49}$ & $-\fB      $ & $-\fB      $ \\
\Mod{F15B} & $15$ & $12.89$ & $3.88$ & $2.01$ & $1.38$ & $323$ & $2.40\E{52}$ & $7.90\E{51}$ & $3.01\E{50}$ & $9.69\E{49}$ & $3.66\E{49}$ & $1.33\E{16}$ \\
\Mod{F20B} & $20$ & $14.76$ & $5.99$ & $2.75$ & $1.36$ & $307$ & $3.75\E{52}$ & $5.45\E{51}$ & $7.42\E{50}$ & $1.71\E{50}$ & $3.71\E{49}$ & $1.37\E{16}$ \\
\hline 
\Mod{H12B} & $12$ & $ 9.77$ & $3.81$ & $1.78$ & $-\fB$ & $474$ & $2.20\E{52}$ & $1.41\E{51}$ & $2.59\E{50}$ & $6.93\E{49}$ & $-\fB      $ & $-\fB      $ \\
\hline
\end{tabular}
\caption{
Properties of the final models of various sequences.  Listed are the
initial and final stellar mass, $\Mi$ and $\Mf$, the final masses of
the helium core $\MHe$, of the carbon/oxygen core $\MCO$ and of the
iron core $\MFe$.  Then, the equatorial surface rotation velocity at
core hydrogen ignition, $\vZAMS$, is given.
Furthermore, the initial stellar angular momentum is given, and the
final angular momentum of the star and the helium, {\CO} and iron cores.
In the last column the average specific angular momentum in the iron
core is shown.
}
\lTab{FinalModels}
\vfill
\rule{10cm}{\arrayrulewidth}\newline
$\fA${\footnotesize dredge-up of helium core}\rule{0.5cm}{0pt}
$\fB${\footnotesize not evolved to pre-collapse stage}\rule{0.5cm}{0pt}
$\fC${\footnotesize star becomes a Wolf-Rayet star during central helium burning}
%
\end{table}
\end{landscape}

\clearpage

\begin{table}
\centering
\begin{tabular}{lccccc}
\hline\hline
evolutionary &
$J(m)/m$ & 
$r$ & 
$\rhoc$ & 
$\omega$ & 
$\omega/\wk$ \\
state & 
$\junit$ & 
$\cm$ & 
$\gccm$ & 
$\rads$ \\
\hline
\El{H}  ignition   & $5.5\E{16}$ & $5.8\E{10} $ &  $  4.8      $ & $5.0\E{-5}$ & $4.6\E{-2}$ \\
\El{H}  exhaustion & $1.0\E{16}$ & $4.3\E{10} $ &  $ 12.5      $ & $1.6\E{-5}$ & $  1\E{-2}$ \\
\El{He} exhaustion & $6.5\E{15}$ & $7.2\E{9}  $ &  $  3\E{3}   $ & $3.8\E{-4}$ & $1.5\E{-2}$ \\
pre-collapse       & $5.6\E{15}$ & $2.2\E{8}  $ &  $3.9\E{9}   $ & $3.7\E{-1}$ & $  0.08   $ \\
neutron star       & $(5\E{15})$ & $(1.2\E{6})$ &  $\sim4\E{14}$ & $1\E{ 4}  $ & $  0.9    $ \\
\hline
\end{tabular}
\caption{
Evolution of the radius, $r$, the angular velocity, $\omega$, and its
ratio to the Keplerian rotational velocity, $\wk=\sqrt{Gm/r^3}$, all
at a mass coordinate of $m=1.7\,\Msun$, and the mass of the iron core
at core collapse of {\ModA{E20}}.  The second column gives the
specific angular momentum $J(m)/m$ of the inner $1.7\,\Msun$.  The
central density $\rhoc$ is given for comparison.  The initial model
has a mass of $20\,\Msun$ and a ZAMS rotational velocity of
$\sim200\,\kms$.  Assuming that the neutron star gets a radius of
about $12\,\km$ and retains about the angular momentum of the iron
core at core collapse, it evolves close to critical rotation.  It is
assumed that the moment of inertia of the rigidly rotating neutron
star of radius $R$ and mass $M$ is given by $0.8\frac{2}{5}MR^2$.  The
geometrical factor $\frac{2}{5}$ corresponds to a solid sphere of
constant density, and the numerical factor $0.8$ is found from a
neutron star model provided by {\cite{Ebe97}}.
\lTab{jev}}
\end{table}

\clearpage

\begin{table}
\centering
\begin{tabular}{lrr}
\hline\hline
& 
\multicolumn{1}{c}{period} & 
$j\;(R=12\,\km)$\\
\raisebox{1.5ex}[0pt]{pulsar} & 
$\ms$ & 
$\junit$ \\
\hline
PSR B0531+21 (Crab)         & $ 33$ & $8.8\E{13}$\\
PSR B0540-69 (LMC)          & $ 50$ & $5.8\E{13}$\\
PSR B1509-58                & $150$ & $1.9\E{13}$\\
PSR J0537-6910 (N157B; LMC) & $ 16$ & $1.8\E{14}$\\
\hline
\end{tabular}
\caption{
Rotational periods of known young pulsars associated with supernova
remnants {\Cite{mas98}} and their specific angular momentum $j=J/M$
for an assumed radius of $R=12\,\km$, if rigid rotation and a moment
of inertia of $I=0.32 MR^2$ is assumed (see {\Tab{jev}}).
\lTab{jNS}}
\end{table}


\clearpage

\begin{figure}
\epsscale{1.0}
\plotone{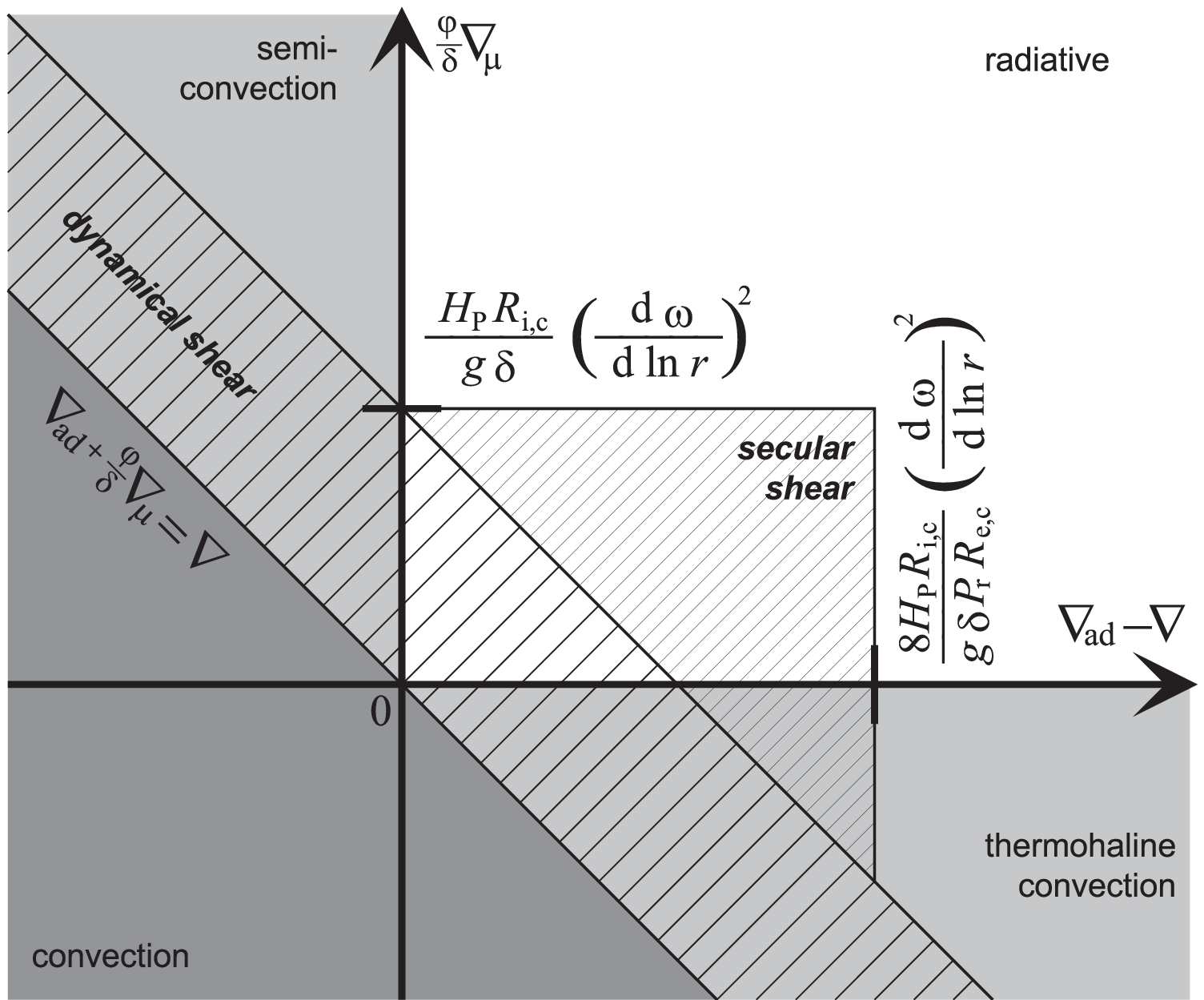}
\caption{
Schematic representation of the regions of different instabilities in
a plane spanned by the stabilizing sub-adiabatic temperature gradient
(horizontal axis), and the stabilizing mean molecular weight gradient
$\Nmu$ (vertical axis).  Both axes have identical units.  Convection
is indicated by dark grey shading, semiconvection (at positive mean
molecular weight gradients) and thermohaline convection (at negative
$\mu$-gradients) are displayed by light grey shading.  The radiatively
stable regime is shown in white.  Rotationally induced instabilities
are indicated by hatched areas.  The dynamical shear instability (wide
thick hatching) acts only up to a certain distance to the
Ledoux-unstable region, which is determined by the amount of shear
(\Fig{DSI}).  The secular shear instability (narrow fine hatching) can
penetrate in regions further away from the convective instability
since it allows for thermal adjustment of displaced mass elements.  It
can also penetrate into the region of stabilizing $\mu$-gradients, but
to a much smaller extent since usually in stars it is
$8/\Brak{\Pr\Rec}\gg 1$ ($\Rec$ is the critical Reynolds number and
$\Pr$ the Prandtl-number; cf. {\Sects{DSI}} and {\Sectff{SSI}}).
\lFig{shear}}
\end{figure}

\clearpage

\begin{figure}
\epsscale{0.5}
\plotone{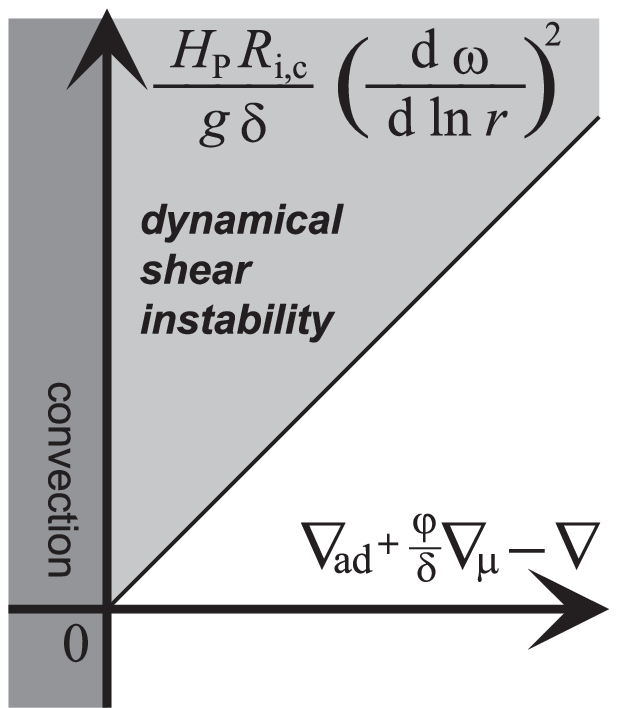}
\caption{
Region of instability due to dynamical shear (light grey) in the plane
of stabilizing temperature and composition stratification (horizontal
axis) and destabilizing shear (vertical axis).  Both axes have the
same units.  Instability due to convection is indicated by dark grey
shading.  See also {\Fig{shear}}.
\lFig{DSI}}
\end{figure}

\clearpage

\begin{figure}
\epsscale{0.5}
\plotone{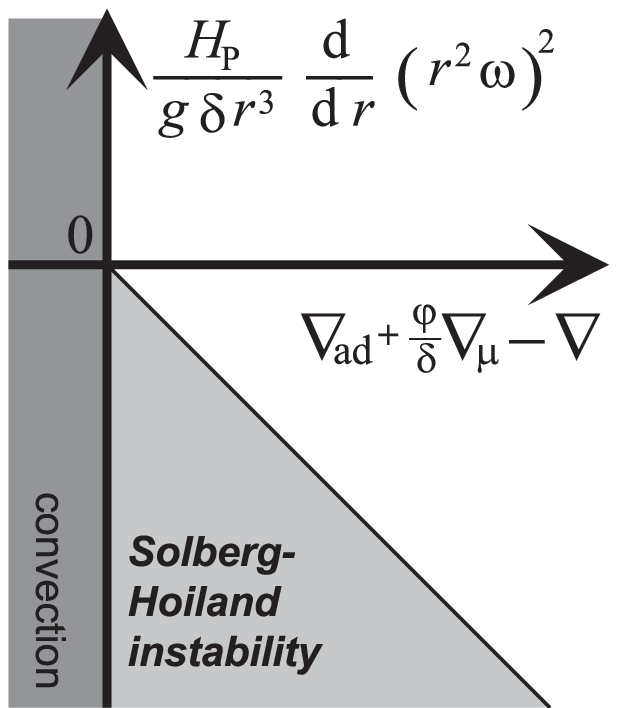}
\caption{
Region of instability according to the Solberg-H{\o}iland criterion
(light grey) in the plane of stabilizing temperature and composition
stratification (horizontal axis) and stabilizing increasing specific
angular momentum ($j\sim r^2\omega$) (vertical axis).  Both axes have
identical units.  Instability due to convection is indicated by dark
grey shading.  The Solberg-H{\o}iland instability can occur in regions
where the specific angular momentum decreases outwards.
\lFig{SHI}}
\end{figure}

\clearpage

\begin{figure}
\epsscale{0.5}
\plotone{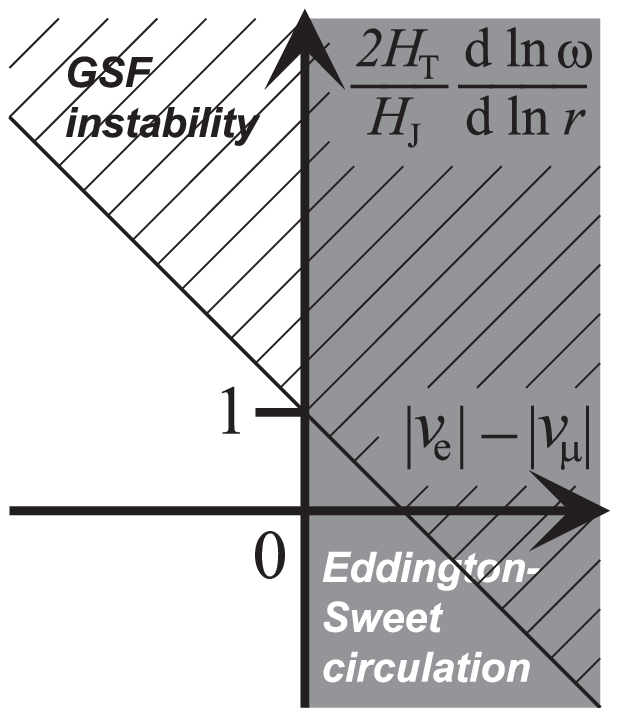}
\caption{
Region of instability due to Eddington-Sweet circulation (dark grey)
and the Goldreich-Schubert-Fricke instability (hatching) in the plane
of residual Eddington-Sweet circulation velocity (horizontal axis) and
the factor $\vGSFO/\vESO$ (vertical axis) of Goldreich-Schubert-Fricke
velocity relative to Eddington-Sweet circulation velocity in the
absence of $\mu$-gradients.  The boundary of the GSF-unstable region
intersects the $x$-axis at $\abs{\vESO}-\abs{\vmu}=\abs{\vESO}$, i.e.,
where $\abs{\vmu}=0$.
\lFig{GSF}}
\end{figure}

\clearpage

\begin{figure}
\epsscale{0.4}
\plotone{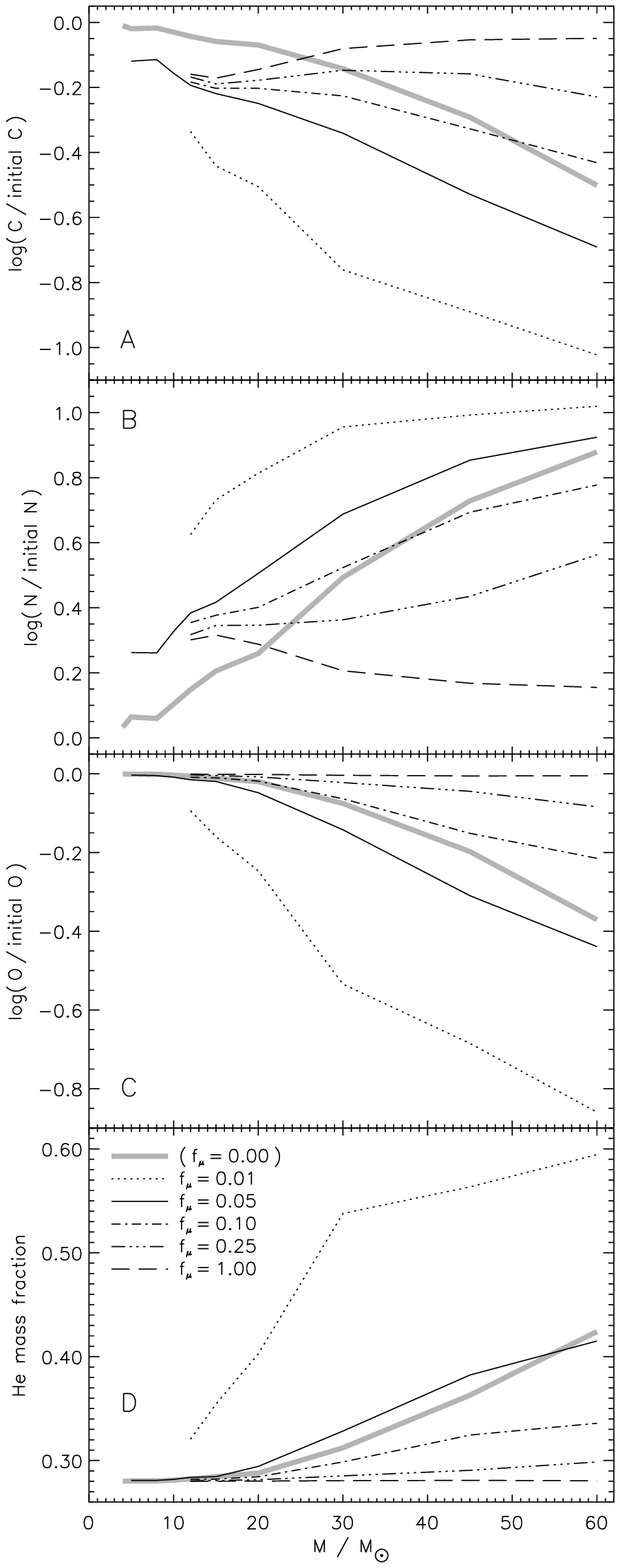}
\caption{
Surface abundances at core hydrogen exhaustion as function of the
initial stellar mass for models with an equatorial surface rotation rate of
$\sim 200\,\kms$ at hydrogen ignition.  Thin lines
correspond to different values of $\fmu$ (\Sect{calib}) for fixed
$\fc=1/30$.  The thick grey line corresponds to $\fc=1/100$ and
$\fmu=0$.  In {\PanRange{A}{C}}, the change of the mass fractions of
carbon, nitrogen, and oxygen, respectively, relative to the initial
values are shown; {\Pan{D}} shows the surface helium mass fractions.  
Models with
initial masses of $12\,\Msun$, $15\,\Msun$, $20\,\Msun$, $30\,\Msun$,
$45\,\Msun$, and $60\,\Msun$ have been calculated with the KEPLER code.
For $\fmu=0.05$ (solid thin line) and
$\fmu=0$ (grey thick line), additionally $5\,\Msun$, $8\,\Msun$, and
$10\,\Msun$ stars are computed, and for the case $\fmu=0$ also a
$4\,\Msun$ model is calculated.
\lFig{M-CNOHe-gauge}}
\end{figure}

\clearpage

\begin{figure}
\epsscale{0.5}
\plotone{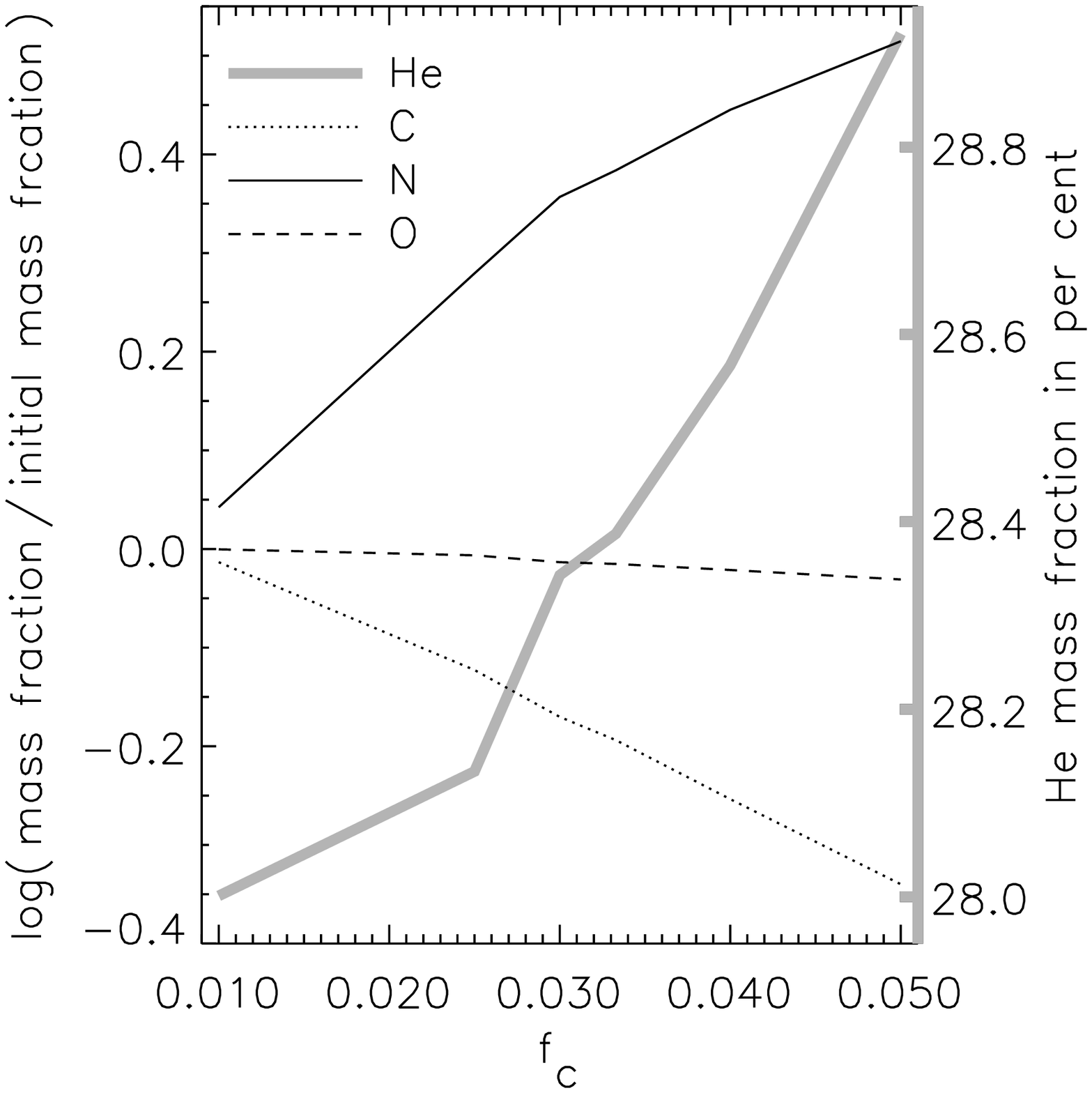}
\caption{
Surface abundances at core hydrogen exhaustion as a function of $\fc$,
for $12\,\Msun$ stars with an equatorial rotational velocity at the
surface of $\sim 200\,\kms$ at hydrogen ignition.
The thin dotted, solid, and dashed lines give the logarithm of
the surface abundance of carbon, nitrogen, and oxygen, respectively,
relative to their initial values (left scale).  The thick grey line
shows the surface mass fraction of helium (right scale).  Model sequences have
been computed for $\fc=0.01$, $0.025$, $0.03$, $1/30$, $0.04$ and
$0.05$ with the KEPLER code.
\lFig{fc-gauge}}
\end{figure}

\clearpage

\begin{figure}
\epsscale{1.0}
\plotone{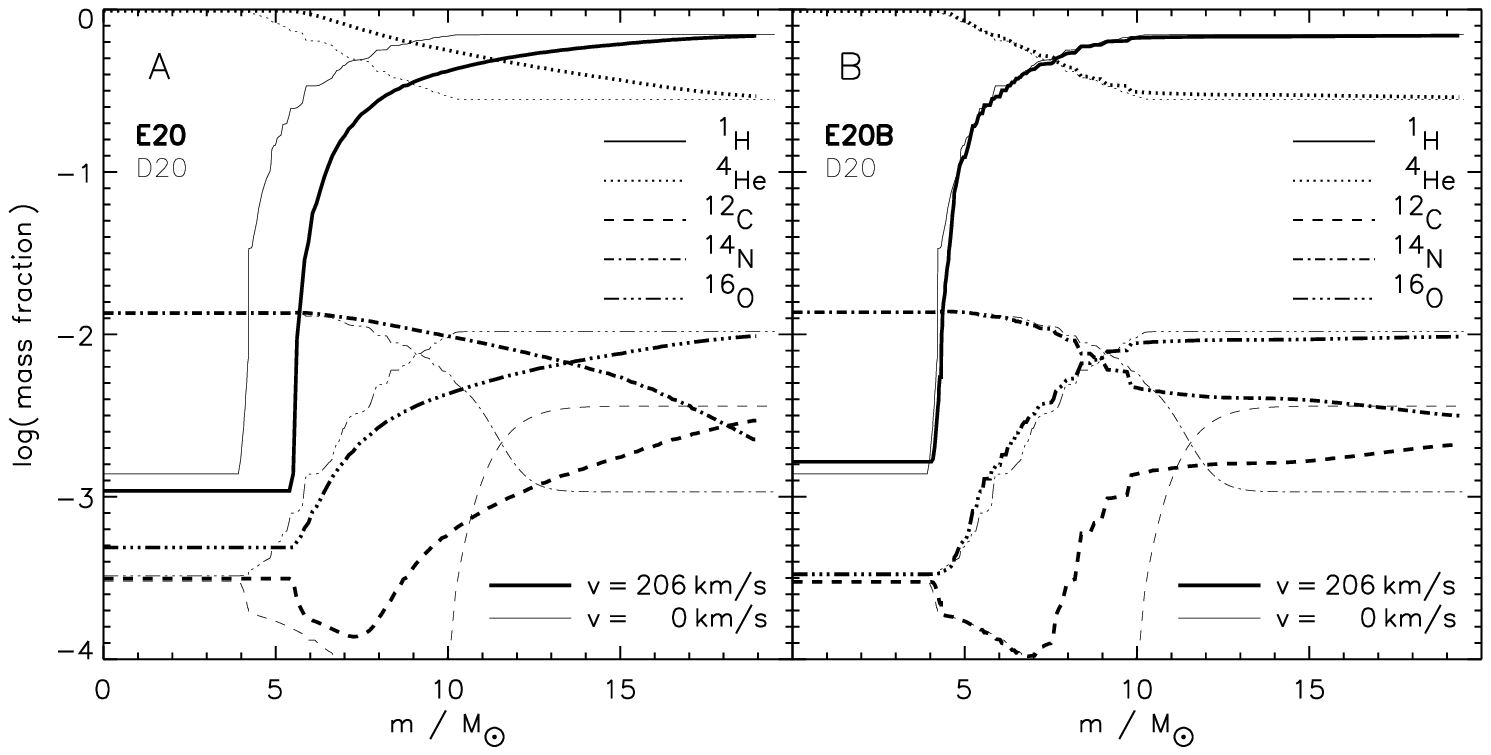}
\caption{
Mass fractions of different isotopes as a function of the mass
coordinate $m$ at core hydrogen exhaustion.  Compared are the chemical
structures of rotating (thick lines) and a non-rotating (thin lines;
same in both Panels; {\ModA{D20}}) $20\,\Msun$ models.  The rotating
models have a ZAMS equatorial rotational velocity of $\sim 200\,\kms$.
\Pan{A}: 
{\ModA{E20}}, where rotationally induced mixing is {\emph{not}}
inhibited by $\mu$-gradients.
\Pan{B}: 
{\ModA{E20B}}, where rotationally induced mixing is 
inhibited by $\mu$-gradients ($\fmu=0.05$).
\lFig{m-X-20AB}}
\end{figure}

\clearpage

\begin{figure}
\epsscale{0.75}
\plotone{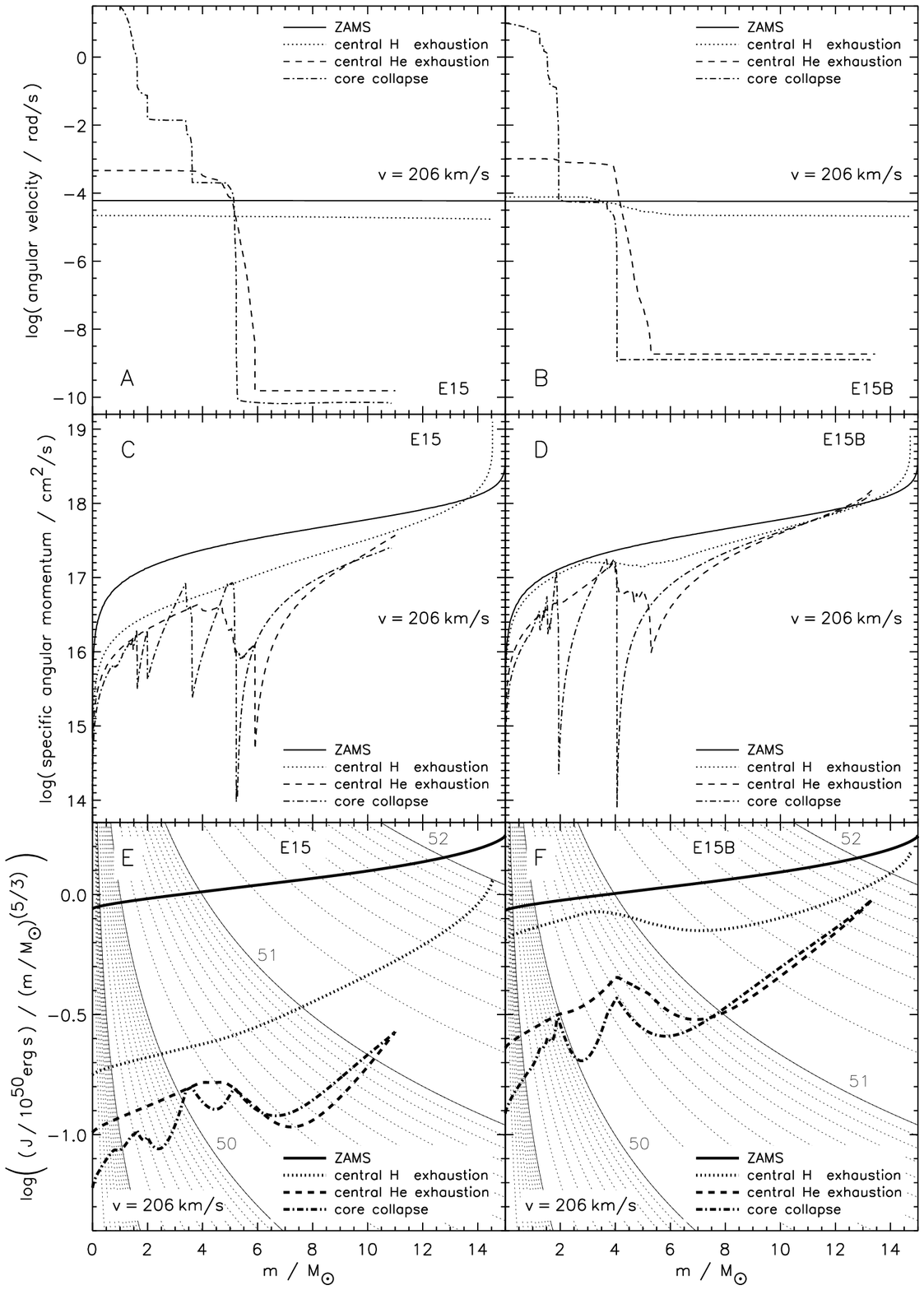}
\caption{
Angular velocity (\PanB{A}{B}), specific angular momentum
(\PanB{C}{D}), and integrated angular momentum, $\Jm=\int_0^m j(m')
\DD m'$, divided by $m^{5/3}$ (thick lines; {\PanB{E}{F}}) as a
function of the mass coordinate $m$ at different evolutionary stages
for two $15\,\Msun$ stars with a ZAMS equatorial rotational velocity
of $\sim 200\,\kms$.  {\PanC{A}{C}{E}} show {\ModA{E15}} (inefficient
$\mu$-barrier), and {\PanC{B}{D}{F}} {\ModA{E15B}} (efficient
$\mu$-barrier).
The thin lines in {\PanB{E}{F}} give a logarithmic scale of levels of
constant $J$, labeled with $\log\Brak{J/\ergs}$.
\lFig{m-wjJ53-E15AB}}
\end{figure}

\clearpage

\begin{figure}
\epsscale{0.75}
\plotone{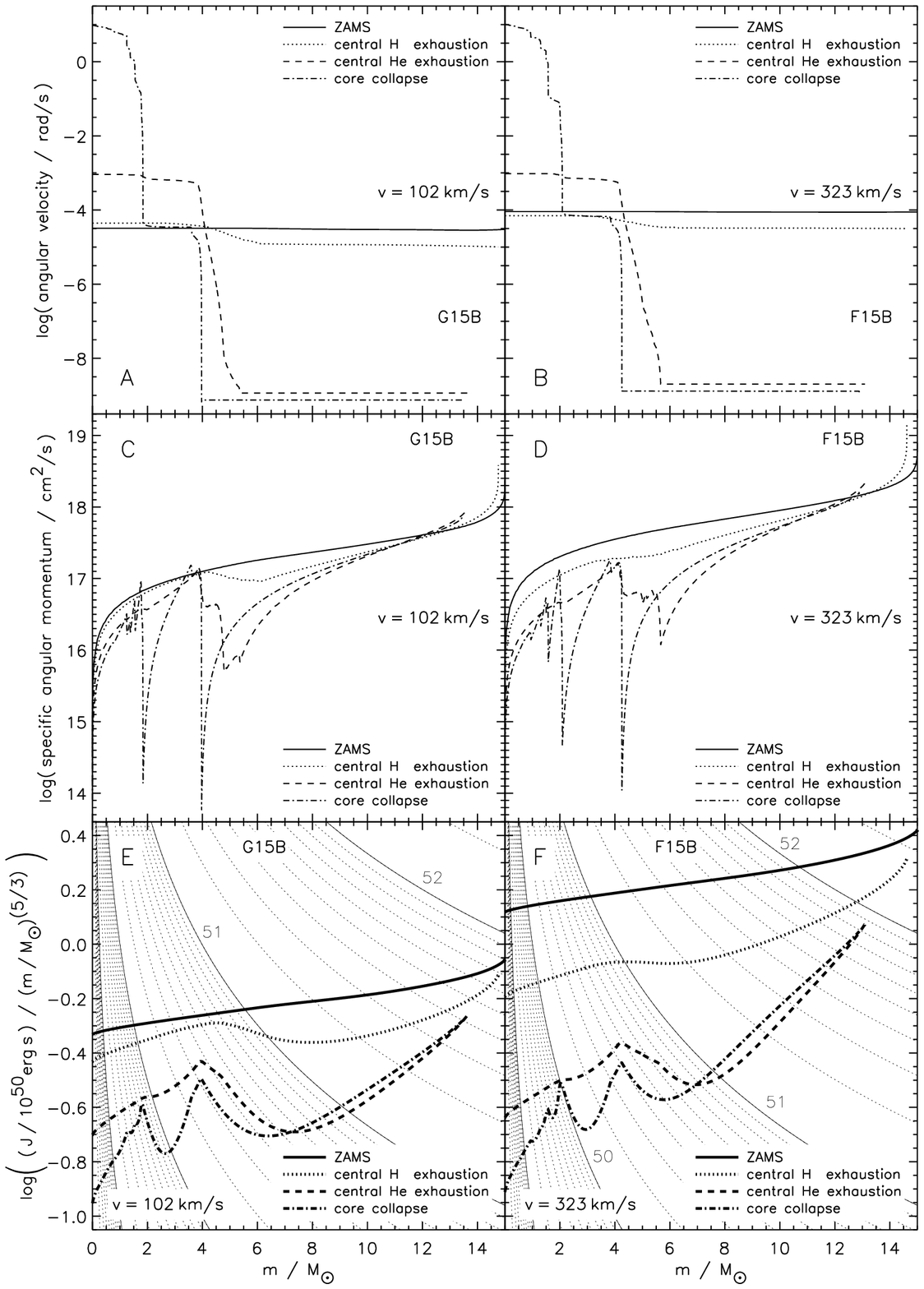}
\caption{ 
Angular velocity (\PanB{A}{B}), specific angular momentum
(\PanB{C}{D}), and integrated angular momentum, $\Jm=\int_0^m j(m')
\DD m'$, divided by $m^{5/3}$ (thick lines; {\PanB{E}{F}}) as a
function of the mass coordinate $m$ at different evolutionary stages
for two $15\,\Msun$ stars.  The evolution of stars with a ZAMS
equatorial rotational velocity of $\sim 100\,\kms$ (left;
{\ModA{G15B}}) and and $\sim 300\,\kms$ (right; {\ModA{F15B}}) are
depicted.  In both models, the effect of $\mu$-gradients on rotational
mixing is taken into account.
The thin lines in {\PanB{E}{F}} give a logarithmic scale of levels of
constant $J$, labeled with $\log\Brak{J/\ergs}$.
\lFig{m-wjJ53-GF15B}}
\end{figure}

\clearpage

\begin{figure}
\epsscale{0.44}
\plotone{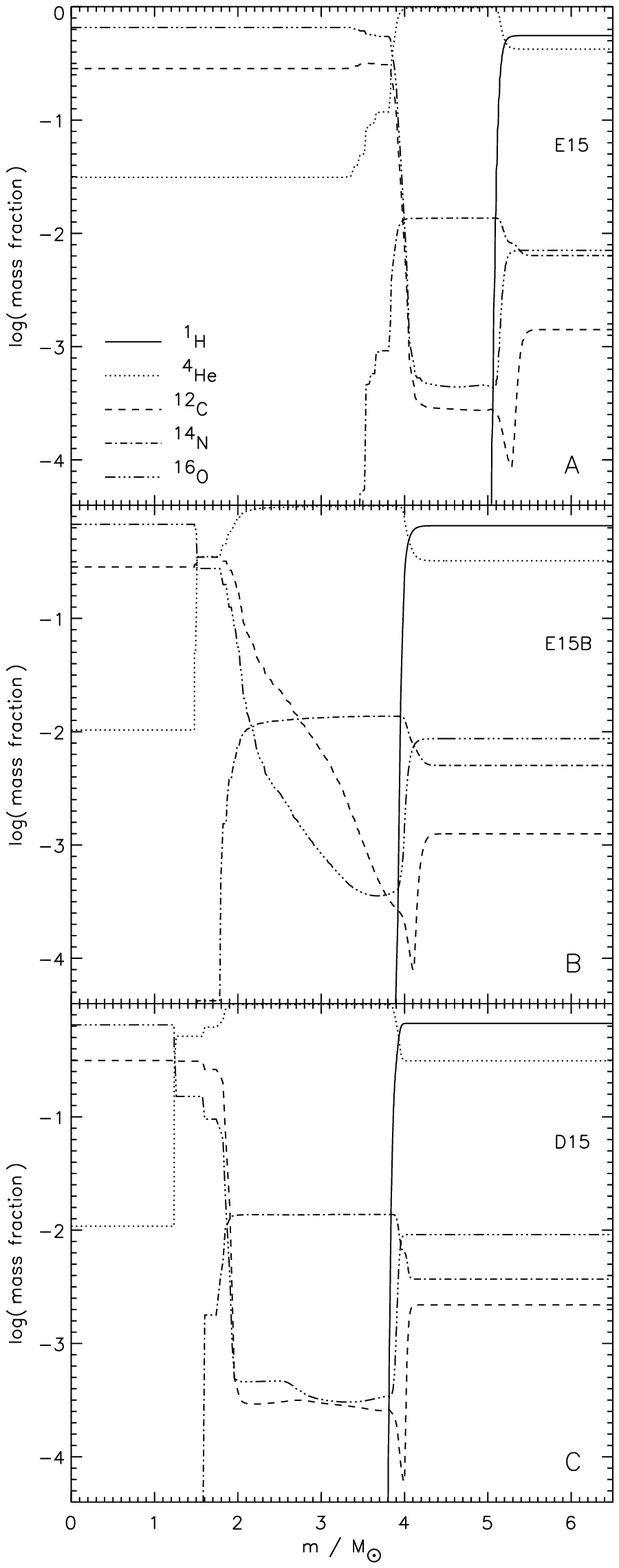}
\caption{
Mass fraction of the dominant species of three $15\,\Msun$ models at
the end of central helium burning as a function of the mass coordinate
$m$.  {\Pans{A}} and {\pan{B}} refer to {\ModB{E15}{E15B}},
respectively, which have a ZAMS equatorial rotational velocity of
$\sim200\,\kms$, but different assumptions for the parameters of
rotationally induced mixing.  {\Pan{C}} shows the non-rotating
{\ModA{D15}}.
\lFig{m-X-DE15AB}}
\end{figure}

\clearpage

\begin{figure}
\epsscale{1.0}
\plotone{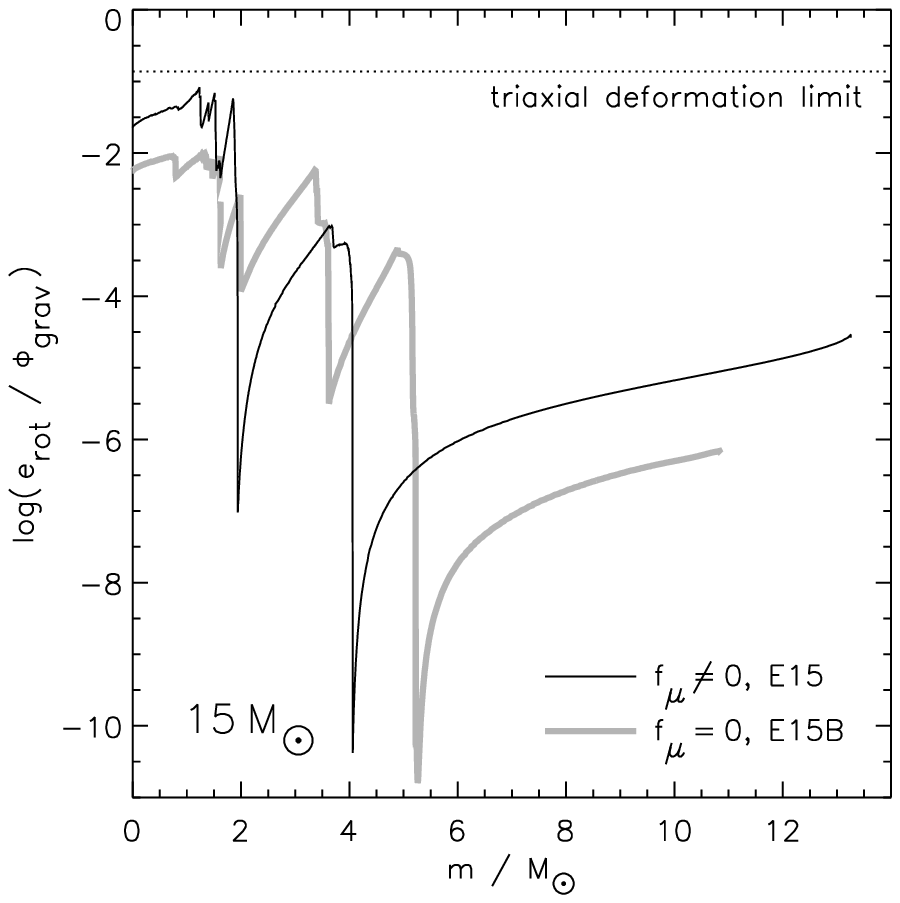}
\caption{ 
Ratio of integrated rotational energy
$\Erot(m)=\frac{1}{2}\int_0^m\omega(m')j(m')\DD m'$ to integrated
gravitational potential energy $\Epot(m)=\int_0^mGm'/r(m')\DD m'$ as a
function of the mass coordinate $m$ for different $15\,\Msun$ stars.
The thin solid line corresponds to {\ModA{E15B}}, the thick grey line
refers to {\ModA{E15}}.
The dotted line shows the limit for secular
instability to triaxial deformations in McLaurin spheroids
{\Cite{OB73}}.  
\lFig{m-phi}}
\end{figure}

\clearpage

\begin{figure}
\epsscale{1.0}
\plotone{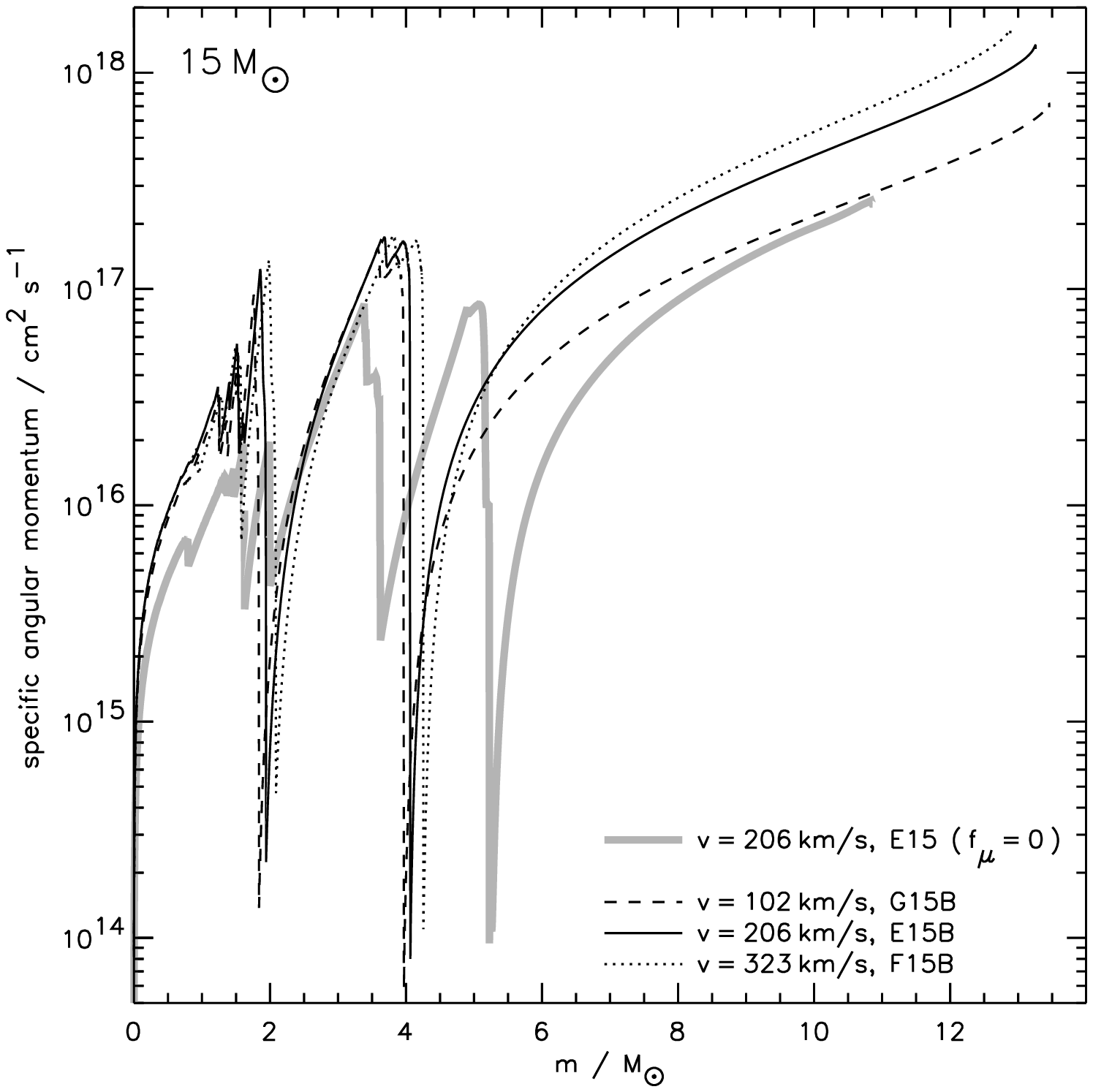}
\caption{ 
Specific angular momentum as a function of the mass coordinate $m$ for
$15\,\Msun$ stars at the onset of core collapse.
The thin lines show {\ModC{G15B}{E15B}{F15B}} with ZAMS
rotational velocity of $\sim 100\,\kms$ (dotted), $\sim 200\,\kms$
(solid), and $\sim 300\,\kms$ (dashed), respectively.  The thick grey
line shows the {\ModA{E15}}. 
\lFig{jj15B}}
\end{figure}

\clearpage

\begin{figure}
\epsscale{1.0}
\plotone{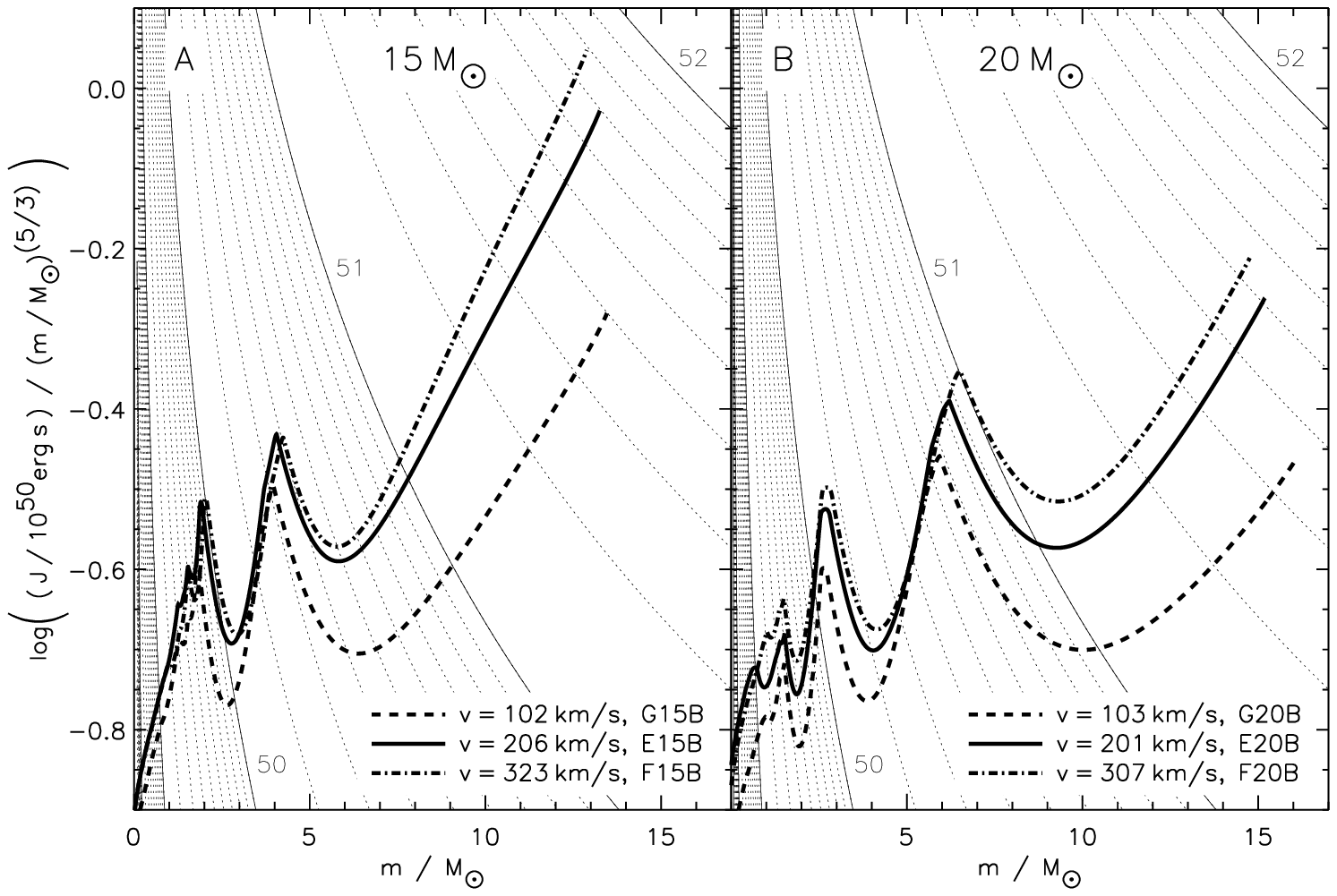}
\caption{ 
Integrated angular momentum $\Jm=\int_0^m j(m') \DD m'$ divided by
$m^{5/3}$ as a function of the mass mass coordinate $m$ for different
ZAMS rotational velocities (thick lines) at core collapse.  The two
Panels show stars of initial masses of $15\,\Msun$ (\Pan{A}:
{\ModC{G15B}{E15B}{F15B}}) and $20\,\Msun$ (\Pan{B}:
{\ModC{G20B}{E20B}{F20B}}).
The thin lines have the same meaning as in
{\FigBb{E}{m-wjJ53-E15AB}{F}}.  
\lFig{m-J53-presn-GEF}}
\end{figure}

\clearpage

\begin{figure}
\epsscale{1.0}
\plotone{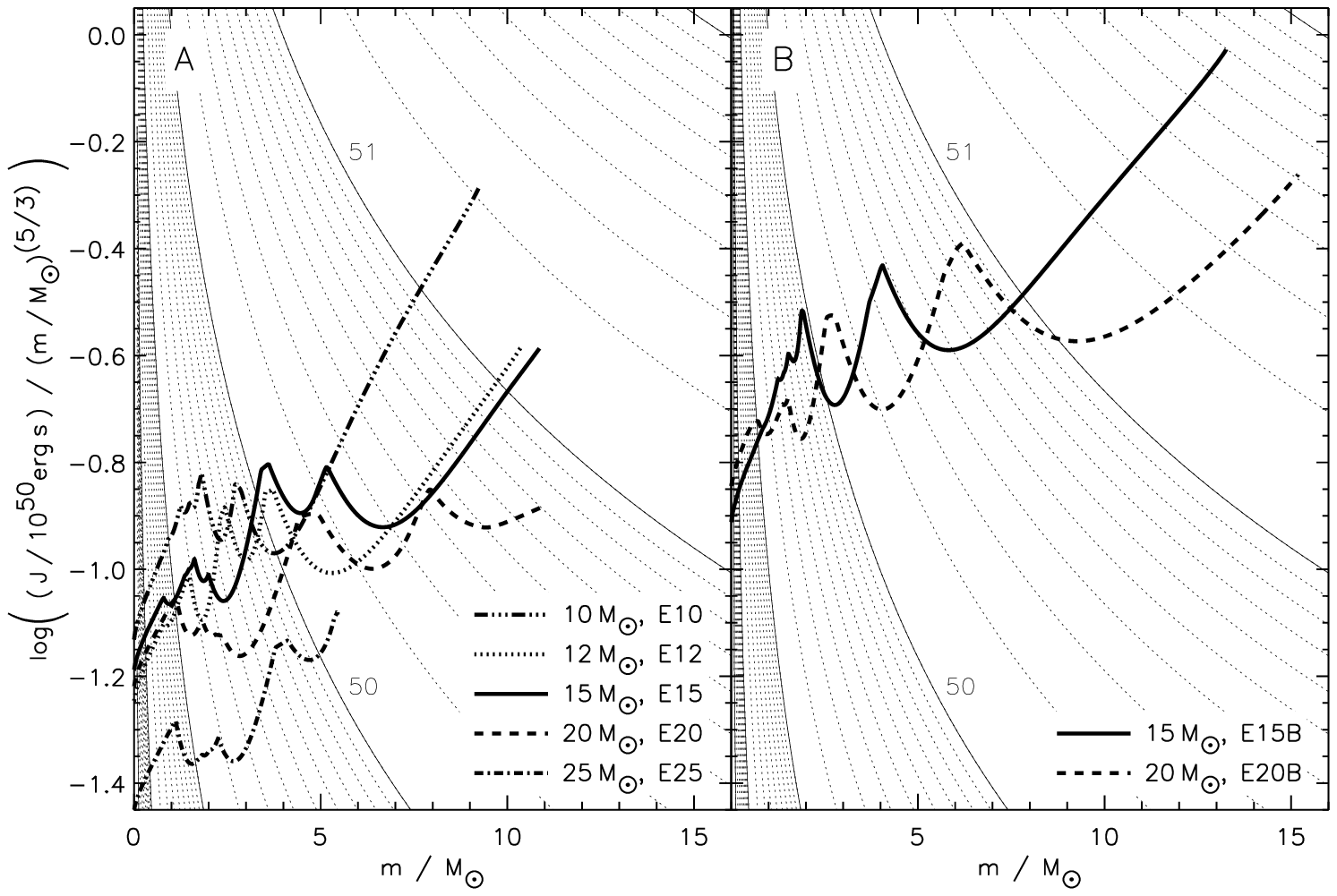}
\caption{ 
Integrated angular momentum $\Jm=\int_0^m j(m') \DD m'$ divided by
$m^{5/3}$ as a function of the mass mass coordinate $m$ for different
initial masses (thick lines) at core collapse.  The stars have a
ZAMS equatorial rotational velocity of $\sim200\,\kms$.  {\Pan{A}}
displays the {\ModE{E10}{E12}{E15}{E20}{E25}}.  {\Pan{B}} gives the
{\ModB{E15B}{E20B}}.
The thin lines have the same meaning as in {\FigBb{E}{m-wjJ53-E15AB}{F}}.
\lFig{m-J53-presn}}
\end{figure}

\clearpage

\begin{figure}
\epsscale{1.0}
\plotone{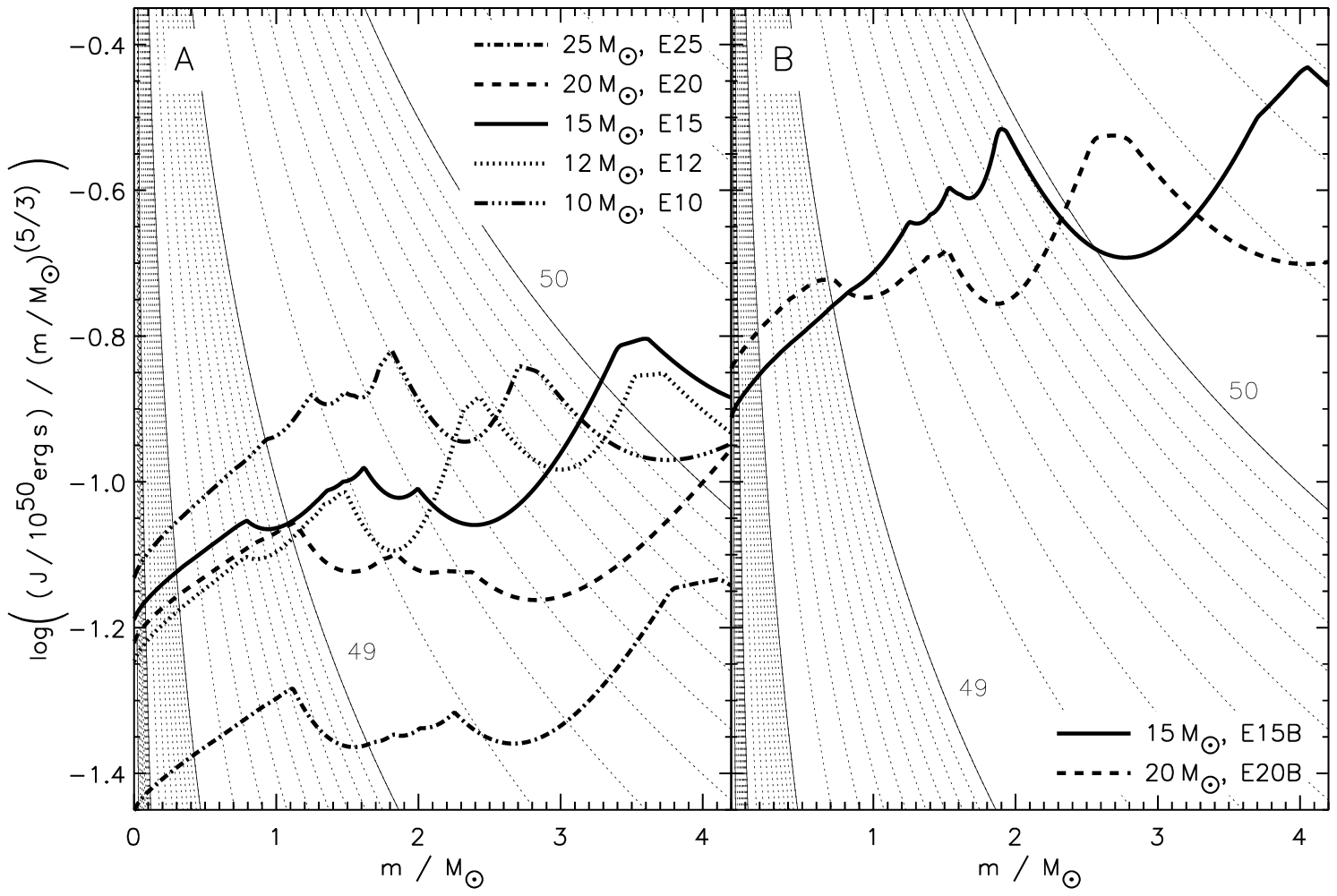}
\caption{ 
Magnification of the innermost $4.2\,\Msun$ of \Fig{m-J53-presn}.
\lFig{m-J53-presn4}}
\end{figure}

\clearpage

\begin{figure}
\epsscale{1.0}
\plotone{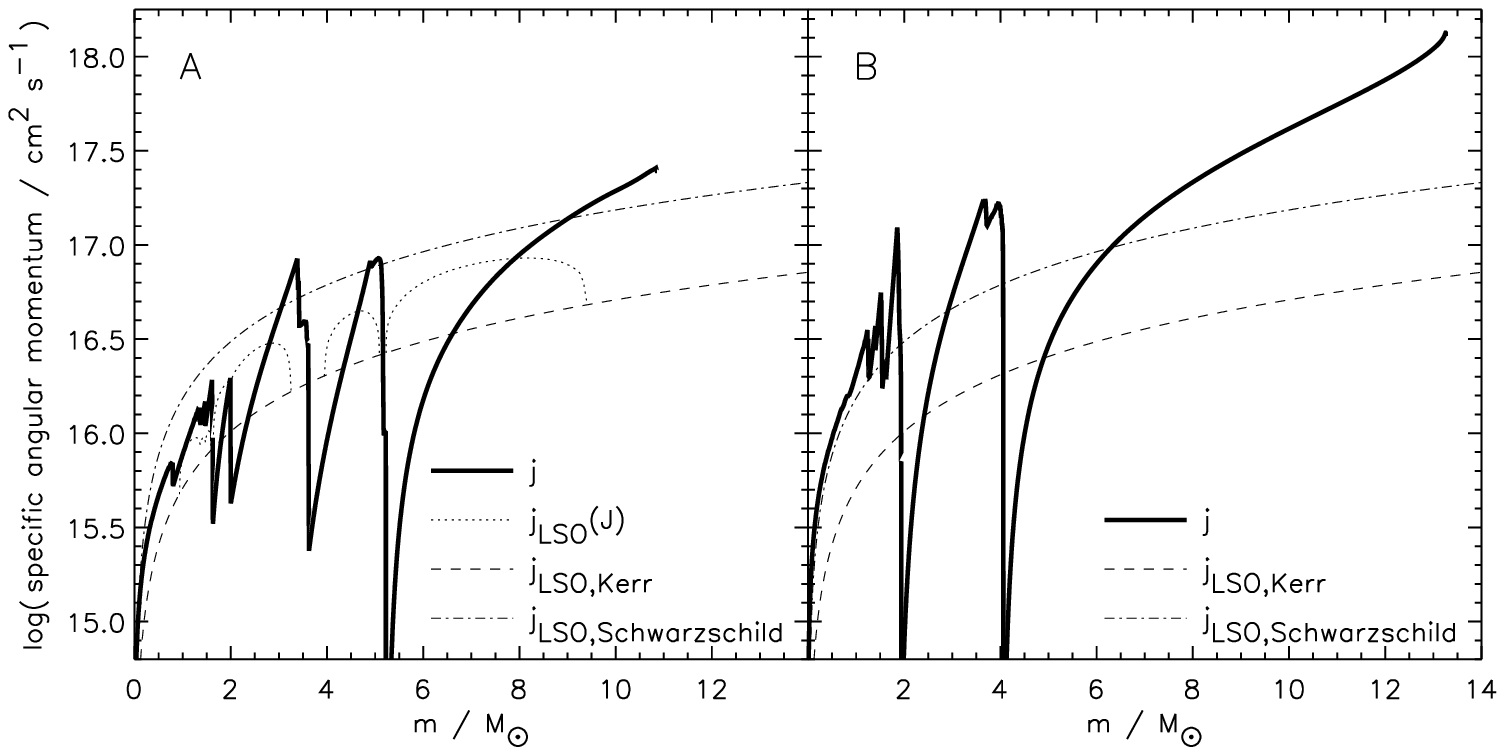}
\caption{ 
Specific angular momentum $j$ in the star (thick line) as a function
of the mass mass coordinate $m$ at the onset of core collapse of a
$15\,\Msun$ star with a ZAMS equatorial rotational velocity of $\sim
200\,\kms$.  {\Pan{A}} and {\pan{B}} show {\ModB{E15}{E15B}},
respectively, and compare the the resulting profiles of stars with
different assumptions about the parameters of rotationally induced
mixing.  
The dash-dotted line gives the specific angular momentum needed to get
into the last stable orbit around a non-rotating Schwarzschild black
hole of (rest-)mass equal to the mass coordinate.  The dashed line
give the same but for a maximum-rotating Kerr black hole
(spin-parameter $Jc/(Gm)=:a=m$).  If assumed that all (rest-)mass
below a given $m$ has fallen into a black hole and added its angular
momentum to it, the dotted line results for the $j$ needed to get into
the last stable orbit; where this approximation would lead to values
of $a>m$ the curve is truncated to the Kerr-limit.  For the
{\ModA{E15B}} shown in {\Pan{B}} this is the case everywhere.  A
derivation of these limits can be found in {\cite{ST83}} or
{\cite{Nov97}}.  It has to be noted, however, that this plot is for
giving a measure of the amount of angular momentum in the pre-collapse
model only.  This shall not imply that these stars actually will form
black holes.
\lFig{m-jLSO-E15AB}}
\end{figure}

\clearpage
\begin{figure}
\epsscale{0.825}
\plotone{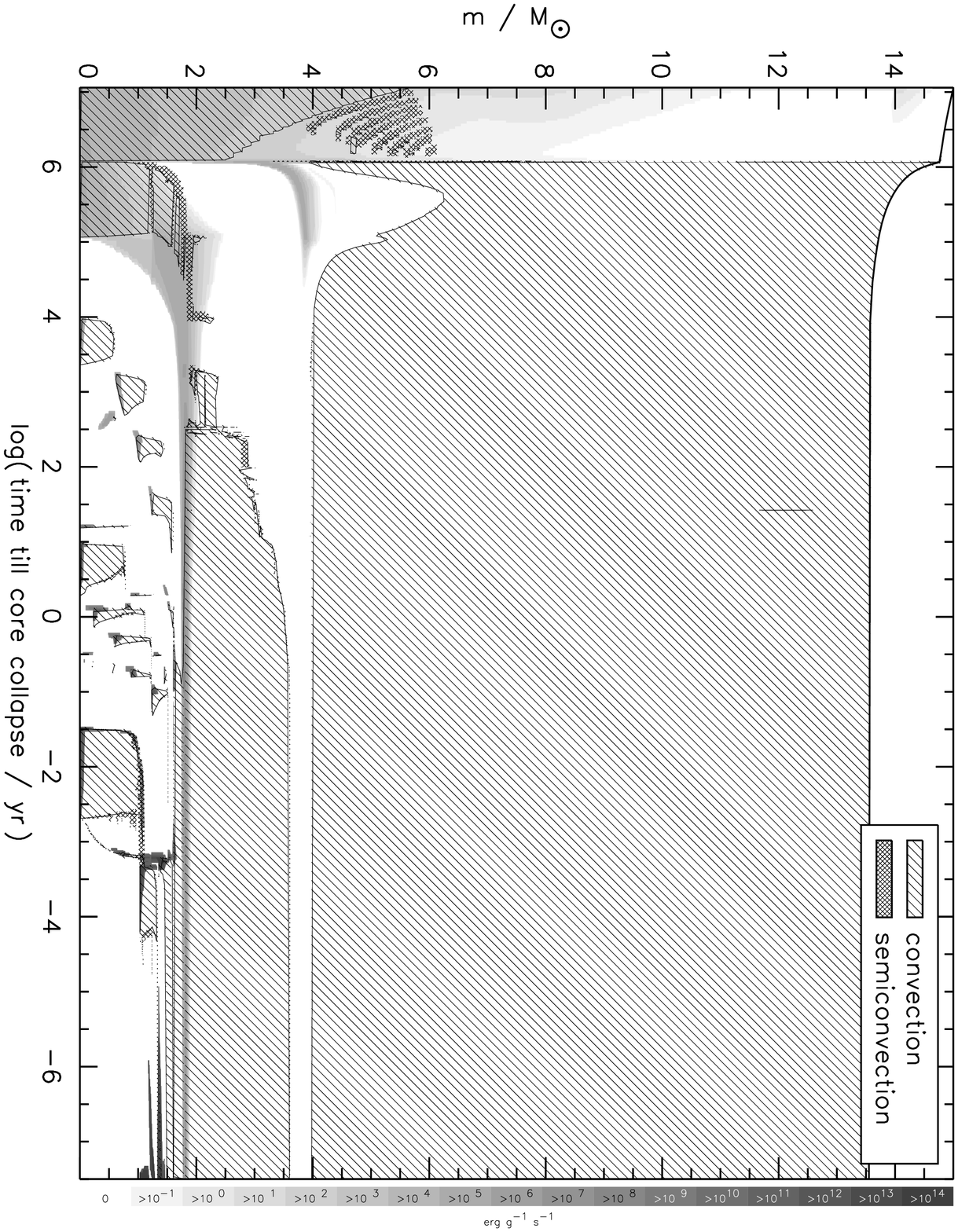}
\caption{
Evolution of {\ModA{D15}} until core collapse.  Convection and net
nuclear energy generation.  See {\App{ExplConv}} for details.
}
\lFig{D15cnv}
\end{figure}


\clearpage
\begin{figure}
\epsscale{0.825}
\plotone{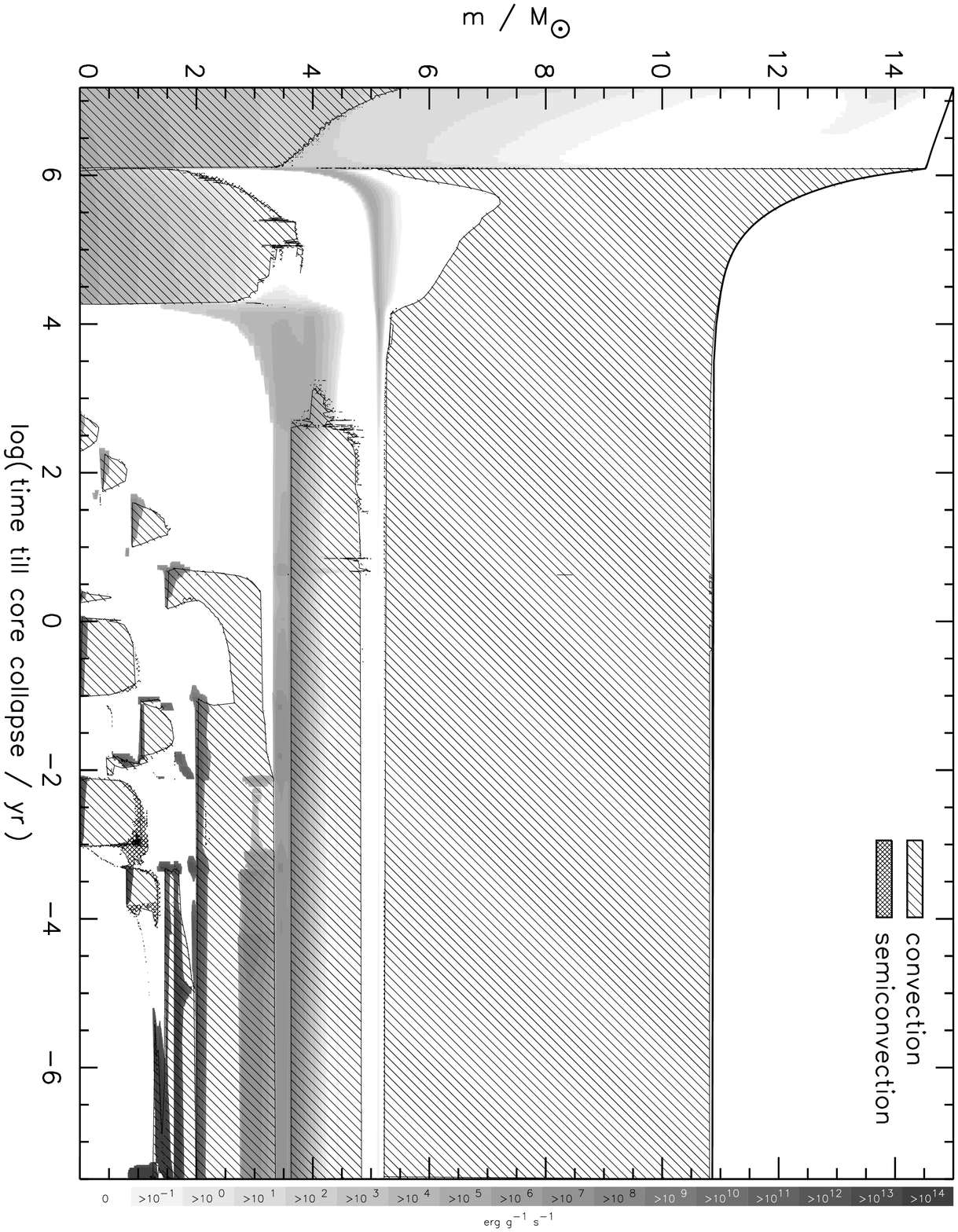}
\caption{
Evolution of {\ModA{E15}} until core collapse.  Convection and net
nuclear energy generation.  See {\App{ExplConv}} for details.
}
\lFig{E15cnv}
\end{figure}


\clearpage
\begin{figure}
\epsscale{0.825}
\plotone{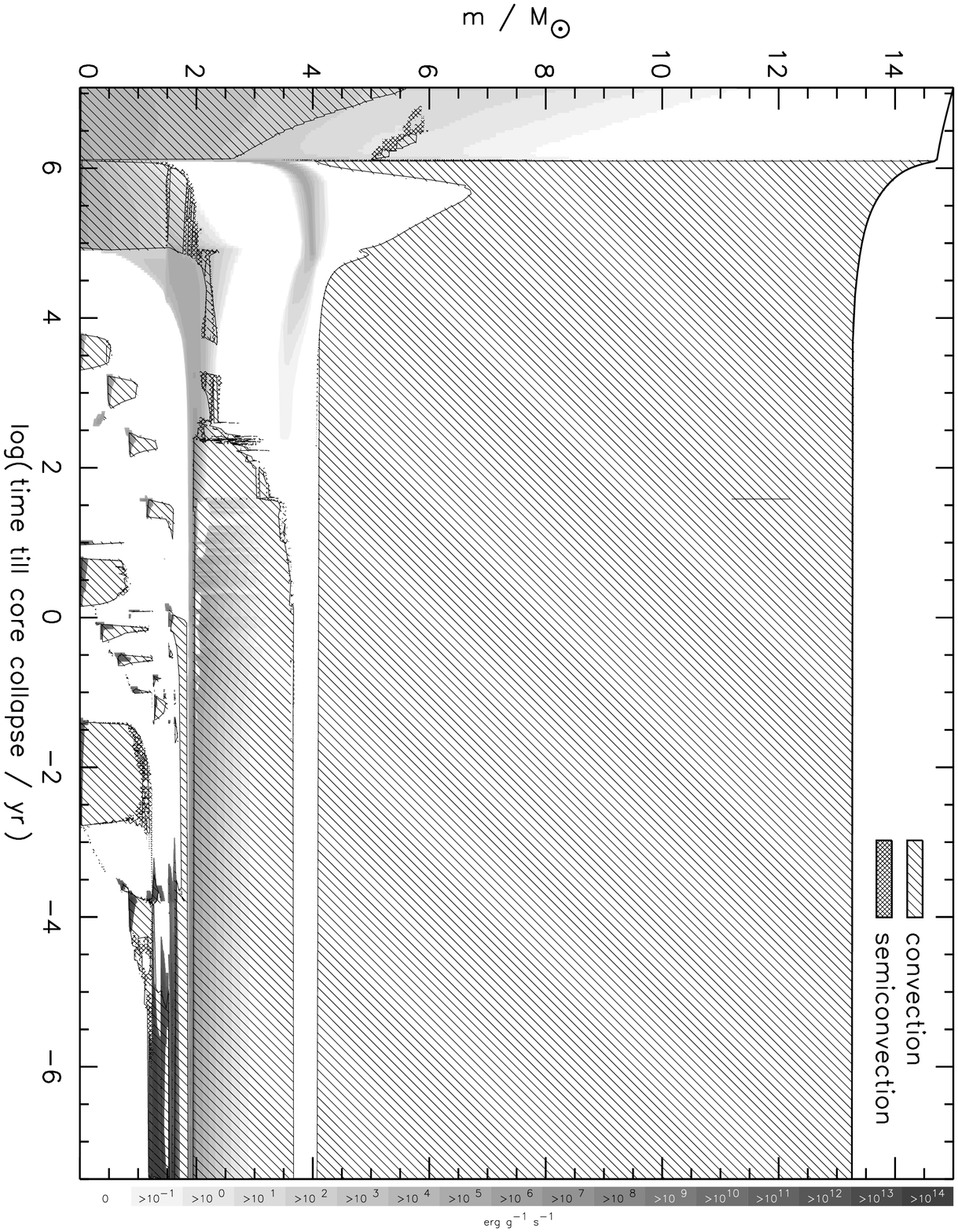}
\caption{
Evolution of {\ModA{E15B}} until core collapse.  Convection and net
nuclear energy generation.  See {\App{ExplConv}} for details.
\lFig{E15Bcnv}}
\end{figure}


\end{document}